\documentstyle[12pt]{article}
\input epsfig.sty

\newcommand{\junk}[1]{}

\font\blackboard=msbm10 at 11pt
\font\blackboards=msbm7
\font\blackboardss=msbm5
\newfam\black
\textfont\black=\blackboard
\scriptfont\black=\blackboards
\scriptscriptfont\black=\blackboardss
\def\bb#1{{\fam\black\relax#1}}

\def\bz{\bb Z}
\def\bc{\bb C}
\def\br{\bb R}

\newcommand{\ba}{\begin{array}}
\newcommand{\ea}{\end{array}}
\newcommand{\be}{\begin{equation}}
\newcommand{\ee}{\end{equation}}
\newcommand{\bea}{\begin{eqnarray}}
\newcommand{\eea}{\end{eqnarray}}
\newcommand{\beas}{\begin{eqnarray*}}
\newcommand{\eeas}{\end{eqnarray*}}

\def\half{{1 \over 2}}
\def\identity{{\rlap{1} \hskip 1.6pt \hbox{1}}}

\def\laplace{{\kern1pt\vbox{\hrule height 1.2pt\hbox{\vrule width 1.2pt\hskip
  3pt\vbox{\vskip 6pt}\hskip 3pt\vrule width 0.6pt}\hrule height 0.6pt}
  \kern1pt}}
\def\scriptlap{{\kern1pt\vbox{\hrule height 0.8pt\hbox{\vrule width 0.8pt
  \hskip2pt\vbox{\vskip 4pt}\hskip 2pt\vrule width 0.4pt}\hrule height 0.4pt}
  \kern1pt}}

\def\roughly#1{\raise.3ex\hbox{$#1$\kern-.75em\lower1ex\hbox{$\sim$}}}

\def\tr{{\rm Tr} \,}
\def\str{{\rm STr} \,}

\def\hatt{{\hat{\cal T}}}
\def\hj{{\hat{\cal J}}}
\def\hm{{\hat{\cal M}}}

\def\st{{\tilde{\cal T}}}
\def\sj{{\tilde{\cal J}}}
\def\sm{{\tilde{\cal M}}}
\def\itt{{\cal  T}}
\def\ijj{{\cal J}}
\def\imm{{\cal M}}

\def\dx{{\dot{X}}}

\newcommand{\NP}{{\em Nucl.\ Phys.\ }}
\newcommand{\PL}{{\em Phys.\ Lett.\ }}
\newcommand{\PR}{{\em Phys.\ Rev.\ }}
\newcommand{\PRP}{{\em Phys.\ Rep.\ }}
\newcommand{\CMP}{{\em Comm.\ Math.\ Phys.\ }}
\newcommand{\MPL}{{\em Mod.\ Phys.\ Lett.\ }}
\newcommand{\PRL}{{\em Phys.\ Rev.\ Lett.\ }}

\newcommand{\JMP}{{\em J.\ Math.\ Phys.\ }}

\newcommand{\gone}[1]{}

\begin{document}
\pagestyle{plain}
\setcounter{page}{1}

\baselineskip16pt

\begin{titlepage}
\begin{flushright}
MIT-CTP-2894\\
hep-th/0002016
\end{flushright}

\vspace*{0.5in}
\begin{center}

{\Large \bf The M(atrix) model of M-theory}\

\end{center}

\vspace{0.5in}

\begin{center}

{\bf Washington Taylor IV}

\vspace{3mm}
{\small \sl Center for Theoretical Physics} \\
{\small \sl MIT, Bldg. 6-306} \\
{\small \sl Cambridge, MA 02139, U.S.A.} \\
{\small \tt wati@mit.edu}\\
\end{center}

\vspace{0.5in}

\begin{abstract}
These lecture notes give a pedagogical and (mostly) self-contained
review of some basic aspects of the Matrix model of M-theory.  The
derivations of the model as a regularized supermembrane theory and as
the discrete light-cone quantization of M-theory are presented.  The
construction of M-theory objects from matrices is described, and
gravitational interactions between these objects are derived using
Yang-Mills perturbation theory.  Generalizations of the model to
compact and curved space-times are discussed, and the current status
of the theory is reviewed.
\end{abstract}

\vspace{0.5in}

\begin{center}
Lecture notes for NATO school\\
``Quantum Geometry''\\
Akureyri, Iceland\\
August 10-20, 1999
\end{center}
\vspace{0.2in}

\vspace{1cm}
\end{titlepage}
\newpage

\section{Introduction}
\label{sec:introduction}

This series of lectures describes the matrix model of M-theory, also
known as M(atrix) Theory.  Matrix theory is a supersymmetric quantum
mechanics theory with matrix degrees of freedom.  It has been known
for over a decade \cite{Goldstone-Hoppe,dhn} that matrix theory arises
as a regularization of the 11D supermembrane theory in light-front
gauge.  It was conjectured in 1996 \cite{BFSS} that when the size of
the matrices is taken to infinity this theory gives a microscopic
second-quantized description of M-theory in light-front coordinates.

These lectures focus on some basic aspects of matrix theory.  We begin
by describing in some detail the two alternative definitions of the
theory in terms of a quantized and regularized supermembrane theory
and as a compactification of M-theory on a lightlike circle.  Given
these definitions of the theory, we then focus on the question of
whether the physics of M-theory and 11-dimensional supergravity can be
described constructively using finite size matrices.  We show that all
the objects of M-theory, including the supergraviton, membrane and
5-brane can be constructed explicitly from configurations of matrices,
although these results are not yet complete in the case of the
5-brane.  We then turn to the gravitational interactions between these
objects, and review what is known about the connection between
perturbative calculations in the matrix quantum mechanics theory and
supergravity interactions.  In the last part of the lectures, some
discussion is given of how the matrix theory formalism may be
generalized to describe compact or curved space-times.

Previous reviews of matrix theory and related work have appeared in
\cite{banks-review,Susskind-review,WT-Trieste,Obers-Pioline,Nicolai-Helling,deWit-review,Banks-TASI}.

In Section \ref{sec:quantized-membrane} we show how matrix theory can
be derived from the light-front quantization of the supermembrane
theory in 11 dimensions.  We discuss in Section \ref{sec:BFSS} the
conjecture of Banks, Fischler, Shenker and Susskind that matrix theory
describes light-front M-theory in flat space, and we review an
argument of Seiberg and Sen showing that finite $N$ matrix theory
describes the discrete light-cone quantization (DLCQ) of M-theory.  In
Section \ref{sec:matrix-objects} we show how the objects of M-theory
(the supergraviton, supermembrane and M5-brane) can be described in
terms of matrix theory degrees of freedom.  Section
\ref{sec:matrix-interactions} reviews what is known about the
interactions between these objects.  We discuss the problem of
reproducing N-body interactions in 11D classical supergravity from
matrix theory, beginning with two-body interactions in the linearized
theory and then discussing many-body interactions and
nonlinear terms as well as quantum corrections to the supergravity
theory.  Section \ref{sec:matrix-general-background} contains a
discussion of the problems of formulating matrix theory on a compact or
curved background geometry.  Finally, we conclude in section
\ref{sec:matrix-summary} with a summary of the current state of
affairs and the outlook for the future of this theory.

Even if in the long run matrix theory turns out not to be the most
useful description of M-theory, there are many features of this theory
which make it well worth studying.  It is the simplest example of a
quantum supersymmetric gauge theory which seems to correspond to a
theory of gravity in a fixed background in some limit.  It is the only
known example of a well-defined quantum theory which has been shown
explicitly to give rise to
long-range interactions which agree with gravity at the linearized
level and which also contain some nonlinearity.  Finally, it provides
simple examples of many of the remarkable connections between D-brane
physics and gauge theory, giving intuition which may be applicable to
a wide variety of situations in string theory and M-theory.

\section{Matrix theory from the quantized supermembrane}
\label{sec:quantized-membrane}

In this section we show that supersymmetric matrix quantum mechanics
arises naturally as a regularization of the supermembrane action in 11
dimensions.  
We begin our discussion with some motivational remarks.  

In retrospect, the supermembrane is a natural place to
begin when trying to construct a microscopic description of M-theory.
There are several distinct 10-dimensional supersymmetric theories of
gravity.  These theories are well-defined classically but, as with all
theories of gravity, are difficult to quantize directly.  Each of these
theories has a bosonic antisymmetric 2-form tensor field $B_{\mu
\nu}$.  This field is analogous to the 1-form field $A_\mu$ of
electromagnetism, but carries an extra space-time index.  Each of
these 10D supergravity theories admits a classical stringlike black
hole solution which is ``electrically'' charged under the 2-form
field, in the sense that the two-dimensional world-volume   $\Sigma$ of
the string couples to the $B$ field through a term 
\[
\int_\Sigma B_{\mu \nu}  \epsilon^{ab} (\partial_a X^\mu)  (\partial_b X^\nu).
\]
where $X^\mu$ are the embedding functions of the string world-volume
in 10 dimensions.  This is the higher-dimensional analog of the usual
coupling of a charged particle to a gauge field through $\int A_\mu
\dot{X}^\mu$.

The quantization of strings in 10-dimensional background geometries
can be carried out consistently in only a limited number of ways.
These constructions lead to the perturbative descriptions of the five
superstring theories known as the type I, IIA, IIB and heterotic $E_8
\times E_8$ and $SO(32)$ theories.  These quantum superstring theories
are first-quantized from the point of view of the target space---that
is, a state in the string Hilbert space corresponds to a single
particle-like state in the target space consisting of a single string.
Although the quantized string spectrum naturally contains states
corresponding to quanta of the supergravity fields (including the
NS-NS field $B_{\mu \nu}$), it is not possible to give a simple
description in terms of the string Hilbert space for extended objects
such as D-branes and the NS 5-brane.  These objects are essentially
nonperturbative phenomena in the superstring theories.

One of the most important developments in the last few years has been the
discovery of a network of duality symmetries which relates the five
superstring theories to each other and to 11-dimensional supergravity.
Of these six theories, the quantized superstring gives a microscopic
description of the five 10-dimensional theories.  It has been
hypothesized that there is a microscopic 11-dimensional theory, dubbed
M-theory, underlying this structure which reduces in the low-energy
limit to 11D supergravity \cite{Witten-various}.  To date, however, a
precise description of this theory is lacking.  Such a theory cannot
be described by a quantized string since there is no antisymmetric
2-form field in the 11D supergravity multiplet and hence no stringlike
solution of the gravity equations.  The 11D supergravity theory
contains, however, an antisymmetric 3-form field $A_{IJK}$, and the
classical theory admits membrane-like solutions which couple
electrically to this field.  It is easy to imagine that a microscopic
description of M-theory might be found by quantizing this
supermembrane.  This idea was explored extensively in the 80's, when
it was first realized that a consistent classical theory of a
supermembrane could be realized in 11 dimensions.  At that time, while
no satisfactory covariant quantization of the membrane theory was
found, it was shown that the supermembrane could be quantized in
light-front coordinates.  In fact, an elegant regularization of this
theory was suggested by Goldstone and Hoppe \cite{Goldstone-Hoppe} in
1982.  They showed that for the bosonic membrane the regularized
quantum theory is a simple quantum-mechanical theory of $N \times N$
matrices which leads to the membrane theory in the large $N$ limit.
This approach was generalized to the supermembrane by de Wit, Hoppe
and Nicolai \cite{dhn}, who showed that the regularized supermembrane
theory is precisely the supersymmetric matrix quantum mechanics now
known as Matrix Theory.  A remarkable feature of the quantum
supermembrane theory is that unlike the quantized string theories, the
membrane theory automatically gives a second quantized theory from the
point of view of the target space.  This issue will be discussed in
more detail in Section 2.

In this section we describe in some detail how matrix theory arises
from the quantization of the supermembrane.  In
\ref{sec:string-review} we review how the bosonic string is quantized
in the light-front formalism.  This will be a useful reference point
for our discussion of membrane quantization.  In
\ref{sec:bosonic-membrane} we describe the theory of the relativistic
bosonic membrane in flat space.  The light-front description of this
theory is discussed in \ref{sec:membrane-light-front}, and the matrix
regularization of the theory is described in
\ref{sec:matrix-regularization}.  In \ref{sec:bosonic-general} we
discuss briefly the description of the bosonic membrane moving in a
general background geometry.  In \ref{sec:supermembrane} we extend the
discussion to the supermembrane.  We discuss the supermembrane in an
arbitrary background geometry.  We discuss the $\kappa$-symmetry of
the supermembrane theory which leads, even at the classical level, to
the condition that the background geometry satisfies the classical 11D
supergravity equations of motion.  The matrix theory Hamiltonian is
derived from the regularized supermembrane theory. The problem of
finding a covariant membrane quantization is discussed in
\ref{sec:covariant-membrane}.

The material in this section roughly follows the original papers
\cite{Goldstone-Hoppe,dhn,bst}.  Note, however, that the original
derivation of the matrix quantum mechanics theory was done in the
Nambu-Goto-type membrane formalism, while we use here the
Polyakov-type approach.
We only consider closed membranes in the discussion here; little is
known about the open membrane which must end on the M-theory 5-brane,
but it would be very interesting to generalize the discussion here to
the open membrane.

\subsection{Review of light-front string}
\label{sec:string-review}

We begin with a brief review of the bosonic string.  This will be a
useful model to compare with
in our discussion of the supermembrane.

The Nambu-Goto action for the relativistic bosonic string moving in a
flat background space-time is
\begin{equation}
S = -T_s  \int d^2 \sigma \sqrt{-\det h_{ab}}
\label{eq:Nambu-Goto}
\end{equation}
where  $T_s = 1/(2 \pi \alpha')$ and
\begin{equation}
h_{ab} = \partial_a X^\mu \partial_b X_\mu.
\label{eq:induced-metric}
\end{equation}

It is convenient to use the Polyakov formalism in which an auxiliary
world-sheet metric $\gamma$ is introduced
\begin{equation}
S = -\frac{1}{4 \pi \alpha'}  \int d^2 \sigma \sqrt{-\gamma}
\gamma^{ab} \partial_a X^\mu \partial_b X_\mu
\label{eq:Polyakov}
\end{equation}
Solving the equation of motion for $\gamma_{ab}$ leads to
\begin{equation}
\gamma_{ab} = h_{ab} = \partial_a X^\mu \partial_b X_\mu
\end{equation}
and replacing this in (\ref{eq:Polyakov}) gives (\ref{eq:Nambu-Goto}).

The action (\ref{eq:Polyakov}) is simplified by going to the gauge
\begin{equation}
\gamma_{ab} = \eta_{ab}.
\label{eq:fix-metric}
\end{equation}
In this gauge we simply have the free field action
\begin{equation}
S =-\frac{1}{4 \pi \alpha'}  \int d^2 \sigma 
\eta^{ab} \partial_a X^\mu \partial_b X_\mu.
\label{eq:simple-action}
\end{equation}
The fields $X^\mu$ satisfy the equation of motion
$\Box X^\mu = 0$ and
are subject to the auxiliary Virasoro constraints
\begin{eqnarray}
\dot{X}^\mu (\partial X_\mu) & = &  0\\
\dot{X}^\mu \dot{X}_\mu & = &  -(\partial X^\mu) (\partial X_\mu) \nonumber
\end{eqnarray}
(we denote $\tau$ derivatives by a dot and $\sigma$ derivatives
by $\partial$).
Because this is a free theory it is fairly straightforward to
quantize.  The approaches to
quantizing this theory include the BRST and light-front formalisms.
The Virasoro constraints can be explicitly solved in light-front gauge
\begin{equation}
 X^+ (\tau, \sigma) = x^+ + p^+ \tau.
\label{eq:light-cone-gauge}
\end{equation}
In the classical theory we have
\begin{eqnarray}
\dot{X}^- & = &  \frac{1}{2p^+} \left(
\dot{X}^i \dot{X}^i + \partial X^i \partial X^i  \right)\\
\partial X^- & = &  \frac{1}{p^+}  \dot{X}^i \partial X^i \nonumber
\end{eqnarray}
The transverse degrees of freedom $X^i$ have Fourier modes with the
commutation relations of simple harmonic oscillators.  These are
straightforward to quantize.  The string spectrum is then given by the
usual mass-shell condition
\begin{equation}
M^2 = 2p^+ p^--p^ip^i = \frac{1}{ \alpha'}  (N -a)
\label{eq:light-cone-mass-shell-open}
\end{equation}

\subsection{The bosonic membrane theory}
\label{sec:bosonic-membrane}

We now discuss the relativistic bosonic membrane moving in an
arbitrary number $D$ of space-time dimensions.    The story begins in a
very similar fashion to the relativistic string.  We want to use a
Nambu-Goto-style action
\begin{equation}
S = -T  \int d^3 \sigma \sqrt{-\det h_{\alpha \beta}}
\label{eq:Nambu-Goto-membrane}
\end{equation}
where  $T$ is the membrane tension 
\begin{equation}
T = \frac{1}{ (2 \pi)^2 l_p^3} 
\end{equation}
and
\begin{equation}
h_{ \alpha \beta} = \partial_\alpha X^\mu \partial_\beta X_\mu
\label{eq:induced-metric-membrane}
\end{equation}
is the pullback of the metric to the three-dimensional membrane
world-volume, with coordinates $\sigma_\alpha, \alpha \in\{0, 1, 2\}$.  We will
use the notation $\tau = \sigma_0$ and use indices $a, b, \ldots$ to
describe ``spatial'' indices $a \in\{1, 2\}$ on the membrane world-volume.

We again wish to use a Polyakov-type formalism in which an auxiliary
world-sheet metric $\gamma_{\alpha\beta}$ is introduced
\begin{equation}
S = - \frac{T}{2}  \int d^3 \sigma \sqrt{-\gamma} \left(
\gamma^{\alpha \beta} \partial_\alpha X^\mu \partial_\beta X_\mu -1\right).
\label{eq:Polyakov-membrane}
\end{equation}
The need for the extra ``cosmological'' term arises from the absence
of scale invariance in the theory.  Computing the equations of motion
from varying $\gamma_{\alpha \beta}$, and using $\delta \sqrt{-\gamma}
=\frac{1}{2}
\sqrt{-\gamma} \gamma^{\alpha \beta} \delta \gamma_{\alpha \beta},
\delta \gamma^{\epsilon \phi} = -\gamma^{\alpha \epsilon} \gamma^{\phi
\beta} \delta \gamma_{\alpha \beta}$,
we get
\begin{equation}
-\gamma^{\alpha \gamma} \gamma^{\beta \delta} h_{\gamma \delta}
+\frac{1}{2} \gamma^{\alpha \beta} t-\frac{1}{2}\gamma^{\alpha \beta}
= 0
\end{equation}
where $t = \gamma^{\alpha \beta} h_{\alpha \beta}$.
Lowering all indices gives
\begin{equation}
\frac{1}{2} \gamma_{\alpha \beta} (t-1) = h_{\alpha \beta}
\end{equation}
or
\begin{equation}
\gamma_{\alpha \beta} = \frac{2h_{\alpha \beta}}{t-1} .
\end{equation}
Contracting indices gives
\begin{equation}
3 = \frac{2t}{t-1} 
\end{equation}
so $t = 3$ and
\begin{equation}
\gamma_{\alpha \beta} = h_{\alpha \beta} = \partial_\alpha X^\mu
\partial_\beta X_\mu.
\end{equation}
Replacing this in (\ref{eq:Polyakov-membrane}) again gives
(\ref{eq:Nambu-Goto-membrane}).
The equation of motion which arises from varying $X^\mu$ is
\begin{equation}
\partial_\alpha \left( \sqrt{-\gamma} \gamma^{\alpha \beta}
\partial_\beta X^\mu \right) = 0.
\end{equation}

To follow the procedure we used for the bosonic string theory, we
would now like to use the symmetries of the theory to gauge-fix the
metric $\gamma_{\alpha \beta}$.  Unfortunately, whereas for the string
we had three components of the metric and three continuous symmetries
(two diffeomorphism symmetries and a scale symmetry), for the membrane
we have six independent metric components and only three diffeomorphism
symmetries.  We can use these symmetries to fix the components
$\gamma_{0\alpha}$ of the metric to be
\begin{eqnarray}
\gamma_{0a} & = &  0 \label{eq:3-gauge}\\
\gamma_{00} & = & -  \frac{4}{\nu^2}  \bar{h} \equiv
-\frac{4}{\nu^2}  \det h_{ab} \nonumber
\end{eqnarray}
where $\nu$ is a constant whose normalization has been chosen to make
the later matrix interpretation transparent.  Once we have chosen this
gauge, no further components of the metric $\gamma_{ab}$ can be fixed.
This gauge can only be chosen when the membrane world-volume is of the
form $\Sigma \times\br$ where $\Sigma$ is a Riemann surface of fixed
topology.  The membrane action becomes
\begin{equation}
S = \frac{T \nu}{4}  \int d^3 \sigma \left(
\dot{X}^\mu \dot{X}_\mu -\frac{4}{ \nu^2}  \bar{h} \right).
 \label{eq:membrane-action-h}
\end{equation}

It is natural to rewrite this theory in terms of a canonical
Poisson bracket on the membrane at constant $\tau$ where
$\{f,g\} \equiv \epsilon^{ab} \partial_a f
\partial_b g$ with $\epsilon^{12} = 1$.
We will assume that the coordinates $\sigma$ are chosen so that with
respect to the symplectic form associated to
this canonical Poisson bracket the volume of the Riemann surface
$\Sigma$ is $\int d^2 \sigma  =4 \pi$.
In terms of this metric we have the handy formulae
\begin{eqnarray}
\bar{h}  =   \det h_{ab}  &=&
\partial_1 X^\mu \partial_1 X^\mu
\partial_2 X^\nu \partial_2 X^\nu
- \partial_1 X^\mu \partial_2 X^\mu
\partial_1 X^\nu \partial_2 X^\nu\nonumber\\
& = &
\frac{1}{2}
\{X^\mu, X^\nu\}\{X_\mu, X_\nu\} 
\\
\partial_a (\bar{h}h^{ab} \partial_b  X^\mu) & = & 
\{\{X^\mu, X^\nu\}, X_\nu\}\\
\bar{h} h^{ab} \partial_a X^\mu \partial_b X^\nu & = &
\{X^\mu, X^\lambda\}\{X_\lambda, X^\nu\}
\end{eqnarray}
In terms of the Poisson bracket, the membrane action becomes
\begin{equation}
S = \frac{T \nu}{4}  \int d^3 \sigma \left(
\dot{X}^\mu \dot{X}_\mu -\frac{2}{ \nu^2}  
\{X^\mu, X^\nu\}\{X_\mu, X_\nu\} \right).
\label{eq:covariant-action}
\end{equation}
The equations of motion for the fields $X^\mu$ are
\begin{eqnarray}
\ddot{X}^\mu &=& {4 \over \nu^2} \partial_a \left(\bar{h} h^{ab}
\partial_b X^\mu\right)   \nonumber\\
           &=& {4 \over \nu^2} \{\{X^\mu,X^\nu\},X_\nu\}
\label{eq:membrane-eom} 
\end{eqnarray}
The auxiliary constraints on the system are
\begin{eqnarray}
\dot{X}^\mu \dot{X}_\mu & = &  -\frac{4}{ \nu^2}
\bar{h} \nonumber\\
 & = &  -\frac{2}{\nu^2} 
\{X^\mu, X^\nu\}\{X_\mu, X_\nu\} \label{eq:membrane-constraint-1}
\end{eqnarray}
and
\begin{equation}
\dot{X}^\mu \partial_a X_\mu = 0.
\label{eq:membrane-constraint-2}
\end{equation}
It follows directly from (\ref{eq:membrane-constraint-2}) that
\begin{equation}
\{\dot{X}^\mu, X_\mu\} = 0.
\end{equation}

We have thus expressed the bosonic membrane theory as a constrained
Hamiltonian system.  The degrees of freedom are $D$ functions $X^\mu$
on the 3-dimensional world-volume of a membrane which has topology
$\Sigma \times\br$ where $\Sigma$ is a Riemann surface.  This theory
is still completely covariant.  It is difficult to quantize, however,
because of the constraints and the nonlinearity of the equations of
motion.  The direct quantization of this covariant theory will be
discussed further in Section \ref{sec:covariant-membrane}.

\subsection{The light-front bosonic membrane}
\label{sec:membrane-light-front}

As we did for the bosonic string, we now consider the theory in
light-front coordinates
\begin{equation}
X^{\pm} = (X^0 \pm X^{D-1})/\sqrt{2}.
\label{eq:xpm}
\end{equation}
Just as in the case of the string, the constraints
(\ref{eq:membrane-constraint-1},\ref{eq:membrane-constraint-2}) can be
explicitly solved in light-front gauge
\begin{equation}
 X^+ (\tau, \sigma_1, \sigma_2) =\tau.
\label{eq:light-cone-gauge-membrane}
\end{equation}
We have
\begin{eqnarray}
\dot{X}^- &=& \half \dot{X}^i \dot{X}^i + {2 \bar{h} \over \nu^2} 
\nonumber\\
          &=& \half \dot{X}^i \dot{X}^i + {1 \over \nu^2}
\{X^i,X^j\}\{X^i,X^j\}  \label{eq:dx}\\
\noalign{\vskip 0.2 cm}
\partial_a X^- &=& \dot{X}^i \partial_a X^i \nonumber
\end{eqnarray}
We can go to a Hamiltonian formalism by computing the canonically conjugate
momentum densities.
\begin{eqnarray}
P^+ & = & - \frac{ \delta {\cal L}}{ \delta ( \dot{X}^-)} 
=  \frac{\nu T}{2}  \\
P^i & = &  \frac{ \delta {\cal L}}{ \delta ( \dot{X}^i)} 
=  \frac{\nu T}{2}  \dot{X}^i \nonumber
\end{eqnarray}
The total momentum in the direction $P^+$ is then
\begin{equation}
p^+ = \int d^2 \sigma P^+ = 2 \pi \nu T.
\end{equation}
The Hamiltonian of the theory is given by
\begin{eqnarray}
H &=&  \int d^2 \sigma \; \left(
P^i \dot{X}^i-P^+ \dot{X}^--{\cal L} \right) \nonumber\\
& = &
{\nu T \over 4} \int d^2 \sigma \left( \dot{X}^i \dot{X}^i
+ {4 \bar{h} \over \nu^2} \right)  \label{eq:light-front-h}\\
  &=& {\nu T \over 4} \int d^2 \sigma \left( \dot{X}^i \dot{X}^i
+ {2 \over \nu^2} \{X^i,X^j\}\{X^i,X^j\} \right)\,. \nonumber
\end{eqnarray}
The only remaining constraint is that the transverse degrees of
freedom must satisfy
\begin{equation}
\{\dot{X}^i,X^i\}=0
\end{equation}
This theory has a residual invariance under time-independent
area-preserving diffeomorphisms.  Such diffeomorphisms do not change
the symplectic form and thus manifestly leave the Hamiltonian
(\ref{eq:light-front-h})

We now have a Hamiltonian formalism for the light-front membrane
theory.  Unfortunately, this theory is still rather difficult to
quantize.  Unlike string theory, where the equations of motion are
linear in this formalism, for the membrane the equations of motion
(\ref{eq:membrane-eom}) are nonlinear and difficult to solve.

\subsection{Matrix regularization}
\label{sec:matrix-regularization}

In 1982 a remarkably clever regularization of the light-front membrane
theory was found by Goldstone and Hoppe in the case where the surface
$\Sigma$ is a sphere $S^2$ \cite{Goldstone-Hoppe}.  According to this
regularization procedure, functions on the membrane surface are mapped
to finite size $N \times N$ matrices.  Just as in the quantization of
a classical mechanical system defined in terms of a Poisson brackets,
the Poisson bracket appearing in the membrane theory is replaced in
the matrix regularization of the theory by a matrix commutator.

The matrix regularization of the theory can be generalized to
membranes of arbitrary topology, but is perhaps most easily understood
by considering the case discussed in \cite{Goldstone-Hoppe}, where the
membrane has the topology of a sphere $S^2$ for all values of $\tau$.
In this case the world-sheet of the membrane surface at fixed time can
be described by a unit sphere with a rotationally invariant canonical
symplectic form.  Functions on this membrane can be described in terms
of functions of the three Cartesian coordinates $\xi_1, \xi_2, \xi_3$
on the unit sphere satisfying
\begin{equation}
\xi_1^2 + \xi_2^2 + \xi_3^2 = 1.
\end{equation}
The Poisson brackets of these functions are
given by
\[
\{\xi_A, \xi_B\} = \epsilon_{ABC} \xi_C.
\]
This is essentially the same algebraic structure as that defined by
the commutation relations of the generators of $SU(2)$.  It is
therefore natural to associate these coordinate functions on $S^2$
with the matrices generating $SU(2)$ in the $N$-dimensional
representation.  In terms of the conventions we are using here, when
the normalization constant $\nu$ is integral, the correct
correspondence is
\[
\xi_A \rightarrow {2 \over N} J_A
\]
where $J_1, J_2, J_3$ are generators of the $N$-dimensional representation of
$SU(2)$ with $N = \nu$,
satisfying the commutation relations
\[
-i [J_A, J_B] = \epsilon_{ABC} J_C \,.
\]

In general, any function on the membrane can be expanded as a sum of
spherical harmonics
\begin{equation}
f (\xi_1, \xi_2, \xi_3) = \sum_{l, m} c_{lm} y_{lm} (\xi_1, \xi_2, \xi_3)
\label{eq:general-function}
\end{equation}
The spherical harmonics can in turn be written as a sum of
monomials in the coordinate functions:
\[
y_{lm} (\xi_1, \xi_2, \xi_3)=
 \sum_k t^{(lm)}_{A_1 \ldots A_l} \xi_{A_1} \cdots \xi_{A_l}\,
\]
where
the coefficients $t^{(lm)}_{A_1 \ldots A_l}$ are symmetric and traceless
(because $\xi_A \xi_A = 1$).  Using the above correspondence, a matrix
approximation to each of the spherical harmonics with $l < N$
can be constructed, which we denote by
$Y$.  
\begin{equation}
Y_{lm} = \left(2 \over N\right)^l
\sum t^{(lm)}_{A_1\ldots A_l}
J_{A_1}
\cdots J_{A_l}
\label{eq:matrix-spherical}
\end{equation}
For a fixed value of $N$ only the spherical harmonics with $l < N$ can
be constructed
because higher order monomials in the generators $J_A$ do not
generate linearly independent matrices.  Note that the number of
independent matrix entries is precisely equal to the number of
independent spherical harmonic coefficients which can be determined
for fixed $N$
\begin{equation}
N^2 = \sum_{l = 0}^{N -1}  (2l + 1)
\end{equation}
The matrix approximations (\ref{eq:matrix-spherical}) of the spherical
harmonics can be used to construct matrix approximations to an
arbitrary function of the form (\ref{eq:general-function})
\be
\label{MMcorrespond}
F =
\sum_{l < N, m}   c_{lm}
Y_{lm}
\ee

The Poisson bracket in the membrane theory is replaced in the matrix
regularized theory with the matrix commutator according to the
prescription
\begin{equation}
\{f, g\} \rightarrow \frac{-i N}{2} [F, G].
\end{equation}
Similarly, an integral over the membrane at fixed $\tau$ is replaced
by a matrix trace through
\begin{equation}
\frac{1}{4 \pi} \int d^2 \sigma f \rightarrow \frac{1}{N}  {\rm
Tr}\; F
\end{equation}

The Poisson bracket of a pair of spherical harmonics takes the form
\begin{equation}
\{y_{lm}, y_{l' m'}\} = g_{lm, l' m'}^{l'' m''} y_{l'' m''}.
\end{equation}
The commutator of a pair of matrix  spherical harmonics
(\ref{eq:matrix-spherical}) can be written
\begin{equation}
\left[Y_{lm}, Y_{l' m'}\right] = G_{lm, l' m'}^{l'' m''} Y_{l'' m''}.
\end{equation}
It can be verified that in the large $N$ limit the structure constant
of these algebras agree
\begin{equation}
\lim_{N\rightarrow \infty} 
\frac{-i N}{ 2}  G_{lm, l' m'}^{l'' m''}
\rightarrow g_{lm, l' m'}^{l'' m''}
\end{equation}
As a result,  it can be shown that
for any smooth functions on the membrane $f, g$ defined
in terms of convergent sums of spherical harmonics, with Poisson
bracket $\{f, g\} = h$ we have
\begin{eqnarray}
\lim_{N \rightarrow \infty}
\frac{1}{N} 
{\rm Tr}\; F & = &  \frac{1}{4 \pi}  \int d^2 \sigma f
\end{eqnarray}
and it is possible to show that
\begin{eqnarray}
\lim_{N \rightarrow \infty} ((\frac{-i N}{2} )[F, G] -H) & = & 0
\end{eqnarray}
This last relation is really shorthand for the statement that
\begin{equation}
\lim_{N \rightarrow \infty}\frac{1}{N} 
 {\rm Tr}\; \left(((\frac{-i N}{2} )[F, G] -H) J \right) = 0
\end{equation}
where $J$ is the matrix approximation to any smooth function $j$ on
the sphere.

We now have a dictionary for transforming between continuum and
matrix-regularized quantities.  The correspondence is given by
\begin{equation}
\label{eq:mm-correspondence}
\xi_A \leftrightarrow {2 \over N} J_A \;\;\;\;\; \;\;\;\;\;
\{\cdot,\cdot\} \leftrightarrow  {-i N \over 2} [\cdot,\cdot]
\;\;\;\;\; \;\;\;\;\;
{1 \over 4 \pi} \int d^2 \sigma \leftrightarrow \frac{1}{N} 
\tr \;
\end{equation}
The matrix regularized membrane Hamiltonian is therefore given by
\begin{eqnarray}
H  & = & (2 \pi l_p^3) \tr \left( \half {\bf P}^i {\bf P}^i
\right)
-  {1 \over  (2 \pi l_p^3)}\tr \left(
{1 \over 4} [{\bf X}^i,{\bf X}^j] [{\bf X}^i,{\bf X}^j]
\right) \nonumber\\
 & = &  {1 \over  (2 \pi l_p^3)} \tr \left( \half \dot{{\bf X}}^i \dot{{\bf
X}}^i - {1 \over 4} [{\bf X}^i,{\bf X}^j] [{\bf X}^i,{\bf X}^j]
\right)\,.\label{MatrixHamiltonian}
\end{eqnarray}
This Hamiltonian gives rise to the matrix equations of
motion
\[
\ddot{\bf X}^i + [[{\bf X}^i,{\bf X}^j],{\bf X}^j] = 0
\]
which must be supplemented with the Gauss constraint
\be
\label{MatrixGauss}
[\dot{\bf X}{}^i, {\bf X}^i] = 0 \,.
\ee
This is a classical theory with a finite number of degrees of
freedom.  The quantization of such a system is straightforward,
although solving the quantum theory can in practice be quite tricky.
Thus, we have found a well-defined quantum theory describing the
matrix regularization of the relativistic membrane theory in
light-front coordinates.

There are a number of rather deep mathematical reasons why the matrix
regularization of the membrane theory works.  One way of looking at
this regularization is in terms of the underlying symmetry of the
theory.  After gauge-fixing, the membrane theory has a residual
invariance under the group of time-independent area-preserving
diffeomorphisms of the membrane world-sheet.  This diffeomorphism
group can be described in a natural mathematical way as a limit of the
matrix group $U(N)$ as $N \rightarrow \infty$.  In the discrete theory
the area-preserving diffeomorphism symmetry thus is replaced by the
$U(N)$ matrix symmetry.  The matrix regularization can also be viewed
in terms of a geometrical quantization of the operators associated
with functions on the membrane.  From this point of view the matrix
membrane is like a ``fuzzy'' membrane which is discrete yet preserves
the $SU(2)$ rotational symmetry of the original smooth sphere.  This
point of view ties into recent developments in noncommutative
geometry.

We will not pursue these points of view in any depth here.  We will
note, however, that from both points of view it is natural to
generalize the construction to higher genus surfaces.  We discuss the
matrix torus explicitly in section \ref{sec:higher-genus-membranes}.

\subsection{The bosonic membrane in a general background}
\label{sec:bosonic-general}

So far we have only considered the membrane in a flat background
Minkowski geometry.  Just as for strings, it is natural to generalize
the discussion to a bosonic membrane moving in a general background
metric $g_{\mu \nu}$ and 3-form field $A_{\mu \nu \rho}$.  The
introduction of a general background metric modifies the Nambu-Goto
action by replacing $h_{\alpha \beta}$ in
(\ref{eq:induced-metric-membrane}) with
\begin{equation}
h_{ \alpha \beta} = \partial_\alpha X^\mu \partial_\beta X^\nu
g_{\mu \nu} (X).
\label{eq:induced-metric-membrane-general}
\end{equation}
The membrane couples to the 3-form field as an electrically charged
object, giving an additional term to the action of the form $\int
A_{\alpha \beta \gamma}$ where $A_{\alpha \beta \gamma}$ is the
pullback to the world-volume of the membrane of the 3-form field.
This gives a total Nambu-Goto-type action for the membrane in a
general background of the form
\begin{equation}
S = -T  \int d^3 \sigma \left( \sqrt{-\det h_{\alpha \beta}}
+6 \dot{X}^\mu \partial_1 X^\nu \partial_2 X^\rho
 A_{\mu \nu \rho}(X)\right).
\end{equation}
With an auxiliary world-volume metric, this action becomes
\begin{eqnarray}
S & = &  - \frac{T}{2}  \int d^3 \sigma 
\left[\sqrt{-\gamma} \left(
\gamma^{\alpha \beta} \partial_\alpha X^\mu \partial_\beta X^\nu
g_{\mu \nu} (X)
-1\right)\right.\label{eq:bosonic-membrane-Polyakov-general}\\
 &  & \hspace{0.9in}
\left.+12 \dot{X}^\mu \partial_1 X^\nu \partial_2 X^\rho
 A_{\mu \nu \rho} (X)\right] \nonumber
\end{eqnarray}

We can gauge fix the action
(\ref{eq:bosonic-membrane-Polyakov-general}) using the same gauge
(\ref{eq:3-gauge}) as in the flat space case.  We can then consider
carrying out a similar procedure for quantizing the membrane in a
general background as we described in the case of the flat background.
We will return to this question in section
\ref{sec:curved-background} when we discuss in more detail the
prospects for constructing matrix theory in a general background.

\subsection{The supermembrane}
\label{sec:supermembrane}

Now let us turn our attention to the supermembrane.  In order to make
contact with M-theory, and indeed to make the membrane theory
well-behaved it is necessary to add supersymmetry to the theory.
Supersymmetric membrane theories can be constructed classically in
dimensions 4, 5, 7 and 11.  These theories have different degrees of
supersymmetry, with 2, 4, 8 and 16 independent supersymmetric
generators respectively.  It is believed that all the supermembrane
theories other than the 11D maximally supersymmetric theory suffer
from anomalies in the Lorentz algebra.  Thus, just as $D = 10$ is the
natural dimension for the superstring, $D = 11$ is the natural
dimension for the supermembrane.

The formalism for describing the supermembrane is rather technically
complicated.  We will not use most of this formalism in the rest of
these lectures, so we restrict ourselves here to a fairly concise
discussion of the structure of the supersymmetric theory.  The reader
not interested in the details of how the supersymmetric form of matrix
theory is derived may wish to skip directly to the result of this
analysis, the supersymmetric matrix theory Hamiltonian
(\ref{eq:matrix-Hamiltonian}), on first reading.  In Section
\ref{sec:supermembrane-action} we describe using superfield notation
the supermembrane action in a general background and its symmetries.
We discuss in particular the fact that the $\kappa$-symmetry of the
theory at the classical level guarantees already that the background
geometry satisfies the equations of motion of 11D supergravity.  In
\ref{sec:supermembrane-flat} we describe in more explicit form the
supermembrane action in a flat background.  We describe the
light-front form of the theory in \ref{sec:supermembrane-quantum},
where we show how the regularized theory gives precisely the
Hamiltonian of the supersymmetric matrix theory.

\subsubsection{The supermembrane action}
\label{sec:supermembrane-action}

In this section we describe the supermembrane action in an arbitrary
background and its symmetries.  In particular,
we describe the $\kappa$-symmetry of
the theory, which implies that the background obeys the classical equations
of 11D supergravity.  For further details see the original paper of
Bergshoeff, Sezgin and Townsend \cite{bst} or the review paper of Duff
\cite{Duff-review}.

The standard NSR description of the superstring gives a theory
which is fairly straightforward to quantize.  This formalism can be
used in a straightforward fashion
to describe the spectra of the five superstring theories.
One disadvantage of this formalism is that the target space
supersymmetry of the theory is  difficult to show
explicitly.  There is another formalism, known as the Green-Schwarz
formalism (\cite{Green-Schwarz-string}, reviewed in \cite{gsw}), in
which the target space supersymmetry of the theory is much more
clear.  In the Green-Schwarz formalism
additional Grassmann degrees of freedom are introduced which transform
as space-time fermions but as world-sheet  vectors.  These correspond
to space-time superspace coordinates for the string.  The
Green-Schwarz superstring action does not have a standard world-sheet
supersymmetry (it can't, since there are no world-sheet fermions).
The theory does, however, have a novel type of supersymmetry known as
a $\kappa$-symmetry.  The existence of the $\kappa$-symmetry in the classical
Green-Schwarz string theory already implies that the theory is
restricted to $D = 3, 4, 6$ or 10.  This is already a much stronger
restriction than can be gleaned from classical superstring with
world-sheet supersymmetry.

Unlike the superstring, there is no known way of formulating the
supermembrane in a world-volume supersymmetric fashion (although there
has been some recent progress in this direction, for further
references see \cite{Duff-review}).  A Green-Schwarz formulation of
the supermembrane in a general background was first found by
Bergshoeff, Sezgin and Townsend \cite{bst}.  We now review this
construction.

We consider an 11-dimensional target space with a general metric
$g_{\mu \nu}$ described by an elfbein $e^a_\mu$, and an arbitrary
background gravitino field $\psi_\mu$ and 3-form field $A_{\mu \nu
\rho}$.  In superspace notation we describe the space as having 11
bosonic coordinates $X^\mu$ and 32 anticommuting fermionic coordinates
$\theta^{\dot{\alpha}}$.  These coordinates are combined into a single
superspace coordinate 
\begin{equation}
Z^M = (X^\mu, \theta^{\dot{\alpha}})
\end{equation}
where $M$ is an index with 43 possible values.  (Space-time spinor
indices $\dot{\alpha}, \dot{\beta}, \ldots$ will carry a dot in this
section to distinguish them from world-volume coordinate indices
$\alpha, \beta, \ldots$).  In superspace the elfbein becomes a 43-bein
$E^A_M$, with $A = (a, \phi)$.  There is also an antisymmetric
superspace 3-form field $B_{MNP}$.  The superspace formulation of 11D
supergravity is written in terms of these two fields.  The
identification of the superspace degrees of freedom with the component
fields $e^a_{\mu}, \psi_\mu$ and $A_{\mu \nu \rho}$ is quite subtle,
and involves a careful analysis of the supersymmetry transformations
in both formalisms as well as gauge choices.  At leading order in
$\theta$ the component fields are identified through
\begin{eqnarray}
E^a_{\mu} & = &  e^a_\mu +{\cal O} (\theta) \nonumber\\
E^\phi_\mu & = &  \psi_\mu^\phi +{\cal O} (\theta)\\
B_{ \mu \nu \rho} & = &A_{\mu \nu \rho} +{\cal O} (\theta)\nonumber
\end{eqnarray}
The identifications of $E^A_M$ and $B_{MNP}$ in terms of component
fields through order $\theta^2$ has only recently been determined
\cite{dpp}.  The identification beyond this order has not been
determined explicitly.

In terms of these superspace fields, the supermembrane action in a
general background is given by
\begin{equation}
S = - \frac{T}{2}  \int d^3 \sigma 
\left[\sqrt{-\gamma} \left(
\gamma^{\alpha \beta} 
\Pi_\alpha^a \Pi_\beta^b \eta_{ab}
-1\right)
+  \epsilon^{\alpha \beta \gamma} 
\Pi_\alpha^A \Pi_\beta^B \Pi_\gamma^C
 B_{ABC}\right]
\label{eq:membrane-Polyakov-general}
\end{equation}
where $\Pi_\alpha^A$ are the components of the pullback of the 43-bein
to the membrane world-volume
\begin{equation}
\Pi_\alpha^A = \partial_\alpha Z^M E^A_M
\label{eq:pi-definition}
\end{equation}
and  $B_{ABC}$ is defined implicitly through
\begin{equation}
B_{MNP} = E^A_ME^B_NE^C_P B_{ABC}
\end{equation}

The action (\ref{eq:membrane-Polyakov-general}) is very closely
related to the superspace formulation of the Green-Schwarz action.
The superstring action differs in that it has no cosmological term and
that the antisymmetric field is a superspace 2-form field.

Let us now review the symmetries of the action
(\ref{eq:membrane-Polyakov-general}).  This action has global
symmetries corresponding to space-time super diffeomorphisms, gauge
transformations and discrete symmetries, as well as the local
symmetries of world-volume diffeomorphisms and $\kappa$ symmetry.
\vspace{0.08in}

\noindent
{\it Super diffeomorphisms:} Under a super diffeomorphism of the
target space generated by a super vector field $\xi^M$ the coordinate
fields, 43-bein and 3-form field transform under
\begin{eqnarray}
\delta Z^M & = &  \xi^M \nonumber\\
\delta E^A_M & = &  \xi^N \partial_NE^A_M + \partial_M \xi^NE^A_N\\
\delta B_{MNP} & = & \xi^Q \partial_Q B_{MNP}
+ (\partial_M \xi^Q) B_{QNP}
- (\partial_N \xi^Q) B_{MQP}
+ (\partial_P \xi^Q) B_{MNQ}\nonumber
\end{eqnarray}
\vspace{0.05in}

\noindent
{\it Super gauge transformations:} This global symmetry transforms the
3-form superfield  by
\begin{equation}
\delta B_{MNP} = \partial_M \Sigma_{NP}
-\partial_N \Sigma_{MP} + \partial_P \Sigma_{MN}.
\end{equation}
\vspace{0.05in}

\noindent
{\it Discrete symmetries:}
There is also a discrete symmetry $\bz_2$
corresponding to taking
\begin{equation}
B_{MNP} \rightarrow -B_{MNP}
\end{equation}
and performing a space-time reflection on a single coordinate.
\vspace{0.05in}

\noindent
{\it World-volume diffeomorphisms:}
Under a world-volume diffeomorphism generated by the vector field
$\eta^\alpha$ the fields transform by
\begin{equation}
\delta Z^M = \eta^\alpha \partial_\alpha Z^M
\end{equation}
\vspace{0.05in}

\noindent
{\it $\kappa$-symmetry:}
The  most interesting symmetry of the theory is the fermionic
$\kappa$-symmetry.  The parameter $\kappa^\psi$ is taken to be an
anticommuting world-volume scalar which transforms as a space-time
32-component spinor.  Under this symmetry the coordinate fields
transform under
\begin{eqnarray}
\delta Z^M E^a_M & = &  0 \label{eq:k-symmetry}\\
\delta Z^M E^\phi_M & = &  (1 + \Gamma)^\phi_\psi \kappa^\psi \nonumber
\end{eqnarray}
where
\begin{equation}
\Gamma= \frac{1}{6 \sqrt{- \gamma}}  \epsilon^{\alpha \beta \gamma}
\Pi^a_\alpha \Pi^b_\beta \Pi^c_\gamma \Gamma_{abc}.
\end{equation}

The $\kappa$-symmetry of the theory has a number of interesting
features.  For one thing, it can be shown that (\ref{eq:k-symmetry})
is only a symmetry of the theory when the background fields $E^a_M,
B_{MNP}$ obey the equations of motion of the classical 11D
supergravity theory.  Thus, 11D supergravity emerges from the membrane
theory even at the classical level.  For the details of this analysis,
see \cite{bst}.  This situation is similar to that which arises in the
Green-Schwarz formulation of the superstring theories.  In the
Green-Schwarz formalism there is a local $\kappa$-symmetry on the
string world-sheet only when the backgrounds satisfy the supergravity
equations of motion.

Another interesting aspect of the $\kappa$-symmetry arises from the
algebraic fact that
\begin{equation}
\Gamma^2 = 1.
\end{equation}
This implies that $(1 + \Gamma)$ is a projection operator.  We can
thus use $\kappa$-symmetry to gauge away half of the fermionic degrees
of freedom $\theta^{\dot{\alpha}}$.  This reduces the number of
propagating fermionic degrees of freedom to 8.  This is also the
number of propagating bosonic degrees of freedom, as can be seen by
going to a static gauge where $X^{0, 1, 2}$ are identified with $\tau,
\sigma_{1, 2}$ so that only the 8 transverse directions appear as
propagating degrees of freedom.

In general, gauge-fixing the $\kappa$-symmetry in any particular way
will break the Lorentz invariance of the theory.  This makes it quite
difficult to find any way of quantizing the theory without breaking
Lorentz symmetry.  This situation is again analogous to the
Green-Schwarz superstring theory, where fixing of $\kappa$-symmetry
also breaks Lorentz invariance and no covariant quantization scheme is
known.

\subsubsection{The supermembrane in flat space}
\label{sec:supermembrane-flat}

To make the connection with matrix theory, we now restrict attention
to a flat Minkowski background space-time with vanishing 3-form field
$A_{\mu \nu \rho}$.  We will return to a discussion of general
backgrounds in section \ref{sec:curved-background}.

In flat space the 43-bein becomes
\begin{eqnarray}
E^a_M & = &  (\delta^a_\mu, (\Gamma^a)_{\dot{\alpha} \dot{\beta}}
 \theta^{\dot{\beta}})\\ 
E^\phi_M & = &  (0, \delta^\phi_{\dot{\alpha}}) \nonumber
\end{eqnarray}
The super 4-form field strength $H_{MNPQ}$ has as its only
nonvanishing components
\begin{equation}
H_{ab \phi \psi} = \frac{1}{3}
(\Gamma_{ab})_{\phi \psi}.
\end{equation}
{}From this and the definition $H = dB$ it is possible to derive the
components of the super 3-form field $B_{MNP}$
\begin{eqnarray}
B_{\mu \nu \rho} & = & 0 \nonumber\\ B_{\mu \nu \dot{\alpha}} & = &
\frac{1}{6} (\Gamma_{\mu \nu} \theta)_{\dot{\alpha}}\\ B_{\mu
\dot{\alpha} \dot{\beta}} & = & \frac{1}{6} (\Gamma_{\mu \nu}
\theta)_{(\dot{\alpha}} (\Gamma^\nu \theta)_{\dot{\beta})} \nonumber\\
B_{ \dot{\alpha} \dot{\beta} \dot{\gamma}} & = & \frac{1}{6}
(\Gamma_{\mu \nu} \theta)_{(\dot{\alpha}} (\Gamma^\mu
\theta)_{\dot{\beta}} (\Gamma^\nu \theta)_{\dot{\gamma})}\nonumber
\end{eqnarray}
{}From (\ref{eq:pi-definition}) it follows that
\begin{equation}
\Pi^\mu_\alpha = \partial_\alpha X^\mu + \bar{\theta} \Gamma^\mu
\partial_\alpha \theta.
\end{equation}
The membrane action (\ref{eq:membrane-Polyakov-general}) reduces in
flat space to
\begin{eqnarray}
S& = & - \frac{T}{2}  \int d^3 \sigma 
\left\{\sqrt{-\gamma} \left(
\gamma^{\alpha \beta} 
\Pi_\alpha^\mu \Pi_\beta^\nu \eta_{\mu \nu}
-1\right) \right. \label{eq:membrane-Polyakov-flat}\\
& &\hspace{0.9in} \left.
-  \epsilon^{\alpha \beta \gamma} 
\left[
\frac{1}{2}\partial_\alpha X^\mu
(\partial_\beta X^\nu + \bar{\theta} \Gamma^\nu \partial_\beta \theta)
\right. \right.\\
& &\hspace{1.3in} \left. \left.
+ \frac{1}{6}  (\bar{\theta} \Gamma^\mu \partial_\alpha \theta)
(\bar{\theta} \Gamma^\nu \partial_\beta \theta) \right]
\bar{\theta} \Gamma_{\mu \nu} \partial_\gamma \theta
\right\} 
\nonumber
\end{eqnarray}
The extra Wess-Zumino type terms which appear in this action are
rather non-obvious from the point of view of the flat space-time
theory, although they have arisen naturally in the superspace
formalism.  These are analogous to terms in the Green-Schwarz
superstring action which were originally found by imposing
$\kappa$-symmetry on the theory.
The equation of motion for $\gamma$ as usual sets $\gamma_{\alpha
\beta}$ to be the pullback of the metric
\begin{equation}
\gamma_{\alpha \beta} = \Pi^\mu_\alpha \Pi^\nu_\beta \eta_{\mu \nu}
\end{equation}

The action (\ref{eq:membrane-Polyakov-flat}) has the target space
supersymmetry
\begin{eqnarray}
\delta \theta & = &  \epsilon\\
\delta X^\mu & = &  -\bar{\epsilon} \Gamma^\mu \theta \nonumber
\end{eqnarray}
This transformation leaves $\Pi^\mu_\alpha$ invariant.  The fact that
it leaves the action invariant follows from the identity
\begin{equation}
\bar{\psi}_{[1} \Gamma^\mu \psi_2 \bar{\psi}_3 \Gamma_{\mu \nu}
\psi_{4]} = 0
\label{eq:4-relation}
\end{equation}
which holds in 11 dimensions (as well as in dimensions 4, 5 and 7).
The relation (\ref{eq:4-relation}) is also necessary to show that the
action is $\kappa$-symmetric.  This relation is analogous to the
relation $\bar{\epsilon} \Gamma_\mu \psi_{[1} \psi_2 \Gamma^\mu
\psi_{3]} = 0$ which holds in 3, 4, 6 and 10 dimensions and which is
necessary for the supersymmetry and $\kappa$-symmetry of the
Green-Schwarz superstring action.

\subsubsection{The quantum supermembrane and supersymmetric matrix theory}
\label{sec:supermembrane-quantum}

We now go to light-front gauge.  As usual we define
\begin{equation}
X^{\pm} = (X^0 \pm X^{D-1})/\sqrt{2}.
\label{eq:xpm-2}
\end{equation}
We write
the  $32 \times 32$ $\Gamma$ matrices in the block forms
\begin{eqnarray}
\Gamma^+ & =& \left(\begin{array}{cc}
0 & 0\\
\sqrt{2} i \identity_{16} & 0\\
\end{array}\right) \nonumber\\
\Gamma^- & =& \left(\begin{array}{cc}
0 &
\sqrt{2} i \identity_{16}\\
0 & 0\\
\end{array}\right)\\
\Gamma_i & = &\left(\begin{array}{cc}
\gamma^i & 0\\
0 & -\gamma^i
\end{array}\right) \nonumber
\end{eqnarray}
where $\gamma^i$ are $16 \times 16$ Euclidean $SO(9)$ gamma matrices.

In addition to setting the gauge
\begin{equation}
X^+ = \tau
\end{equation}
We can also use $\kappa$-symmetry to fix
\begin{equation}
\Gamma^+ \theta = 0
\end{equation}
{}From the above form of the matrices $\Gamma^{\mu}$ it is clear that
this projects onto the 16 Grassmann degrees of freedom $(0, \theta)$,
and that as a consequence
all expressions of the forms
\begin{equation}
\bar{\theta} \Gamma^\mu \partial_\alpha \theta, \;\;\;\;\; \mu \neq -
\end{equation}
and
\begin{equation}
\bar{\theta} \Gamma^{ij} \partial_\alpha \theta \;\;\;\;\;
{\rm or} \;\;\;\;\;
\bar{\theta} \Gamma^{+\mu} \partial_\alpha \theta
\end{equation}
must vanish in this gauge.  This simplifies the theory in this gauge
considerably.  First, we have
\begin{equation}
\Pi^\mu_\alpha = \partial_\alpha X^\mu, \;\;\;\;\; \mu \neq -
\end{equation}
Second, we find that the terms on the second
line of (\ref{eq:membrane-Polyakov-flat}) simplify to
\begin{equation}
- \bar{\theta} \Gamma_{+i} \{X^i, \theta\}
\end{equation}

Solving for the derivatives $\partial_\gamma X^-$ as in (\ref{eq:dx})
we get
\begin{eqnarray}
\dot{X}^-
          &=& \half \dot{X}^i \dot{X}^i + {1 \over \nu^2}
\{X^i,X^j\}\{X^i,X^j\} + \bar{\theta} \Gamma_+\dot{\theta} \nonumber
\\
&=  &\Pi^-_0 + \bar{\theta} \Gamma_+ \dot{\theta}\\
\noalign{\vskip 0.2 cm}
\partial_a X^- &=& \dot{X}^i \partial_a X^i +
\bar{\theta} \Gamma_+ \partial_a \theta \nonumber\\
&=  &\Pi^-_a + \bar{\theta} \Gamma_+ \partial_a \theta
\label{eq:dxs}
\end{eqnarray}

Combining these observations, we find that the light-front
supermembrane Hamiltonian becomes
\begin{equation}
H = {\nu T \over 4} \int d^2 \sigma \left( \dot{X}^i \dot{X}^i
+ {2 \over \nu^2} \{X^i,X^j\}\{X^i,X^j\} -\frac{2}{\nu}  
\theta^T \gamma_i\{X^i, \theta\}
\right)
\end{equation}
where $\theta$ is a 16-component Majorana spinor of $SO(9)$.

It is straightforward to apply the matrix regularization procedure
discussed in section \ref{sec:matrix-regularization} to this
Hamiltonian.  This gives the supersymmetric form of matrix theory
\be
H = {1 \over  (2 \pi l_p^3)} \tr \left( \half \dot{{\bf X}}^i \dot{{\bf
X}}^i - {1 \over 4} [{\bf X}^i,{\bf X}^j] [{\bf X}^i,{\bf X}^j] +
\half
{\bf \theta}^T \gamma_i[ {\bf X}^i, {\bf \theta}] \label{eq:matrix-Hamiltonian}
\right)\,.  \ee
This is the matrix quantum mechanics theory which will play a central
role in these lectures.  This theory was derived in \cite{dhn} 
from the regularized supermembrane action, but had been previously
found and studied as a particularly simple example of a supersymmetric
theory with gauge symmetry \cite{Claudson-Halpern,Flume,brr}.

\subsection{Covariant membrane quantization}
\label{sec:covariant-membrane}

It is natural to think of generalizing the matrix regularization
approach to the covariant formulation of the bosonic and
supersymmetric membrane theories (\ref{eq:covariant-action}) and
(\ref{eq:membrane-Polyakov-flat}).  Some progress was made in this
direction by Fujikawa and Okuyama in \cite{Fujikawa-Okuyama}.  For the
bosonic membrane it is straightforward to implement the matrix
regularization procedure.  The only catch is that the BRST charge
needed to implement the gauge-fixing procedure cannot be simply
expressed in terms of the Poisson bracket on the membrane.  For the
supermembrane, there is a more serious complication related to the
$\kappa$-symmetry of the theory.  Essentially, as mentioned above, any
gauge-fixing of the $\kappa$-symmetry will break the 11-dimensional
Lorentz invariance of the theory.  This is the same difficulty as one
encounters when trying to construct a covariant quantization of the
Green-Schwarz superstring.  The approach taken in the second paper of
\cite{Fujikawa-Okuyama} is to fix the $\kappa$-symmetry in a way which
breaks the 32 of SO(10, 1) into $16_R + 16_L$ of SO(9, 1).  Thus, they
end up with a matrix formulation of a theory with explicit SO(9, 1)
Lorentz symmetry.  Although this theory does not have the desired
complete SO(10, 1) Lorentz symmetry of M-theory, there are many
questions which might be addressed by this theory with limited Lorentz
invariance.  It would be interesting to study the quantum mechanics of
this alternative matrix formulation of M-theory in further detail.

\section{The BFSS conjecture}
\label{sec:BFSS}

\setcounter{equation}{0}

As we have already discussed,
the fact that the light-front supermembrane theory can be regularized
and described as a supersymmetric quantum mechanics theory has been
known for over a decade.  At the time that this theory was first
developed, however, it was believed that the quantum supermembrane
theory suffered from instabilities which would make the low-energy
interpretation as a theory of quantized gravity impossible.  In 1996
the supersymmetric Yang-Mills quantum mechanics theory was brought
back into currency as a candidate for a microscopic description of an
11-dimensional quantum mechanical theory containing gravity by Banks,
Fischler, Shenker and Susskind (BFSS).  The BFSS proposal, which
quickly became known as the ``Matrix Theory Conjecture'' was motivated
not by the quantum supermembrane theory, but by considering the
low-energy theory of a system of many D0-branes  as a partonic
description of light-front M-theory.

In this section we discuss the apparent instability of the quantized
membrane theory and the BFSS conjecture.  We describe the membrane
instability in subsection \ref{sec:instability}.  We give a brief
introduction to M-theory in section \ref{sec:M-theory}, and describe
the BFSS conjecture in subsection \ref{sec:BFSS-s}.  In subsection
\ref{sec:second-quantized} we describe the resolution of the apparent
instability of the membrane theory by an interpretation in terms of a
second-quantized gravity theory.  Finally, in subsection
\ref{sec:proof} we review an argument due to Seiberg and Sen which shows that
matrix theory should be equivalent to a discrete light-front
quantization of M-theory, even at finite $N$.

\subsection{Membrane ``instability''}
\label{sec:instability}

At the time that de Wit, Hoppe and Nicolai wrote the paper \cite{dhn}
showing that the regularized supermembrane theory could be described
in terms of supersymmetric matrix quantum mechanics, the general hope
seemed to be that the quantized supermembrane theory would have a
discrete spectrum of states.  In string theory the spectrum of 
states in the Hilbert space of the string can be put into one-to-one
correspondence with elementary particle-like states in the target
space.  The facts that the massless particle spectrum contains a
graviton and that there is a mass gap separating the massless states
from massive excitations are crucial for this interpretation.  For the
supermembrane theory, however, the spectrum does not seem to have
these properties.  This can be seen in both the classical and quantum
membrane theories.  In this section we discuss this apparent
difficulty with the membrane theory, which was first described in
detail in \cite{dln}.

\begin{figure}[ht]
\begin{center}
\epsfig{figure=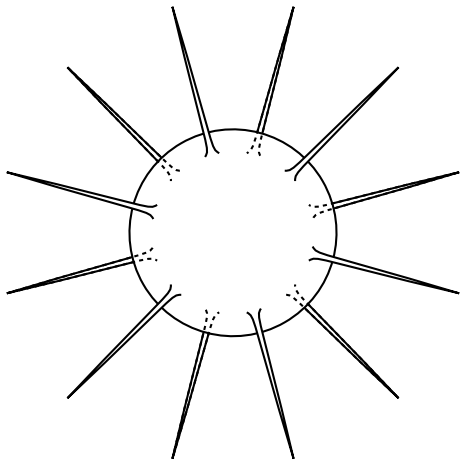, width=0.8\textwidth}
\end{center}
\caption{Classical membrane instability arises from spikes of
infinitesimal area\label{fig:spikes}}
\end{figure}

The simplest way to see the instability of the membrane theory is to
consider a classical bosonic membrane whose energy is proportional to
the area of the membrane times a constant tension.  Such a membrane
can have long narrow spikes at very low cost in energy (See
Figure~\ref{fig:spikes}).  If the spike is roughly cylindrical and has a
radius $r$ and length $L$ then the energy is $2 \pi T rL$.  For a
spike with very large $L$ but a small radius $r\ll 1/TL$ the energy
cost is small but the spike is very long.  This indicates that a
quantum membrane will tend to have many fluctuations of this type,
making it difficult to think of the membrane as a single pointlike
object.  Note that the quantum string theory does not have this
problem since a long spike in a string always has energy proportional
to the length of the string.

In the matrix regularized version of the membrane theory, this
instability appears as a set of flat directions in the classical
theory.  For example,
if we have a pair of matrices with nonzero entries of the form
\begin{equation}
X^1 = \left(\begin{array}{cc}
x & 0\\0 & 0
\end{array}\right)
\;\;\;\;\;
X^2 = \left(\begin{array}{cc}
0 &  y\\ y & 0
\end{array}\right)\label{eq:}
\end{equation}
then a potential term $[X^1, X^2]^2$ corresponds to a term $x^2 y^2$.
If either $x = 0$ or $y = 0$ then the other variable is unconstrained,
giving flat directions in the moduli space of solutions to the
classical equations of motion.  This corresponds classically to a
marginal instability in the matrix theory with $N > 1$.
(Note that in the previous section we distinguished matrices ${\bf X}^i$ from
related functions $X^i$ by using bold font for matrices.  We will
henceforth drop this font distinction as long as the difference can
easily be distinguished from context.)

In the quantum bosonic membrane theory, the apparent instability from
the flat directions is cured because of the 0-modes of off-diagonal
degrees of freedom.  In the above example, for instance, if $x$ takes
a large value then $y$ corresponds to a harmonic oscillator degree of
freedom with a large mass.  The zero point energy of this oscillator
becomes larger as $x$ increases, giving an effective confining
potential which removes the flat directions of the classical theory.
This would seem to resolve the instability problem.  Indeed, in the
matrix regularized quantum bosonic membrane theory, there is a
discrete spectrum of energy levels for the system of  $N \times N$
matrices.

When we consider the supersymmetric theory, on the other hand, the
problem reemerges.  The zero point energies of the fermionic
oscillators associated with the extra Grassmann degrees of freedom in
the supersymmetric theory conspire to precisely cancel the zero point
energies of the bosonic oscillators.  This cancellation gives rise to
a continuous spectrum in the supersymmetric matrix theory.  This
result was formally
proven by de Wit, L\"uscher and Nicolai in \cite{dln}.  They showed that
for any $\epsilon > 0$ and any energy $E\in [0, \infty)$ there exists
a state $\psi$ in the $N = 2$ maximally supersymmetric matrix model
which is normalizable ($\int  | \psi |^2 = | | \psi | |^2 =1$) and
which has
\[
|| (H-E) \psi | |^2 < \epsilon.
\]
This implies that the spectrum of the supersymmetric matrix quantum
mechanics theory is continuous\footnote{Note that \cite{dln} did not
resolve the question of whether a state existed with identically vanishing
energy $H = 0$.  This question was not resolved until the much later
work of Sethi and Stern \cite{Sethi-Stern} showed that such a
marginally bound state does indeed exist in the maximally
supersymmetric theory}.  This result indicated that it would not be
possible to have a simple interpretation of the states of the theory
in terms of a discrete particle spectrum.  After this work there was
little further development on the supersymmetric matrix quantum
mechanics theory as a theory of membranes or gravity until almost a
decade later.

\subsection{M-theory}
\label{sec:M-theory}

The concept of M-theory has played a fairly central role in the
development of the web of duality symmetries which relate the five
string theories to each other and to supergravity
\cite{Hull-Townsend,Witten-various,dlm,Schwarz-m,Horava-Witten}.
M-theory is a conjectured eleven-dimensional theory whose low-energy
limit corresponds to 11D supergravity.  Although there are
difficulties with constructing a quantum version of 11D supergravity,
it is a well-defined classical theory with the following field content
\cite{cjs}:
\vspace{0.03in}

\noindent
$e^a_I$: a vielbein field (bosonic, with 44 components)\\
\noindent
$\psi_I$: a Majorana fermion gravitino (fermionic, with 128
components)\\
\noindent
$A_{I J K}$: a 3-form potential (bosonic, with 84 components).
\vspace{0.02in}

In addition to being a local theory of gravity with an extra 3-form
potential field, M-theory also contains extended objects.  These
consist of a two-dimensional supermembrane and a 5-brane, which couple
electrically and magnetically to the 3-form field.

One way of defining M-theory is as the strong coupling limit of the
type IIA string.  The IIA string theory is taken to be equivalent to
M-theory compactified on a circle $S^1$, where the radius of
compactification $R$ of the circle in direction 11 is related to the
string coupling $g$ through $R= g^{2/3}l_p = g l_s$, where $l_p$ and
$l_s = \sqrt{\alpha'}$ are the M-theory Planck length and the string
length respectively.  The
decompactification limit $R \rightarrow \infty$ corresponds then to
the strong coupling limit of the IIA string theory.  (Note that we
will always take the eleven dimensions of M-theory to be labeled $0,
1, \ldots, 8, 9, 11$; capitalized roman indices $I, J, \ldots$ denote
11-dimensional indices).

Given this relationship between compactified M-theory and  IIA
string theory, a correspondence can be constructed between various objects in
the two theories.  For example, the Kaluza-Klein photon associated
with the components $g_{\mu 11}$ of the 11D metric tensor can be
associated with the R-R gauge field $A_\mu$ in IIA string theory.  The
only object which is charged under this R-R gauge field in IIA string
theory is the D0-brane; thus, the D0-brane can be associated with a
supergraviton with
momentum $p_{11}$ in the compactified direction.  The membrane and
5-brane of M-theory can be associated with different IIA objects
depending on whether or not they are wrapped around the compactified
direction;  the correspondence between
various M-theory and IIA objects is given in Table~\ref{tab:m2}.
\begin{table}[bth]
\caption{Correspondence between objects in M-theory and
IIA string theory\label{tab:m2}}\vspace{0.4cm}
\begin{center}
\begin{tabular}{|l|l|}
\hline
M-theory & IIA\\
\hline
KK photon ($g_{\mu 11}$) & RR gauge field $A_\mu$\\
supergraviton with $p_{11}= 1/R$ & D0-brane\\
wrapped membrane & IIA string\\
unwrapped membrane & IIA D2-brane\\
wrapped 5-brane & IIA D4-brane\\
unwrapped 5-brane & IIA NS5-brane\\
\hline
\end{tabular}
\end{center}
\end{table}

\subsection{The BFSS conjecture}
\label{sec:BFSS-s}

In 1996, motivated by recent work on D-branes and string dualities,
Banks, Fischler, Shenker and Susskind (BFSS) proposed that the large
$N$ limit of the supersymmetric matrix quantum mechanics model
(\ref{eq:matrix-Hamiltonian}) should describe all of M-theory in a
light-front coordinate system \cite{BFSS}.  Although this conjecture
fits neatly into the framework of the quantized membrane theory, the
starting point of BFSS was to consider M-theory compactified on a
circle $S^1$, with a large momentum in the compact direction.  As we
have just discussed, when M-theory is compactified on $S^1$ the
corresponding theory in 10D is the type IIA string theory, and the
quanta corresponding to momentum in the compact direction are the
D0-branes of the IIA theory.  In the limits where the radius of
compactification $R$ and the compact momentum $p_{11}$ are both taken
to be large, this correspondence relates M-theory in the ``infinite
momentum frame'' (IMF) to the nonrelativistic theory of many D0-branes
in type IIA string theory.

The low-energy Lagrangian for a system of many type IIA D0-branes is
the matrix quantum mechanics Lagrangian arising from the dimensional
reduction to 0 + 1 dimensions of the 10D super Yang-Mills Lagrangian
\begin{equation}
{\cal L} = \frac{1}{2gl_s}  {\rm Tr}\; \left[
\dot{X}^a \dot{X}^a+\frac{1}{2}[X^a, X^b]^2 + \theta^T (i \dot{\theta}
-\Gamma_a[X^a, \theta]) \right]
\label{eq:D0-Lagrangian}
\end{equation}
(the gauge has been fixed to $A_0 = 0$.)
The corresponding Hamiltonian is
\be
H = {1 \over  2gl_s} \tr \left( \dot{ X}^i \dot{
X}^i - {1 \over  2} [X^i,X^j] [X^i,X^j] +
\theta^T \gamma_i[X^i, \theta] \label{eq:matrix-Hamiltonian-2}
\right)\,.  \ee
Using the relations $R = g^{2/3}l_{11} = gl_s$, we see that in string
units  ($2 \pi l_s^2 = 1$) we can replace
$gl_s = R = 2 \pi l_{11}^3$.  So the Hamiltonian
(\ref{eq:matrix-Hamiltonian-2}) arising in the matrix quantum
mechanics picture is in fact precisely equivalent to the matrix
membrane Hamiltonian (\ref{eq:matrix-Hamiltonian}).
This connection and its possible physical significance was first
pointed out by Townsend \cite{Townsend}.
The matrix theory Hamiltonian is often written, following BFSS, in the form
\begin{eqnarray}
H & = &   \frac{R}{2}    \tr \left( P^i P^i
 - {1 \over  2} [X^i,X^j] [X^i,X^j] +
 \theta^T \gamma_i[X^i, \theta]\right)
\end{eqnarray}
where we have rescaled $X/g^{1/3} \rightarrow X$ and written the
Hamiltonian in Planck units $l_{11} = 1$.

The original BFSS  conjecture was made in the context of the large $N$
theory.  It was later argued by Susskind that the finite $N$ matrix
quantum mechanics theory should be equivalent to the discrete
light-front quantized (DLCQ) sector of M-theory with $N$ units of
compact momentum \cite{Susskind-DLCQ}.
We describe in section (\ref{sec:proof}) below an argument
due to Seiberg and Sen which  makes this connection more precise and which
justifies the use of the low-energy D0-brane action in the BFSS
conjecture.

While the BFSS conjecture was based on a different framework from the
matrix quantization of the supermembrane theory we have discussed
above, the fact that the membrane naturally appears as a coherent
state in the matrix quantum mechanics theory was a substantial piece
of additional evidence given by BFSS for the validity of their
conjecture.  Two additional pieces of evidence were given by BFSS
which extended their conjecture beyond the previous work on the matrix
membrane theory.

One important point made by BFSS is that the Hilbert space of the matrix
quantum mechanics theory naturally contains multiple particle states.
This observation, which we discuss in more detail in the following
subsection, resolves the problem of the continuous spectrum discussed above.
Another piece of evidence given by BFSS for their conjecture is the
fact that quantum effects in matrix theory give rise to long-range
interactions between a pair of gravitational quanta (D0-branes) which
have precisely the correct form expected from light-front
supergravity.  This result was first shown by a calculation of
Douglas, Kabat, Pouliot and Shenker \cite{DKPS}; we will discuss this
result and its generalization to more general matrix theory
interactions in Section \ref{sec:matrix-interactions}.

\subsection{Matrix theory as a second quantized theory}
\label{sec:second-quantized}

\begin{figure}[ht]
\begin{center}
\epsfig{figure=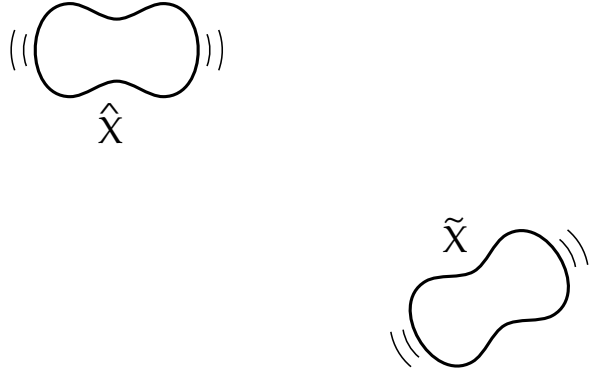, width=0.8\textwidth}
\end{center}
\caption{Two matrix theory objects described by block-diagonal
matrices\label{fig:multiple-objects}}
\end{figure}

The classical equations of motion for a bosonic matrix configuration
with the Hamiltonian (\ref{eq:matrix-Hamiltonian}) are (up to an
overall constant)
\begin{equation}
\ddot{X}^i = -[[X^i, X^j], X^j].
\label{eq:classical-eom}
\end{equation}
If we consider a block-diagonal set of matrices
\[
X^i = \left(\begin{array}{cc}
\hat{X}^i & 0\\
0 & \tilde{X}^i
\end{array} \right)
\]
with first time derivatives $\dot{X}^i$ which are also of
block-diagonal form, then the classical equations of motion for the
blocks are separable
\begin{eqnarray*}
\ddot{\hat{X}}^i & = & -[[\hat{X}^i, \hat{X}^j], \hat{X}^j]\\
\ddot{\tilde{X}}^i & = & -[[\tilde{X}^i, \tilde{X}^j], \tilde{X}^j]
\end{eqnarray*}
If we think of each of these blocks as describing a matrix theory
object with center of mass
\[
 \hat{x}^i = \frac{1}{ \hat{N}}  {\rm Tr}\; \hat{X}^i
\]
\[
 \tilde{x}^i = \frac{1}{ \tilde{N}}  {\rm Tr}\; \tilde{X}^i
\]
then we have two objects obeying classically independent equations of
motion (See Figure~\ref{fig:multiple-objects}).  It is straightforward
to generalize this construction to a block-diagonal matrix
configuration describing $k$ classically independent objects.  This
gives a simple indication of how matrix theory can encode, even in
finite $N$ matrices, a configuration of multiple objects.  In this
sense it is natural to think of matrix theory as a second quantized
theory from the point of view of the target space.

Given the realization that matrix theory should describe a second
quantized theory, the puzzle discussed above regarding the continuous
spectrum of the theory is easily resolved.  If there is a state in
matrix theory corresponding to a single graviton of M-theory (as we
will discuss in more detail in section \ref{sec:gravitons}) with $H =
0$ which is roughly a localized state, then by taking two such states
to have a large separation and a small relative velocity $v$ it should
be possible to construct a two-body state with an arbitrarily small
total energy.  Since the D0-branes of the IIA theory correspond to
gravitons in M-theory with a single unit of longitudinal momentum, we
would therefore naturally expect to have a continuous spectrum of
energies even in the theory with $N = 2$.  This resolves the puzzle
found by de Wit, L\"uscher and Nicolai in a very pleasing fashion, which
suggests that matrix theory is perhaps even more powerful than string
theory, which only gives a first-quantized theory in the target space.

\begin{figure}[ht]
\begin{center}
\epsfig{figure=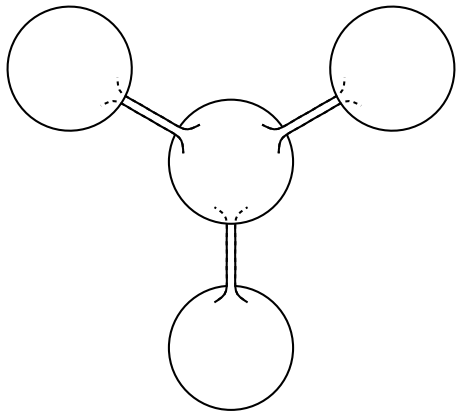, width=0.8\textwidth}
\end{center}
\caption{Membrane of fixed (spherical) topology mapped to multiple
membranes  connected by
tubes in the target space\label{fig:multiple-membranes}}
\end{figure}

The second quantized nature of matrix theory can also be seen
naturally in the continuous membrane theory.  Recall that the
instability of membrane theory appears in the classical theory of a
continuous membrane when we consider the possibility of long thin
spikes of negligible energy, as discussed in section
\ref{sec:instability}.  In a similar fashion, it is possible for a
classical smooth membrane of fixed topology to be mapped to a
configuration in the target space which looks like a system of
multiple distinct macroscopic membranes connected by infinitesimal
tubes of negligible energy (See Figure~\ref{fig:multiple-membranes}).
In the limit where the tubes become very small, their effect on the
classical dynamics of the multiple membrane configuration disappears
and we effectively have a system of multiple independent membranes
moving in the target space.  At the classical level, the sum of the
genera of the membranes in the target space must be equal to or
smaller than the genus of the single world-sheet membrane, but when
quantum effects are included handles can be added to the membrane as
well as removed \cite{Prezas-Taylor}.  These considerations seem to
indicate that any consistent quantum theory which contains a
continuous membrane in its effective low-energy theory must contain
configurations with arbitrary membrane topology and must therefore be
a ``second quantized'' theory from the point of view of the target
space.

\subsection{Matrix theory and DLCQ M-theory}
\label{sec:proof}

A theory which has been compactified on a lightlike circle can be
viewed as a limit of a theory compactified on a spacelike circle
where the size of the spacelike circle becomes vanishingly small in
the limit.  This point of view was used by Seiberg and Sen in
\cite{Seiberg-DLCQ,Sen-DLCQ} to argue that light-front compactified M-theory
is described through such a limiting process by the low-energy
Lagrangian for many D0-branes, and hence by matrix theory.  In this
section we go through this argument in detail.

Consider a space-time which has been compactified on a lightlike
circle by identifying
\begin{equation}
\left(\begin{array}{c}
x\\
t
\end{array} \right) \sim
\left(\begin{array}{c}
x-R/\sqrt{2}\\
t+ R/\sqrt{2}
\end{array} \right) 
\label{eq:light-front-compact}
\end{equation}
This theory has a quantized momentum in the compact direction
\begin{equation}
P^+ = \frac{N}{R} 
\end{equation}
The compactification (\ref{eq:light-front-compact}) can be described
as a limit of a family of spacelike compactifications
\begin{equation}
\left(\begin{array}{c}
x\\
t
\end{array} \right) \sim
\left(\begin{array}{c}
x-\sqrt{R^2/2 + R_s^2}\\
t+ R/\sqrt{2}
\end{array} \right) 
\label{eq:spacelike}
\end{equation}
parameterized by the size $R_s \rightarrow 0$ of the spacelike
circle, which is taken to vanish in the limit.

The system satisfying (\ref{eq:spacelike}) is related through a boost
to a system with the identification
\begin{equation}
\left(\begin{array}{c}
x'\\
t'
\end{array} \right) \sim
\left(\begin{array}{c}
x'- R_s\\
t'
\end{array} \right) 
\label{eq:space-compact}
\end{equation}
where
\begin{equation}
\left(\begin{array}{c}
x'\\
t'
\end{array} \right)  =
\left(\begin{array}{cc}
\frac{1}{ \sqrt{1-\beta^2}} &\frac{\beta}{ \sqrt{1-\beta^2}}\\
\frac{\beta}{ \sqrt{1-\beta^2}}&\frac{1}{ \sqrt{1-\beta^2}}
\end{array} \right) 
\left(\begin{array}{c}
x\\
t
\end{array} \right) 
\label{eq:transform}
\end{equation}
The boost parameter $\beta$ is given by
\begin{equation}
\beta = \frac{1}{ \sqrt{1 + \frac{2 R_s^2}{R^2} }}  \equiv
1-\frac{R_s^2}{ R^2} .
\end{equation}

In the context of matrix theory we are interested in understanding
M-theory compactified on a lightlike circle.  This is related through
the above limiting process to a family of spacelike compactifications
of M-theory, which we know can be identified with the IIA string
theory.  At first glance, it may seem that the limit we are
considering here is difficult to analyze from the IIA point of view.
The IIA string coupling and string length are related to the
compactification radius and 11D Planck length as usual by
\begin{eqnarray}
g & = &  (\frac{R_s}{l_{11}} )^{3/2}\\
l_s^2 & = &  \frac{l_{11}^3}{R_s}  \nonumber
\end{eqnarray}
Thus, in the limit $R_s \rightarrow \infty$ the string coupling $g$ becomes small as
desired; the string length $l_s$, however, goes to $\infty$.  Since
$l_s^2 = \alpha'$, this corresponds to a limit of vanishing string
tension.  Such a limiting theory is very complicated and would not
seem to provide a useful alternative description of the theory.

Let us consider, however, how the energy of the states we are
interested in behaves in the class of limiting theories with
spacelike compactification.  If we want to describe the behavior of a
state which has light-front energy $P^-$ and compact momentum 
$P^+ = N/R$ then the spatial momentum in the theory with spatial $R_s$
compactification is $P' = N/R_s$.  The energy in the spatially
compactified theory is 
\begin{equation}
E' = N/R_s + \Delta E,
\end{equation}
where $\Delta E$  has the energy scale we are interested in
understanding.  The term $N/R_s$ in the energy is simply the
mass-energy of the $N$ D0-branes which correspond to the momentum in
the compactified M-theory direction.  Relating back to the near
lightlike compactified theory we have
\begin{eqnarray}
\left(\begin{array}{c}
P\\
E
\end{array} \right)   & = & 
\left(\begin{array}{cc}
\frac{1}{ \sqrt{1-\beta^2}} &-\frac{\beta}{ \sqrt{1-\beta^2}}\\
-\frac{\beta}{ \sqrt{1-\beta^2}}&\frac{1}{ \sqrt{1-\beta^2}}
\end{array} \right) 
\left(\begin{array}{c}
P'\\
E'
\end{array} \right)
\end{eqnarray}
so
\begin{equation}
P^-=\frac{1}{ \sqrt{2}}  (E-P) =
 \frac{1}{ \sqrt{2}}  \frac{1+\beta}{ \sqrt{1-\beta^2}}  \Delta E
 \approx \frac{R}{R_s}  \Delta E
\end{equation}
As a result we see that the energy $ \Delta E$ of the IIA
configuration needed to approximate the light-front energy $P^-$
is given by
\begin{equation}
 \Delta E \approx P^-\frac{R_s}{R} 
\end{equation}
We know that the string scale $1/l_s$ becomes small as $R_s
\rightarrow 0$.  We can compare the energy scale of interest to this
string scale, however, and find
\begin{equation}
\frac{\Delta E}{(1/l_s)}  = \frac{P^-}{R}  R_sl_s
=\frac{P^-}{R}  \sqrt{R_sl_{11}^3}
\end{equation}
This ratio vanishes in the limit $R_s \rightarrow 0$, which implies
that although the string scale vanishes, the energy scale of interest
is smaller still.  Thus, it is reasonable to study the lightlike
compactification through a limit of spatial compactifications in this
fashion.

To make the correspondence between the light-front compactified theory
and the spatially compactified limiting theories more transparent,
we perform a change of units to a new Planck length
$\tilde{l}_{11}$ in the spatially compactified theories in such a way
that the energy of the states of interest is independent of $R_s$.
For this condition to hold we must have
\begin{equation}
\Delta E \tilde{l}_{11}  = P^-\frac{R_s {l}_{11}^2}{ R \tilde{l}_{11}} 
\end{equation}
where $E, R$ and $P^-$ are independent of $R_s$ and all units have
been explicitly included.  This requires us to keep the quantity
\begin{equation}
\frac{R_s}{  \tilde{l}_{11}^2}
\end{equation}
fixed in the limiting process.  Thus, in the limit $\tilde{l}_{11}
\rightarrow 0$.

We can now summarize the discussion with the following story: to
describe the sector of M-theory corresponding to light-front
compactification on a circle of radius $R$ with light-front momentum
$P^+ = N/R$ we may consider the limit $R_s \rightarrow 0$
of a family of IIA
configurations with $N$ D0-branes where the string coupling and string
length
\begin{eqnarray}
\tilde{g} & = & (R_s/\tilde{l}_{11})^{3/2}\rightarrow 0 \nonumber\\
\tilde{l}_s & = &  \sqrt{ \tilde{l}_{11}^3/R_s} \rightarrow 0
\end{eqnarray}
are defined in terms of a Planck length $\tilde{l}_{11}$ and
compactification length $R_s$ which satisfy
\begin{equation}
R_s / \tilde{l}_{11}^2 = R/l_{11}^2
\end{equation}
All transverse directions scale normally through
\begin{equation}
\tilde{x}^i/\tilde{l}_{11} = x^i/l_{11}
\end{equation}

To give a very concrete example of how this limiting process works,
let us consider a system with a single unit of longitudinal momentum 
\begin{equation}
P^+ =  \frac{1}{R} 
\end{equation}
We know that in the corresponding IIA theory, we
have a single D0-brane whose Lagrangian has the Born-Infeld form
\begin{equation}
{\cal L} = -\frac{1}{\tilde{g} \tilde{l}_s} 
\sqrt{1-\dot{\tilde{x}}^i \dot{\tilde{x}}^i}
\end{equation}
Expanding the square root we have
\begin{equation}
{\cal L} = -\frac{1}{\tilde{g} \tilde{l}_s} 
\left( 1
-\frac{1}{2}\dot{\tilde{x}}^i \dot{\tilde{x}}^i +{\cal O}
(\dot{\tilde{x}}^4)
\right).
\end{equation}
Replacing $\tilde{g} \tilde{l}_s \rightarrow R_s$ and
$\tilde{x}\rightarrow x \tilde{l}_{11}/l_{11}$ gives
\begin{equation}
{\cal L} = -\frac{1}{R_s}  + \frac{1}{2 R}  \dot{x}^i \dot{x}^i +{\cal
O} (R_s/R).
\end{equation}
Thus, we see that all the higher order terms in the Born-Infeld action
vanish in the $R_s \rightarrow 0$ limit.  The leading term is the
D0-brane energy $1/R_s$ which we subtract to compare to the M-theory
light-front energy $P^-$.  Although we do not know the full form of
the nonabelian Born-Infeld action describing $N$ D0-branes in IIA, it
is clear that an analogous argument shows that all terms in this
action other than those in the nonrelativistic supersymmetric matrix
theory action will vanish in the limit $R_s \rightarrow 0$.

This argument apparently demonstrates that matrix theory gives a
complete description of the dynamics of DLCQ M-theory.  There are
several caveats which should be taken into account, however, with
respect to this discussion.  First, in order for this argument to be
correct, it is necessary that there exists a well-defined theory with
the properties expected of M-theory, and that there exist a
well-defined IIA string theory which arises as the compactification of
M-theory.  Neither of these statements is at this point definitely
established.  Thus, this argument must be taken as contingent upon the
definition of these theories.  
Second, although we know that 11D supergravity arises as
the low-energy limit of M-theory, this argument does not necessarily
indicate that matrix theory describes DLCQ supergravity in the
low-energy limit.  It may be that to make the connection to
supergravity it is necessary to deal with subtleties of the large $N$
limit.

In the remainder of these lectures we will discuss some more explicit
approaches to connecting matrix theory with supergravity.  In
particular, we will see how far it is possible to go in demonstrating
that 11D supergravity arises from calculations in the finite $N$
version of matrix theory, which is a completely well-defined theory.
In the last sections we will return to a more general discussion of
the status of matrix theory.


\vspace{0.2in}

\section{M-theory objects from matrix theory}
\label{sec:matrix-objects}
\setcounter{equation}{0}

In this section we discuss how the matrix theory degrees of freedom
can be used to construct the various objects of M-theory: the
supergraviton, supermembrane and 5-brane.  We discuss each of these
objects in turn in subsections \ref{sec:gravitons},
\ref{sec:matrix-membranes}, \ref{sec:5-branes}, after which we give a
general discussion of the structure of extended objects and their
charges in subsection \ref{sec:extended}

\subsection{Supergravitons}
\label{sec:gravitons}

Since in DLCQ M-theory there should be a pointlike state corresponding
to a longitudinal graviton with $p^+ = N/R$ and arbitrary transverse
momentum $p^i$, we expect from the massless condition $m^2 = -p^Ip_I =
0$ that such an object will have matrix theory energy
\[
E = \frac{p_i^2}{ 2p^+} 
\]
We discuss such states first classically and then in the quantum theory.

\subsubsection{Classical supergravitons}
\label{sec:classical-gravitons}

The classical matrix theory potential is $-[X^i, X^j]^2$, from which
we have the classical equations of motion 
\[
\ddot{X}^i = -[[X^i, X^j], X^j].
\]
One simple class of solutions to these equations of motion can be
found when the matrices minimize the potential at all times and
therefore all commute.  Such solutions are of the form
\[
X^i = \left(\begin{array}{cccc}
x^i_1 + v^i_1 t & 0 &0 &  \ddots\\
0 & x^i_2 +v^i_2 t &\ddots & 0\\
0 & \ddots & \ddots & 0\\
 \ddots & 0 & 0 & x^i_N +v^i_N t
\end{array}\right)
\]
This corresponds to a classical $N$-graviton solution, where each
graviton has
\[
p_a^+ = 1/R \;\;\;\;\; p^i_a = v^i_a/R \;\;\;\;\; E_a = v_a^2/(2 R) =
(p_a^i)^2/2p^+
\]
A single classical graviton with $p^+ = N/R$ can be formed by setting
\[
x_1^i = \cdots = x_N^i, \;\;\;\;\; v_1^i = \cdots = v_N^i
\]
so that the trajectories of all the components are identical.
Although this may seem like a very simple model for a graviton, it is
precisely such matrix configurations which are used as a background in
most computations of quantum effects in matrix theory corresponding to
gravitational interactions, as will be discussed further in the following.

\subsubsection{Quantum supergravitons}
\label{sec:quantum-gravitons}

The picture of a supergraviton in quantum matrix theory is somewhat
more subtle than the simple classical picture just discussed.  Let us
first consider the case of a single supergraviton with $p^+ = 1/R$.
This corresponds to the U(1) case of the super Yang-Mills quantum
mechanics theory.  The Hamiltonian is simply 
\[
H = \frac{1}{2 R}  \dot{X}^2
\]
since all commutators vanish in this theory.  The bosonic part of the
theory is simply a free nonrelativistic particle.  In the fermionic
sector there are 16 spinor variables with anticommutation relations
\[
\{\theta_\alpha, \theta_\beta\} = \delta_{\alpha \beta}.
\]
By using the standard trick of writing these as 8 fermion creation
and annihilation operators
\[
\theta_i^{\pm} = \frac{1}{ \sqrt{2}}  (\theta_i \pm \theta_{i + 8}) \;
\;\;\;\;\; 1 \leq i \leq 8
\]
we see that the Hilbert space for the fermions is a standard fermion
Fock space of dimension $2^8 = 256$.  Indeed, this is precisely the
number of states needed to represent all the polarization states of
the graviton (44), the antisymmetric 3-tensor field (84) and the
gravitino (128).  For details of how the polarization states are
represented in terms of the fermionic Fock space, see
\cite{dhn,Plefka-Waldron}.

The case when $N > 1$ is much more subtle.  We can factor out the
overall $U(1)$ so that every state in the $SU(N)$ quantum mechanics
theory has 256 corresponding states in the full theory.  For the
matrix theory conjecture to be correct, as BFSS pointed out, it should
then be the case that for every $N$ there exists a unique threshold
bound state in the $SU(N)$ theory with $H = 0$.  As mentioned before,
no definitive answer as to the existence of such a state was given in
the early work on matrix theory.
This result was finally proven by Sethi and Stern for $N = 2$ in
\cite{Sethi-Stern}.  Progress towards proving the result for arbitrary
values of $N$ was made in
\cite{Porrati-Rozenberg,Green-Gutperle-d,mns}, and the result for a
general gauge group was given in \cite{Kac-Smilga} (see also \cite{hkr}).

\subsection{Membranes}
\label{sec:matrix-membranes}

In this section we discuss the description of M-theory membranes in
terms of the matrix quantum mechanics degrees of freedom. It is clear
from the derivation of matrix theory as a regularized supermembrane
theory that there must be matrix configurations which in the large
$N$ limit give arbitrarily good descriptions of any membrane
configuration.  It is instructive, however, to study in detail the
structure of such membrane configurations.  In subsection
\ref{sec:D2-branes-D0-branes} we discuss the significance of the
matrix representation of membranes in the language of type IIA
D0-branes.  In subsection \ref{sec:spherical-membranes} we discuss in
some detail how a spherical membrane can be very accurately described
by matrices even with small values of $N$.  In subsection
\ref{sec:higher-genus-membranes} we discuss higher genus matrix
membranes.  In subsection \ref{sec:infinite-membranes} we discuss
noncompact matrix membranes, and finally in subsection
\ref{sec:matrix-strings} we discuss M-theory membranes which are
wrapped on the longitudinal direction and appear as strings in the IIA theory.

\subsubsection{D2-branes from D0-branes}
\label{sec:D2-branes-D0-branes}

As we have mentioned, it is clear from inverting the matrix membrane
regularization procedure that smooth membranes can be approximated by
finite size matrices.  This construction may seem less natural in the
language of type IIA string theory, where it corresponds to a
construction of a IIA D2-brane out of the degrees of freedom
describing a system of $N$ D0-branes.  In fact, however, the fact that
this construction is possible is simply the T-dual of the familiar
statement that D0-branes are described by the magnetic flux of the
gauge field living on a set of $N$ D2-branes \cite{Douglas}.  Both of
these statements can in turn be seen by performing T-duality on a
diagonally wrapped D1-brane on a 2-torus.

To see this explicitly,
consider a set of $N$ D2-branes on a torus $T^2$ with $k$  units of
magnetic flux
\begin{equation}
\frac{1}{2 \pi} \int F =  k
\label{eq:flux-unit}
\end{equation}
Under a T-duality transformation on one direction of the torus, the
gauge field component $A_{2}$ is replaced by an infinite matrix
\[
X^{2} = i \partial_2 + A_2
\]
representing a transverse scalar field for a set of $N$ D1-branes
living on the dual torus $(T^2)^{*}$.  These matrices are infinite
because they contain information about winding strings connecting the
infinite number of copies of each brane which live on the infinite
covering space of the dual torus.  (This construction is described in
more detail in section \ref{sec:T-duality}.)  This T-dual
configuration corresponds to a single D-string which is diagonally
wound $N$ times around the $x^1$ direction and $k$ times around the
$x^2$ direction; this can be seen from the fact that the T-dual of
(\ref{eq:flux-unit}) is $\partial_1 X^2 = (k L_2^*/N L_1) \identity$.
Since under T-duality in the $x^2$ direction a D1-brane wrapped in the
$x^2$ direction becomes a D0-brane, we can identify the flux
(\ref{eq:flux-unit}) with $k$ D0-branes in the original theory.

Further T-dualizing in the direction $x^1$, we replace
\[
X^{1} = i \partial_1 + A_1.
\]
where $X^1, X^2$ are now infinite matrices describing transverse
fields of a system of $N$ D0-branes on the dual torus $(T^2)^{**}$.
When the normalization constants
are treated carefully, the flux condition (\ref{eq:flux-unit}) now becomes
the condition on the D0-brane matrices
\begin{equation}
{\rm Tr}\;[X^1, X^2] = \frac{iAk}{2 \pi} 
\label{eq:D2-brane-charge}
\end{equation}
where $A$ is the area of the dual torus.  Since the T-dual in the
$x^1$ direction of the D-string wrapped in the $x^2$ direction is a
D2-brane, we interpret $k$ in (\ref{eq:D2-brane-charge}) as the D2-brane
charge of a system of $N$ D0-branes.

This construction can be interpreted more generally, so that in
general a pair of matrices $X^a, X^b$ describing a D0-brane
configuration satisfying
\begin{equation}
{\rm Tr}\;[X^a, X^b] = \frac{iA}{2 \pi} 
\label{eq:local-membrane}
\end{equation}
should be interpreted as giving rise to a piece of a D2-brane of area
$A$.  Of course, for finite matrices the trace of the commutator must
vanish.  This is simply a consequence of the fact that the net
D2-brane charge of any compact object must vanish.  However, not only is it
possible to have a nonzero membrane charge when the matrices are
infinite, but it is also possible to treat (\ref{eq:local-membrane})
as a local expression by restricting the trace to a subset of the
diagonal elements.  We will see a specific example of this in the next
subsection.  The local relation (\ref{eq:local-membrane}) will also be
useful in constructing higher moments of the membrane charge, which
can be nonzero even for finite size configurations, as we shall
discuss later.

\subsubsection{Spherical membranes}
\label{sec:spherical-membranes}

One extremely simple example of a membrane configuration which can be
approximated very well even at finite $N$ by simple matrix
configurations is the symmetric spherical membrane \cite{Dan-Wati}.
Imagine that we wish to construct a membrane embedded in an isotropic
sphere
\[
x_1^2 + x_2^2 + x_3^2 = r^2
\]
in the first three dimensions of $\br^{11}$.  The embedding functions
for such a continuous membrane can be written as linear functions
\[
X^i = r \xi^i \; \;\;\;\;\; 1 \leq i \leq 3
\]
of the three Euclidean coordinates $\xi^i$ on the spherical
world-volume.  Using the matrix-membrane correspondence
(\ref{eq:mm-correspondence}) we see that the matrix approximation to
this membrane will be given by the $N \times N$ matrices
\begin{equation}
{\bf X}^i = \frac{2r}{ N}  J^i \; \;\;\;\;\; 1 \leq i \leq 3
\label{eq:matrix-sphere}
\end{equation}
where $J^i$ are the generators of $SU(2)$ in the
$N$-dimensional representation.

It is quite interesting to see how many of the geometrical and
physical properties of the sphere can be extracted from the algebraic
structure of these matrices, even for small values of $N$.  We list
here some of these properties.

\noindent {\bf i) Spherical locus:} The matrices
(\ref{eq:matrix-sphere}) satisfy
\[
{\bf X}_1^2 + {\bf X}_2^2 + {\bf X}_3^2 = \frac{4r^2}{ N^2}  C_2 (N)\identity
= r^2 (1-1/N^2)\identity
\]
where $C_2 (N) = (N^2 -1)/4$ is the quadratic Casimir of $SU(2)$ in
the $N$-dimensional representation.  This shows that the D0-branes are
in a noncommutative sense ``localized'' on a sphere of radius $r
+{\cal O} (1/N^2)$.

\noindent {\bf ii)  Rotational invariance:} The matrices
(\ref{eq:matrix-sphere}) satisfy
\[
R_{ij} {\bf X}_j = U (R) \cdot {\bf X}_i \cdot U (R^{-1})
\]
where $R \in SO(3)$ and $U (R)$ is the $N$-dimensional representation
of $R$.  Thus, the spherical matrix configuration is rotationally
invariant up to a gauge transformation.

\noindent {\bf iii)   Spectrum:} The matrix ${\bf X}^3 = 2rJ_3/N$ (as well
as the other matrices) has a spectrum of eigenvalues which are
uniformly distributed in the interval $[-r, r]$.  This is precisely
the correct distribution if we imagine a perfectly symmetric sphere
with D0-branes distributed uniformly on its surface and project this
distribution onto a single axis.

\noindent {\bf iv)  Local membrane charge:} As discussed above, the
expression (\ref{eq:local-membrane}) gives an area for a piece of a
membrane described by a pair of matrices.  We can use this formula to
check the interpretation of the matrix sphere.  We do this by
computing the membrane charge in the 1-2 plane of the half of the
configuration with eigenvalues $X^3 > 0$.  This should correspond to
the projected area of the ``upper hemisphere'' of the sphere.  We compute
\[
A_h =-2 \pi i {\rm Tr}_{1/2} \;[{\bf X}^1, {\bf X}^2]
\]
where the trace is restricted to the set of eigenvalues where $X^3 >
0$ in the standard representation.  This is possible since $[{\bf X}^1, {\bf X}^2]
\sim {\bf X}^3$.  We find
\[
A_h = 2 \pi \frac{4}{ N^2}  r^2 {\rm Tr}_{1/2}\; J_3
= \pi r^2 (1 +{\cal O} (1/N^2))
\]
thus, we find precisely the expected area of the projected hemisphere.

\noindent {\bf v)  Energy:}
In M-theory we expect the tension energy of a (momentarily) stationary
membrane sphere to be
\[
e = \frac{4 \pi r^2}{(2 \pi)^2 l_{11}^3}  = \frac{r^2}{\pi l_{11}^3} 
\]
Using $p^I p_I = -e^2$ we see that the light-front energy should be
\begin{equation}
E = \frac{e^2}{ 2p^+} 
\label{eq:membrane-energy}
\end{equation}
in 11D Planck units.  Let us compute the matrix membrane energy.  It
is given by
\[
E = -\frac{1}{4 R} [{\bf X}^i, {\bf X}^j]^2 = \frac{2r^4}{ N R}  +{\cal O} (N^{-3})
\]
in string units.  This is easily seen to agree with
(\ref{eq:membrane-energy}).

It is also straightforward to verify that the equations of motion for
the membrane are correctly reproduced in matrix theory.

Thus, we see that many of the geometrical and physical properties of
the membrane can be extracted from algebraic information about the
structure of the appropriate membrane configuration.  The discussion
we have carried out here has only applied to the simple case of the
rotationally invariant spherically embedded membrane.  It is
straightforward to extend the discussion to a membrane of spherical
topology and arbitrary  shape, however, simply by using the
matrix-membrane correspondence (\ref{eq:mm-correspondence}) to
construct matrices approximating an arbitrary smooth spherical
membrane.  We now turn to the question of membranes with non-spherical
topology.

\subsubsection{Higher genus membranes}
\label{sec:higher-genus-membranes}

So far we have only discussed membranes of spherical topology.  It is
possible to describe compact membranes of arbitrary genus by
generalizing this construction, although an explicit construction is
only known for the sphere and torus.  In this section we give a brief
description of the matrix torus, following the work of Fairlie,
Fletcher and Zachos \cite{ffz1,Fairlie-Zachos,ffz2}.

We consider a torus defined by two coordinates $x_1, x_2 \in[0, 2
\pi]$ with symplectic form $\omega_{ij} = \epsilon_{ij}/\pi$
corresponding to a total volume $\int d^2 x \; \omega = 4 \pi$ as in the
case of the sphere  discussed in section
\ref{sec:matrix-regularization}.
As in the case of the sphere we wish to find a map from functions
on the torus to matrices which is compatible with the correspondence
\begin{equation}
\{\cdot,\cdot\} \leftrightarrow  {-i N \over 2} [\cdot,\cdot]
\;\;\;\;\; \;\;\;\;\;
{1 \over 4 \pi^2} \int d^2 x \leftrightarrow \frac{1}{N} 
\tr
\end{equation}
A natural (complex)
basis for the functions on $T^2$ is given by the Fourier
modes
\begin{equation}
y_{nm} (x_1, x_2) = e^{inx_1 + imx_2}
\end{equation}
The real functions on $T^2$ are given by the linear combinations
\begin{equation}
\frac{1}{2} \left( y_{nm} + y_{-n\, -m} \right), \;\;\;\;\;
\frac{-i}{2} \left( y_{nm} - y_{-n\, -m} \right).
\end{equation}
The Poisson bracket algebra of the functions $y_{nm}$ is
\begin{equation}
\{y_{nm}, y_{n' m'}
\}=-\pi  (nm' -mn') y_{n + n',m + m'}
\label{eq:torus-Poisson}
\end{equation}

To describe the matrix approximations for these functions we use the
't Hooft matrices
\begin{equation}
U = 
\pmatrix{
1& & & & \cr
&q && & \cr
& &q^2 & &\cr
& & & \ddots & \cr
& & &&  q^{N-1}
} 
\end{equation}
and
\begin{equation}
V =
\pmatrix{
 & 1 & && \cr
 &   & 1 && \cr
 &   &   & \ddots & \cr
& & & & 1\cr
1&   &   &  & 
} 
\end{equation}
where
\begin{equation}
q = e^{2\pi i\over N}.
\end{equation}
The matrices $U, V$ satisfy
\begin{equation}
UV = q^{-1}VU.
\end{equation}
In terms of these matrices we can define
\begin{equation}
Y_{nm} = q^{nm/2}
U^nV^m = q^{-nm/2} V^mU^n
\end{equation}
so that the matrix approximation to an arbitrary function 
\begin{equation}
f(x_1, x_2) = \sum_{n, m}c_{nm} y_{nm} (x_1, x_2)
\end{equation}
is given by
\begin{equation}
F = \sum_{n, m}c_{nm} Y_{nm} .
\end{equation}
By computing
\begin{eqnarray*}
[Y_{nm}, Y_{n' m'}] &= & (q^{(mn' -nm')/2}
-q^{(nm' -mn')/2}) Y_{n + n', m + m'}\\
& 
\rightarrow & \frac{2 \pi i}{N}  (mn' -nm')Y_{n + n', m + m'}
\end{eqnarray*}
We see that for fixed $n, m, n', m'$ in the large $N$ limit the matrix
commutation relations correctly reproduce (\ref{eq:torus-Poisson})
just as in the case of the sphere.

As a concrete example let us consider embedding a torus into $\br^4
\subset\br^9$ so that the membrane fills the locus of points
satisfying
\begin{equation}
X_1^2 + X_2^2 = r^2 \;\;\;\;\; \;\;\;\;\;
X_3^2 + X_4^2 = s^2.
\label{eq:torus-locus}
\end{equation}
Such a membrane configuration can be realized through the following
matrices
\begin{eqnarray}
 {\bf X}_1 & = &  \frac{ r}{2} (U + U^{\dagger}) \nonumber\\
 {\bf X}_2 & = &  \frac{-ir}{2} (U - U^{\dagger})\label{eq:symmetric-torus}\\
 {\bf X}_3 & = &  \frac{s}{2} (V + V^{\dagger})\nonumber\\
 {\bf X}_4 & = &  \frac{-is}{2} (V - V^{\dagger})\nonumber
\end{eqnarray}
It is straightforward to check that this matrix configuration has
geometrical properties analogous to those of the matrix membrane
sphere discussed in the previous subsection.  In particular, the
equation (\ref{eq:torus-locus}) is satisfied identically as a matrix
equation.  Note, however that this configuration is not gauge
invariant under $U(1)$ rotations in the 12 and 34 planes---only under a
$\bz_N$ subgroup of each of these $U(1)$'s.

\subsubsection{Infinite membranes}
\label{sec:infinite-membranes}

So far we have discussed compact membranes, which can be described in
terms of finite-size $N \times N$ matrices.  In the large $N$ limit
it is also possible to construct membranes with infinite spatial extent.
The matrices $X^i$ describing such configurations are
infinite-dimensional matrices which correspond to operators on a
Hilbert space.  Infinite membranes are of particular interest because
they can be BPS states which solve the classical equations of motion
of matrix theory.  Extended compact membranes cannot be static
solutions of the equations of motion since their membrane tension
always causes them to contract and oscillate, as in the case of the
spherical membrane.

The simplest infinite membrane is the flat planar membrane
corresponding in IIA theory to an infinite D2-brane.  This solution
can be found by looking at the limit of the spherical membrane at
large radius.  It is simpler, however, to simply directly construct
the solution by regularizing the flat membrane of M-theory.  As in the
other cases we have studied, we wish to quantize the Poisson bracket
algebra of functions on the brane.  Functions on the infinite membrane
can be described in terms of two coordinates $x_1, x_2$ with a
symplectic form $\omega_{ij} = \epsilon_{ij}$
giving a Poisson bracket
\begin{equation}
\{f (x_1, x_2),g (x_1, x_2)\} = \partial_1f \partial_2g-\partial_1g
\partial_2f.
\end{equation}
This algebra of functions can be ``quantized'' to the algebra of
operators generated by $Q, P$ satisfying
\begin{equation}
[Q, P] = \frac{i \epsilon^2}{2 \pi} \identity
\end{equation}
where $\epsilon$ is a constant parameter.  As usual in the
quantization process there are operator-ordering ambiguities which
must be resolved in determining a general map from functions expressed
as polynomials in $x_1, x_2$
to operators expressed as polynomials of $Q, P$.

This gives a map from functions on $\br^2$ to operators which allows
us to describe fluctuations around a flat membrane geometry with a
single unit of $P^+ = 1/R$ in each region of area $\epsilon^2$ on the
membrane.  Configurations of this type were discussed in the original
BFSS paper \cite{BFSS} and their existence used as additional evidence
for the validity of their conjecture.  Note that this configuration
only makes sense in the large $N$ limit.

In addition to the flat membrane solution there are other infinite
membranes which are static solutions of M-theory in flat space.  In
particular, there are BPS solutions corresponding to membranes which
are holomorphically embedded in $\bc^4 =\br^8 \subset\br^9$.  These
are static solutions of the membrane equations of motion.  Finding a
matrix theory description of such membranes is possible but involves
some somewhat subtle issues related to choosing a regularization which
preserves the complex structure of the brane.  The details of this
construction for a general holomorphic membrane are discussed in
\cite{Cornalba-Taylor}.

\subsubsection{Wrapped membranes as matrix strings}
\label{sec:matrix-strings}

So far we have discussed M-theory membranes which are unwrapped in the
longitudinal direction and which therefore appear as D2-branes in the
IIA language of matrix theory.  It is also possible to describe
wrapped M-theory membranes which correspond to strings in the IIA
picture.  The charge in matrix theory which measures the number of
strings present is proportional to
\begin{equation}
 \frac{i}{ R}  {\rm Tr}\;
\left([X^i, X^j] \dot{X}^j +[[X^i, \theta^{\dot{\alpha}}],
\theta^{\dot{\alpha}}] \right)
\label{eq:string-charge}
\end{equation}
This result can be understood in several ways.  It was found in
\cite{bss} as a central charge in the matrix theory SUSY algebra
corresponding to string charge; we will discuss this algebra further
in the subsection \ref{sec:extended}.  An intuitive way of understanding why
(\ref{eq:string-charge}) measures string charge is by a T-duality
argument analogous to that used in \ref{sec:D2-branes-D0-branes} to
derive the D2-brane charge of a system of D0-branes.  If we compactify
on a 2-torus in the $i$ and $j$ directions, the D0-branes become
D2-branes and the bosonic part of (\ref{eq:string-charge}) becomes
\begin{equation}
\frac{1}{R} 
F^{ij} F_{j0}.
\end{equation}
This is the part of the energy-momentum tensor usually referred to as
the Poynting vector in the 4D theory, and which describes momentum in
the $i$ direction.  Such momentum is of course T-dual to string
winding in the original picture, so we understand the identification
of the original charge (\ref{eq:string-charge}) as counting
fundamental IIA strings corresponding to wound M-theory membranes.
Configurations with nonzero values of this charge were considered by
Imamura in \cite{Imamura-string}.

To realize a classical configuration  in matrix theory which contains
fundamental strings it is clear from the form of the charge that we
need to construct configuration with local membrane charge extended in
a pair of directions $X^i, X^j$ and to give the D0-branes velocity in
the $X^j$ direction.  For example, we could consider an infinite
planar membrane (as discussed in the previous subsection) sliding
along itself according to the equation
\begin{eqnarray}
X^1 & = &  Q+ t \identity\\
X^2 & = &  P 
\end{eqnarray}
This corresponds to an M-theory membrane which has a projection onto
the $X^1, X^2$ plane and which wraps around the compact direction as a
periodic function of $X^1$ so that the IIA system contains a D2-brane
with infinite strings extended in the $X^2$ direction since
\begin{equation}
 \dot{X}^1 [X^1, X^2]  \sim  \identity.
\end{equation}

Another example of a matrix theory system containing fundamental
strings can be constructed by spinning the torus from
(\ref{eq:symmetric-torus}) in
the 12
plane to stabilize it.  This gives the system some fundamental
strings wrapped around the 34 circle.  By taking the radius $r$ to be
very small we can construct a configuration of a single fundamental
string wrapped in a circle of radius $s$.  As $s\rightarrow \infty$
this becomes an infinite fundamental string.

It is interesting to note that there is no classical matrix theory
solution corresponding to a classical string which is truly
1-dimensional and has no local membrane charge.  This follows from the
appearance of the commutator $[X^i, X^j]$ in the string charge, which
vanishes unless the matrices describe a configuration with at least
two dimensions of spatial extent.  We can come very close to a
1-dimensional classical string configuration by considering a
one-dimensional array of D0-branes at equal intervals on the $X^1$
axis
\begin{equation}
X^1 = a \left(\begin{array}{ccccc}
\ddots & \ddots & \ddots & &\\
\ddots & 1 & 0 &\ddots &\\
\ddots & 0 & 0 & 0 & \ddots\\
\ddots & \ddots & 0 & -1 & \ddots\\
& & \ddots & \ddots & \ddots
\end{array}\right)
\end{equation}
We can now construct an excitation of the off-diagonal elements
of $X^2$ corresponding to a string threading through the line of
D0-branes
\begin{equation}
X^2 = b  \left(\begin{array}{ccccc}
\ddots & \ddots & \ddots & &\\
\ddots & 0 &  e^{i \omega t} &\ddots &\\
\ddots & e^{-i \omega t} & 0 & e^{i \omega t} & \ddots\\
 \ddots & \ddots & e^{-i \omega t} & 0 & \ddots\\
& & \ddots & \ddots & \ddots
\end{array}\right)
\label{eq:string-mode}
\end{equation}
where $\omega = a$.  In the classical theory, this configuration can
have arbitrary string charge.  If the mode (\ref{eq:string-mode}) is
quantized then the string charge is quantized in the
correct units.
This string is almost 1-dimensional but has a small additional extent
in the $X^2$ direction corresponding to the extra dimension of the
M-theory membrane.  From the M-theory point of view this extra dimension
must appear because the membrane cannot have momentum in a direction
parallel to its direction of extension since it has no internal
degrees of freedom.  Thus, the momentum in the
compact direction represented by the D0-branes must appear on the
membrane as a  fluctuation in some transverse direction.

\subsection{5-branes}
\label{sec:5-branes}

The M-theory 5-brane can appear in two possible
guises in type IIA string theory.  If the 5-brane is wrapped around
the compact direction it becomes a D4-brane in the IIA theory, while if
it is unwrapped it appears as an NS 5-brane.  
We will refer to these two configurations as ``longitudinal'' and
``transverse'' 5-branes in matrix theory.  We begin by discussing the
transverse 5-brane.

{\it A priori}, one
might think that it should be possible to see both types of 5-branes in
matrix theory.    
Several calculations, however, indicate that the transverse 5-brane
does not carry a conserved charge which can be described in terms of
the matrix degrees of freedom.  In principle, if this charge existed
we would expect it to appear both in the supersymmetry algebra of
matrix theory  (discussed in the next subsection) and in the set of
supergravity currents whose interactions are described by perturbative
matrix theory calculations (discussed in section
\ref{sec:2-body-linear}).  In fact, no charge or current with the
proper tensor structure for a transverse 5-brane appears in either of
these calculations.

One way of understanding this apparent puzzle is by comparing to the
situation for D-branes in light-front string theory \cite{bss}.
Due to the Virasoro
constraints, strings in the light-front
formalism must have Neumann boundary conditions in both the
light-front directions $X^+,X^-$.  Thus, in light-front string theory
there are no transverse D-branes which can be used as boundary
conditions for the string.  A similar situation holds for
membranes in M-theory, which can end on M5-branes.  The
boundary conditions on the bosonic membrane fields which can be
derived from the action (\ref{eq:membrane-action-h})
state that
\begin{equation}
(\bar{h} h^{ab} \partial_b X^i) \delta X^i = 0
\end{equation}
Combined with the Virasoro-type constraint
\begin{equation}
\partial_a X^-= \dot{X}^i \partial_a X^i
\end{equation}
we find that, just as in the string theory case, membranes must have
Neumann boundary conditions in the light-front directions.

These considerations would seem to lead to the conclusion that
transverse 5-branes simply cannot be constructed in matrix theory.  On
the other hand, it was argued in \cite{grt} that there
may be a way to construct a transverse 5-brane using S-duality, at
least when the theory has been compactified on a 3-torus.  
To construct an infinite extended transverse 5-brane in this fashion
would require performing an S-duality on $(3 + 1)$-dimensional ${\cal
N} = 4$ supersymmetric Yang-Mills theory with gauge group $U(\infty)$,
which is a poorly understood procedure to say the least.  In
\cite{Mark-Wati-5}, however, a finite size transverse 5-brane with
geometry $T^3 \times S^2$ was constructed using S-duality of the
four-dimensional $U(N)$ with finite $N$.  Furthermore, it was shown
that this object couples correctly to the supergravity fields even
in the absence of an explicit transverse 5-brane charge.  This
seems to indicate that transverse 5-branes in matrix theory can be
constructed locally, but that they are essentially solitonic objects
and do not carry independent conserved quantum numbers.  It would be
nice to have a more explicit construction of a general class of such
finite size transverse 5-branes, particular in the noncompact version
of matrix theory.

We now turn to the wrapped, or ``longitudinal'', M5-brane which we
will refer to as the ``L5-brane''.  This object appears as a D4-brane
in the IIA theory.  An infinite D4-brane was considered as a matrix
theory background in \cite{Berkooz-Douglas} by including extra fields
corresponding to strings stretching between the D0-branes of matrix
theory and the background D4-brane.  As in the case of the membrane,
however, we would like to find a way to explicitly describe a
dynamical L5-brane using the matrix degrees of freedom.  Just as for
the D2-brane, it may be surprising that a D4-brane can be constructed
from a configuration of D0-branes.  This can be seen from the same
type of T-duality argument we used for the D2-brane in
\ref{sec:D2-branes-D0-branes}.  By putting D4-branes and D0-branes on
a torus $T^4$ we find that the charge-volume relation analogous to
(\ref{eq:D2-brane-charge}) for a D4-brane is 
\cite{grt,WT-Trieste}
\begin{equation}
{\rm Tr}\;\epsilon_{ijkl} X^i X^j X^k X^l = \frac{V}{2 \pi^2} 
\end{equation}
This is the T-dual of the instanton number in a 4D gauge theory which
measures D0-brane charge on D4-branes.

Unlike the case of the membrane, there is no general theory describing
an arbitrary L5-brane geometry in matrix theory language.  In fact,
the only L5-brane configurations which have been explicitly
constructed to date are those corresponding to the highly symmetric
geometries $S^4,\bc P^2$ and $\br^4$.  We now make a few brief
comments about these configurations.

The L5-brane with isotropic $S^4$ geometry is similar in many ways to
the membrane with $S^2$ geometry discussed in section
\ref{sec:spherical-membranes}.  There are a number of unusual features
of the $S^4$  system, however, which deserve mention.  For full
details of the construction see \cite{clt}.

A rotationally invariant spherical L5-brane can only be constructed
for those values of $N$ which are of the form
\begin{equation}
N = \frac{(n + 1) (n + 2) (n + 3)}{3} 
\end{equation}
where $n$ is integral.  For $N$ of this form we define the
configuration by
\begin{equation}
X_i = \frac{r}{n}  G_i, \;\;\;\;\;  i \in \{1,  \ldots,  5\}.
\label{eq:sphere}
\end{equation}
where $G_i$ are the generators of the $n$-fold symmetric tensor
product representation of the five four-dimensional Euclidean gamma matrices
$\Gamma_i$ satisfying $\Gamma_i \Gamma_j + \Gamma_j \Gamma_i = 2 \delta_{ij}$
\[
G^{(n)}_i = \left(
\Gamma_i \otimes \identity \otimes \cdots \otimes \identity+
\identity \otimes \Gamma_i \otimes \identity \otimes \cdots \otimes
\identity+ \cdots + 
 \identity \otimes \cdots \otimes \identity
\otimes\Gamma_i\right)_{{\rm S}}
\]
where the subscript $S$ indicates that only the completely symmetric
representation is used.
For any $n$ this configuration has the geometrical properties expected
of $n$ superimposed L5-branes contained in the locus of points
describing a 4-sphere.  As for the spherical membrane discussed in
\ref{sec:spherical-membranes} the configuration is confined to the
appropriate spherical locus
\begin{equation}
X_1^2 + X_2^2 + X_3^2 + X_4^2 + X_5^2 \approx r^2 \identity.
\end{equation}
The configuration is symmetric under $SO(5)$ and has the correct
spectrum and the local D4-brane charge of $n$ spherical branes.  The
energy and equations of motion of this system agree with those
expected from M-theory.

Although the system can only be defined in a completely symmetric
fashion for certain values of $n, N$, this does not seem like a
fundamental issue.  This constraint is a consequence of the imposition
of exact rotational symmetry on the system.  It may be that for large
and arbitrary $N$ it is possible to construct a very good
approximation to a spherical L5-brane which breaks rotational
invariance to a very small degree.  A more fundamental problem,
however, is that there is no obvious way of including small
fluctuations of the membrane geometry around the perfectly isotropic
sphere in a systematic way.  In the case of the membrane, we know that
for any particular geometry the fluctuations around that geometry can
be encoded into matrices which form an arbitrarily good approximation
to a smooth fluctuation through the procedure of replacing functions
described in terms of an orthonormal basis by appropriate matrix
analogues.  In the case of the L5-brane we have no such procedure.  In
fact, there seems to be an obstacle to including all degrees of
freedom corresponding to local fluctuations of the brane.  It is
natural to speculate by analogy with the membrane case that arbitrary
fluctuations should be encoded in symmetric polynomials in the
matrices $G_i$.  It can be shown, however, that this is not possible.
This geometry has been discussed in a related context in the
noncommutative geometry literature \cite{gkp} as a noncommutative
version of $S^4$.  There also, it was
found that not all functions on the sphere could be consistently
quantized.  

As for the infinite membrane,
the infinite L5-brane with geometry of a flat $\br^4 \subset\br^9$ can
be viewed as a local limit of a large spherical geometry or it can be
constructed directly.  We need to find a set of operators $X^{1-4}$ on
some Hilbert space
satisfying
\begin{equation}
\epsilon_{ijkl} X^i X^j X^k X^l  = \frac{\epsilon^4}{2 \pi^2}  \identity.
\end{equation}
Such a configuration can be constructed using matrices which are
tensor products of the form $\identity \otimes Q, P$ and $Q, P \otimes
\identity$.  This gives a ``stack of D2-branes'' solution with
D2-brane charge as well as D4-brane charge \cite{bss}.  It is also
possible to construct a configuration with no D2-brane charge by
identifying $X^a$ with the components of the covariant derivative
operator for an instanton on $S^4$
\begin{equation}
X^i = i \partial^i + A_i.
\end{equation}
This construction is known as the Banks-Casher
instanton \cite{Banks-Casher}.  Just as for the spherical L5-brane,
it is not known how to construct small fluctuations of the membrane
geometry around any of these flat solutions.

The only other known configuration of an L5-brane in matrix theory
corresponds to a brane with geometry $\bc P^2$.  This configuration
was constructed by Nair and Randjbar-Daemi as a particular example of
a coset space $G/H$ with $G = SU(3)$ and $H = U(2)$ \cite{Nair-R}.
They choose the matrices
\begin{equation}
X_i = \frac{rt_i}{\sqrt{N}} 
\end{equation}
where $t_i$ are generators spanning ${\bf g}/{\bf h}$
in a particular representation of $SU(3)$.  The geometry
defined in this fashion seems to be in some ways better behaved than
the $S^4$ geometry.  For one thing, configurations of a single brane
with arbitrarily large $N$ can be constructed.  Furthermore, it seems
to be possible to include all local fluctuations as symmetric
functions of the matrices $t_i$.  This configuration is also somewhat
confusing, however, as it extends in only four spatial dimensions,
which makes the geometrical interpretation somewhat unclear.

Clearly there are many aspects of the L5-brane in matrix theory which
are not understood.  The principal outstanding problem is to find a
systematic way of describing an arbitrary L5-brane geometry including
its fluctuations.  One approach to this might be to find a
way of regularizing the world-volume theory of an M5-brane in a
fashion similar to the matrix regularization of the supermembrane.
It is also possible that understanding the structure of noncommutative
4-manifolds might help clarify this question.  This is one of many
places where noncommutative geometry seems to tie in closely with
matrix theory.  We will discuss other such connections with
noncommutative geometry later in these lectures.

\subsection{Extended objects from matrices}
\label{sec:extended}

We have seen that not only pointlike graviton states, but also
objects extended in one, two, and four transverse directions can be
constructed from matrix degrees of freedom.  In this subsection we
make some general comments about the appearance of these extended
objects and their structure.

One systematic way of understanding the conserved charges associated
with the longitudinal and transverse membrane and the longitudinal 5-brane in
matrix theory arises from considering the supersymmetry algebra of the
theory.  The 11-dimensional supersymmetry algebra takes
the form
\begin{equation}
\{Q_\alpha, Q_\beta\} \sim P^I (\gamma_I)_{\alpha \beta} + Z^{I_1 I_2}
(\gamma_{I_1 I_2})_{\alpha \beta}+ Z^{I_1 \ldots I_5}
(\gamma_{I_1 \ldots I_5})_{\alpha \beta}
\label{eq:11D-SUSY}
\end{equation}
where the central terms $Z$ correspond to 2-brane and 5-brane charges.
The supersymmetry algebra of Matrix theory was explicitly computed by
Banks, Seiberg and Shenker \cite{bss}.  Similar calculations had been
performed previously \cite{Claudson-Halpern,dhn}; however, in these
earlier analyses terms such as ${\rm Tr}\;[X^i, X^j]$ and ${\rm
Tr}\;X^{[i} X^{j} X^k X^{l]}$ were dropped since they vanish for
finite $N$.  The full supersymmetry algebra of the theory takes the
schematic form
\begin{equation}
\{Q, Q\} \sim P^I + z^i + z^{ij} + z^{ijkl},
\end{equation}
as we would expect for the light-front supersymmetry algebra
corresponding to (\ref{eq:11D-SUSY}).
The charge
\begin{equation}
z^i \sim i {\rm Tr}\; \left(\{P^j,[X^i, X^j]\} +
{}[{}[X^i, \theta^{\alpha}], \theta^{\alpha}] \right)
\end{equation}
corresponds to longitudinal membranes (strings),
the charge
\begin{equation}
z^{ij} \sim -i {\rm Tr}\;[X^i, X^j]
\end{equation}
corresponds to transverse membranes and
\begin{equation}
z^{ijkl} \sim {\rm Tr}\;X^{[i} X^{j} X^k X^{l]}
\end{equation}
corresponds to longitudinal 5-brane charge.
For all the extended objects we have
described in the preceding subsections,
these results agree with the charges we motivated by T-duality arguments.

Note that the charges of all the extended objects in the theory vanish
when the matrix size $N$ is finite.  Physically, this corresponds to
the fact that any finite-size configuration of strings, 2-branes and
4-branes must have net charges which vanish.  

Another approach to understanding the charges associated with the
extended objects of matrix theory arises from the study of the
coupling of these objects to supergravity fields, which we will
discuss in the next section.  From this point of view, perturbative
matrix theory calculations can be used to determine not only the
conserved charges of the theory, but also the higher multipole moments
of all the components of the supercurrent describing the matrix
configuration.  For example \cite{Mark-Wati,Dan-Wati-2}, the
multipole moments of the membrane charge $z^{ij}= -2 \pi i {\rm
Tr}\;[X^i, X^j]$ can be written in terms of
the matrix moments
\begin{equation}
z^{ij (k_1 \cdots k_n)} = -2 \pi i \;{\rm STr}\; \left([X^i, X^j] X^{k_1}
\cdots X^{k_n} \right)
\end{equation}
which are the matrix analogues of the moments
\begin{equation}
\int d^2 \sigma \; \{X^i, X^j\} X^{k_1} \cdots X^{k_n}
\label{eq:membrane-moments}
\end{equation}
for the continuous membrane.  The symbol ${\rm STr}$ indicates a
symmetrized trace, wherein the trace is averaged over all possible
orderings of the terms $[X^i, X^j]$ and $X^{k_\nu}$ appearing inside
the trace.  This corresponds to a particular ordering prescription in
applying the matrix-membrane correspondence to
(\ref{eq:membrane-moments}).  There is no {\it a priori} justification
for this ordering prescription, but it is a consequence of explicit
calculations of interactions between general matrix theory objects as
described in the next section.
The same prescription can be used to define the multipole moments of
the longitudinal membrane and 5-brane charges.

Although as we have mentioned, the conserved charges in matrix theory
corresponding to extended objects all vanish at finite $N$, the same
is not true of the higher moments of these charges.  For example, the
isotropic spherical matrix membrane configuration
discussed in section \ref{sec:spherical-membranes} has nonvanishing
membrane dipole moments
\begin{eqnarray}
z^{12 (3)} = z^{23 (1)} = z^{31 (2)}  & = & -2 \pi i {\rm Tr}\;
\left([X^1, X^2] X^3 \right) \nonumber\\
& = & \frac{4 \pi r^3}{3}  (1-1/N^2)
\end{eqnarray}
which agrees with the membrane dipole moment $4 \pi r^3/3$ of the
smooth spherical membrane up to terms of order $1/N^2$.  Using the
multipole moments of a fixed matrix configuration we can essentially
reproduce the complete spatial dependence of the matter configuration
to which the matrices correspond.  This higher moment structure
describing higher-dimensional extended objects through
lower-dimensional objects is very general, and has a precise analog in
describing the supercurrents and charges of Dirichlet $(p +
2k)$-branes in terms of the world-volume theory of a system of
D$p$-branes \cite{Mark-Wati-5}.  This structure has many possible
applications to D-brane physics as well as to matrix theory.  For
example, it was recently pointed out by Myers \cite{Myers-dielectric}
that putting a system of D$p$-branes in a constant background $(p
+4)$-form flux will produce a dielectric effect in which spherical
bubbles of D$(p + 2)$-branes will be formed with dipole moments which
screen the background field.

\section{Interactions in matrix theory}
\label{sec:matrix-interactions}
\setcounter{equation}{0}

In this section we discuss interactions in matrix theory between block
matrices describing general time-dependent matrix theory
configurations which may include gravitons, membranes and 5-branes.
We begin by reviewing the perturbative Yang-Mills formalism in
background field gauge.  This formalism can be used to carry out loop
calculations in matrix theory, giving results which can be related to
supergravity interactions.  We carry out two explicit examples of this
calculation at one-loop order: first for a pair of 0-branes with
relative velocity $v$, following \cite{DKPS}, then for the leading
order term in the interaction between an arbitrary pair of bosonic
background configurations, following \cite{Dan-Wati-2}.  Following
these examples, we summarize the extent to which perturbative
Yang-Mills calculations of this kind have been shown to agree with
classical supergravity.  At the level of linearized supergravity, it has been
found that there is an infinite series of terms in the one-loop matrix
theory effective potential which precisely reproduce all tree-level
supergravity interactions arising from the exchange of a single
graviton, 3-form quantum or gravitino.  There is limited information
about the extent to which nonlinear supergravity effects are
reproduced by higher-loop matrix theory calculations, however. While it has
been shown that the nonlinear structure of 3-graviton scattering is
correctly reproduced by a two-loop matrix theory calculation, there is
not a clear picture of what should be expected beyond this.  We
discuss these results and how they are related to supersymmetric
nonrenormalization theorems which protect some terms in the
perturbative Yang-Mills expansion from higher-loop corrections.

In this section we primarily focus on the problem of deriving
classical 11-dimensional supergravity from matrix theory.  A very
interesting, but more difficult, question is whether matrix theory can
also successfully reproduce string/M-theory corrections to classical
supergravity.  The first such corrections would be ${\cal R}^4$
corrections to the Einstein-Hilbert action.  Some work has been done
investigating the question of whether these terms can be seen in
matrix theory
\cite{Susskind-talk,Berglund-Minic,Serone,Esko-Per-short,Beckers-graviton,hpsw}.
While more work needs to be done in this direction, the results of
\cite{Esko-Per-short,Beckers-graviton,hpsw} indicate that the
perturbative loop expansion in matrix theory probably does not
correctly reproduce quantum effects in M-theory.  The most likely
explanation for this discrepancy is that such terms are not subject to
nonrenormalization theorems, and are only reproduced in the large $N$
limit.  We discuss these issues again briefly in the last section.

In subsection \ref{sec:2-body}  we
describe two-body interactions in matrix theory, and in subsection 
\ref{sec:N-body}
we discuss interactions between more than two objects.  Section
\ref{sec:longitudinal} contains a brief discussion of interactions
involving longitudinal momentum transfer, which correspond to
nonperturbative processes in matrix theory.

\subsection{Two-body interactions}
\label{sec:2-body}

The background field formalism \cite{Abbott} for describing matrix theory
interactions between block matrices which are widely separated in
eigenvalue space was first used by Douglas, Kabat, Pouliot and Shenker
in \cite{DKPS} to describe interactions between a pair of D0-branes in
type IIA string theory moving with relative velocity $v$.
In this
subsection we discuss their result and the generalization  to general
bosonic background configurations.  The matrix theory Lagrangian is
\begin{equation}
{\cal L}= \frac{1}{2 R}  {\rm Tr}\; \left[
D_0 X^i\;D_0 X^i
+\frac{1}{2} [X^i, X^j]^2+  \theta^T (i\dot{\theta}
- \gamma_i[X^i, \theta]) \right]
\label{eq:matrix-Lagrangian}
\end{equation}
where
\begin{equation}
D_0 X^i = \partial_t X^i-i[A, X^i].
\end{equation}

We wish to expand each of the matrix theory fields around a classical
background.  We will assume here for simplicity that the background
has a vanishing gauge field and vanishing fermionic fields.  For a
discussion of the general situation with background fermionic fields
as well as bosonic fields see \cite{Mark-Wati-3}.  We expand the
bosonic field in terms of a background plus a fluctuation
\begin{eqnarray*}
X^i & = &  B^i + Y^i.
\end{eqnarray*}
We choose the background field gauge
\begin{equation}
D^{{\rm bg}}_{\mu} A^\mu =
\partial_tA-i[B^i, X^i] = 0.
\end{equation}
This gauge can be implemented by adding a term $-(D^{{\rm bg}}_{\mu}
A^\mu )^2$ to the action and including the appropriate ghosts.  The
nice feature of this gauge is that the terms quadratic in the bosonic
fluctuations simplify to the form
\begin{equation}
\dot{Y}^i \dot{Y}^i - [B^i, Y^j]^2 -[B^i, B^j][Y^i, Y^j]
\end{equation}

The complete gauge-fixed action including ghosts is written in
Euclidean time $\tau = it$ as
\begin{equation}
S = S_0 + S_2 + S_3 + S_4
\end{equation}
where
\begin{eqnarray}
S_0 & = &\frac{1}{2 R}   \int d \tau {\rm Tr}\;
 \left[ \partial_\tau B^i \partial_\tau B^i+\frac{1}{2}[B^i, B^j]^2
\right]\nonumber
\\
S_2 & = &\frac{1}{2 R}   \int d \tau {\rm Tr}\;
 \left[ \partial_\tau Y^i \partial_\tau Y^i
-[B^i, Y^j][B^i, Y^j]  -[B^i, B^j][Y^i, Y^j]\right.
\nonumber\\
& &\left.
\hspace{0.6in}
+ \partial_\tau A \partial_\tau A
 -[B^i, A][B^i, A]-2i \dot{B}^i[A, Y^i]\right.
\label{eq:s-2}\\
& &\left.
\hspace{0.6in}
 + \partial_\tau \bar{C} \partial_\tau C
  -[B^i,\bar{C}][B^i, C] +
 \theta^T \dot{\theta} -\theta^T \gamma_i[B^i, \theta]
\right] \nonumber
\end{eqnarray}
and where $S_3$ and $S_4$ contain terms cubic and quartic in the
fluctuations $Y^i, A, C, \theta$.  Note that we have taken $A
\rightarrow -iA$ as appropriate for the Euclidean formulation.

We now wish to use this gauge-fixed action to compute the effective
potential governing the interaction between a pair of matrix theory
objects.  In general, to calculate the interaction potential to
arbitrary order it is necessary to include the terms $S_3$ and $S_4$
in the action.  The propagators for each of the fields can be computed
from the quadratic term $S_2$.  A systematic diagrammatic expansion
will then yield the effective potential to arbitrary high order.  We
begin our discussion of matrix theory interactions, however, with the
simplest case: the interaction of two objects at leading order in the
inverse separation distance.  In \ref{sec:2-graviton} we discuss the
simplest case of this situation, the scattering of a pair of
gravitons.  In \ref{sec:2-body-linear} we discuss the situation of two
general matrix theory objects, giving an explicit calculation for the
leading term in the case where both objects are purely bosonic.  After
working out these explicit examples we review what is known about the
scattering of a general pair of matrix theory objects to arbitrary
order in section \ref{sec:2-body-general}.  We review the special case
of a pair of gravitons in section \ref{sec:2-graviton-general}.  We
discuss the N-body problem in \ref{sec:N-body}.

\subsubsection{Two graviton interactions at leading order}
\label{sec:2-graviton}

As we have discussed in \ref{sec:classical-gravitons}, a classical
background describing a pair of gravitons with relative velocity $v$
and impact parameter $b$
(and no polarization information) is given in the center of mass frame
by
\begin{eqnarray}
B^1 & = & \frac{-i}{2}   \left(\begin{array}{cc}
v\tau & 0\\
0 & -v\tau
\end{array} \right)\\
B^2 & = & \frac{1}{2}   \left(\begin{array}{cc}
b & 0\\
0 & - b
\end{array} \right)\\
B^i & = &  0, \;\;\;\;\; \;\;\;\;\; i > 2
\end{eqnarray}

Inserting these backgrounds into (\ref{eq:s-2}) we see that at a fixed
value of time the Lagrangian at quadratic order for the 10 complex
bosonic off-diagonal components of $A$ and $Y^i$ is that of a system
of 10 harmonic oscillators with mass matrix
\begin{equation}
(\Omega_b)^2 = \left(\begin{array}{ccccc}
r^2 & -2 iv& 0 & \cdots & 0\\
2iv & r^2 & 0 & \ddots & 0\\
0 & 0 & r^2 & \ddots & \vdots\\
\vdots & \ddots &  \ddots & \ddots & 0\\
0 & 0 & \cdots & 0 & r^2
\end{array} \right)
\end{equation}
where $r^2 = b^2 + (vt)^2$ is the instantaneous separation between the
gravitons.

There are two complex off-diagonal ghosts with $\Omega^2 = r^2$.

There are 16 fermionic oscillators with a mass-squared matrix 
\begin{equation}
(\Omega_f)^2 = r^2 \identity_{16 \times 16} + v \gamma_1
\end{equation}
This matrix can be found by writing
\begin{equation}
P^{\dagger} P = -\partial^2 + (\Omega_f)^2
\end{equation}
where
\begin{equation}
P = \partial -v \tau \gamma_1 -b \gamma_2
\end{equation}

To perform a completely general calculation of the two-body
effective interaction potential to all orders in $1/r$ it is necessary
to perform a diagrammatic expansion using the exact propagator for the
bosonic and fermionic fields.  For example, the bosonic propagator
satisfying
\begin{equation}
(-\partial^2 + b^2 + v^2 \tau^2) \Delta_B
(\tau, \tau' | b^2 + v^2 \tau^2) = \delta (\tau -\tau')
\end{equation}
is given by the expression \cite{Becker-Becker}
\begin{eqnarray}
\Delta_B
(\tau, \tau' | b^2 + v^2 \tau^2)  & = & 
\int_0^\infty ds \; e^{-b^2 s} \sqrt{\frac{v}{2 \pi \sinh 2sv} }
\times
\label{eq:exact-propagator}\\
& &\hspace{0.2in}
\exp \left( -\frac{v}{2 \sinh 2sv}((\tau^2 + \tau'^2) \cosh 2sv-2 \tau
\tau') \right).\nonumber
\end{eqnarray}
In general, even for a simple 2-graviton calculation there is a fair
amount of algebra involved in extracting the effective potential using
propagators of the form (\ref{eq:exact-propagator}).  If, however, we
are only interested in calculating the leading term in the long-range
interaction potential we can simplify the calculation by making the
quasi-static assumption that all the oscillator frequencies $\omega$
of interest
are large compared to the ratio $v/r$ of the relative velocity divided
by the separation scale.  In this regime, we can make the
approximation that all the oscillators stay in their ground state over
the interaction time-scale, so that the effective potential between
the two objects is simply given by the sum of the ground-state
energies of the boson, ghost and fermion oscillators
\begin{equation}
V_{ {\rm qs}} = \sum_{b}\omega_b-\sum_{g}  \omega_g -\frac{1}{2}
\sum_{f} \omega_f.
\end{equation}
Note that the bosonic and ghost oscillators are complex so that no
factor of 1/2 is included.

In the situation of two-graviton scattering we can therefore calculate
the effective potential by diagonalizing the frequency matrices
$\Omega_b, \Omega_g$ and $\Omega_f$.  We find that 
the bosonic oscillators have frequencies
\begin{eqnarray*}
\omega_b & = &   r \;\;\;\;\; {\rm with\ multiplicity\ 8}\\
\omega_b & = &  \sqrt{r^2 \pm 2v} \;\;\;\;\;
{\rm with\ multiplicity\ 1\ each}.
\end{eqnarray*}
The 2 ghosts have frequencies 
\begin{equation}
\omega_g = r,
\end{equation}
and the 16 fermions have frequencies
\begin{equation}
\omega_f = \sqrt{r^2 \pm v} \;\;\;\;\;
{\rm with\ multiplicity\ 8\ each}.
\end{equation}

The effective potential for a two-graviton system with instantaneous
relative velocity $v$ and separation $r$ is thus given by the leading
term in a $1/r$ expansion of the expression
\begin{equation}
V  = \sqrt{r^2 + 2v} + \sqrt{r^2 -2v} + 6r
-4 \sqrt{r^2 + v} + 4 \sqrt{r^2 -v}.
\end{equation}
Expanding in $v/r^2$ we see that the terms of order $r, v/r, v^2/r^3$
and $v^3/r^5$ all cancel.  The leading term is
\begin{equation}
V = -\frac{15}{16}  \frac{v^4}{r^7}  +{\cal O} (\frac{v^6}{r^{11}} )
\end{equation}

This potential was first computed by Douglas, Kabat, Pouliot and
Shenker \cite{DKPS}.  This result agrees with the leading term in the
effective potential between two gravitons with $P^+ = 1/R$ in
light-front 11D supergravity.  We will discuss the supergravity side
of this calculation in more detail in the following section, where we
generalize this calculation to an arbitrary pair of matrix theory objects.

\subsubsection{General 2-body systems and linearized supergravity
at leading order}
\label{sec:2-body-linear}

We now generalize the discussion to an arbitrary pair of matrix theory
objects, which are described by a block-diagonal background
\begin{equation}
B^i = \left(\begin{array}{cc}
\hat{X}^i & 0\\
0 & \tilde{X}^i
\end{array}\right)
\end{equation}
where $\hat{X}^i$ and $\tilde{X}^i$ are $\hat{N} \times \hat{N}$ and
$\tilde{N} \times \tilde{N}$ matrices describing the
two objects.  The separation distance between the objects, which we
will use as an expansion parameter, is given by
\begin{equation}
r^i = \frac{1}{ \hat{N}}  {\rm Tr}\; \hat{X}^i
-\frac{1}{ \tilde{N}}  {\rm Tr}\; \tilde{X}^i
\end{equation}
There are $\hat{N} \tilde{N}$ independent complex off-diagonal
components of the fluctuation matrices $Y^i$.  We will find it useful
to treat these components as an $\hat{N} \tilde{N}$-component vector
$Z^i$.  We now construct a $\hat{N} \tilde{N}\times \hat{N} \tilde{N}$
matrix which acts on the $Z^i$ vectors
\begin{equation}
K_i \equiv  \hat{X}_i \otimes \identity_{\tilde{N} \times \tilde{N}} 
-  \identity_{N \times N} \otimes \tilde{X}_i^T\,.
\label{eq:k-definition}
\end{equation}
It is convenient to extract the centers of mass explicitly so that
$K^i$ can be rewritten as
\begin{equation}
K^i = r^i \identity + \bar{K}^i
\end{equation}
where $\bar{K}^i$ is of order $1$ in terms of the separation scale
$r$.  The matrices $K$ encode the adjoint action of the background $B$ on the
fluctuations $Y$ so that we can extract the part of $[B, Y]$ depending
on the off-diagonal fields $Z$ through
\begin{equation}
[B^i, Y^j] \rightarrow K^iZ^j.
\end{equation}

This formalism allows us to write the quadratic terms from
(\ref{eq:s-2}) in the action for the off-diagonal fields in a simple
form
\begin{eqnarray}
\lefteqn{\dot{Y}^i \dot{Y}^i-[B^i, Y^j][B^i, Y^j]-[B^i, B^j][Y^i, Y^j]
}\nonumber \\ & & \rightarrow
\dot{Z}_i^{\dagger} \dot{Z}^i-
Z^{\dagger}_j K^iK_iZ^j-2 Z^{\dagger}_i [K_i, K_j]Z^j
\end{eqnarray}
Performing a similar operation for the terms quadratic in fluctuations
of the $A$ field, we find that the full frequency-squared matrices for
the bosonic, ghost and fermionic fields can be written
\begin{eqnarray}
\Omega_b^2 & = &  K^2 \identity_{10 \times 10}-2iF_{\mu \nu} \nonumber \\
\Omega_g^2 & = &  K^2 \identity_{2 \times 2} \\
\Omega_f^2 & = &  K^2 \identity_{16 \times 16} \nonumber
- iF_{\mu \nu} \gamma^\mu \gamma^\nu
\end{eqnarray}
where $\gamma^0 = \identity$ and the field strength matrix $F_{\mu
\nu}$ is given by
\begin{eqnarray}
F_{0i} & = &  \dot{K}^i \label{eq:f-definition}\\
F_{ij} & = &  i[K^i, K^j] \nonumber
\end{eqnarray}

Note that each of the frequencies has a leading term $r$ and
subleading terms of order 1.
Expanding the frequency matrices in powers of $1/r$ we find that for a
completely arbitrary pair of objects described by the background
matrices $\hat{X}^i$ and $\tilde{X}^i$ the potential vanishes to order
$1/r^6$.  At order $1/r^7$ we find that the potential is
\begin{eqnarray}
V_{\rm leading}& = & \tr \left(\Omega_b\right) - \half \tr
\left(\Omega_f\right) - 2 \tr \left(\Omega_g\right)\,\\ & = &
-\frac{5}{128r^7} {\rm STr}\;{\cal F}
\end{eqnarray}
where
\begin{equation}
{\cal F} = 
    24 F^\mu{}_\nu F^\nu{}_\lambda F^\lambda{}_\sigma F^\sigma{}_\mu
-  6  F_{\mu\nu} F^{\mu \nu} F_{\lambda \sigma} F^{\lambda \sigma}
\end{equation}
and ${\rm STr}\;$ indicates that the trace is symmetrized over all
possible orderings of $F$'s in the product $F^4$.

{}From the definition (\ref{eq:k-definition})
it is clear that the field strength $F_{\mu \nu}$ decomposes into a
piece from each of the two objects
\begin{equation}
F_{\mu \nu} = \hat{F}_{\mu \nu} - \tilde{F}_{\mu \nu}
\end{equation}
where $\hat{F}_{\mu \nu}$ and $\tilde{F}_{\mu  \nu}$ are defined
through (\ref{eq:f-definition}) in terms of $\hat{X}$ and
$\tilde{X}$.  We can therefore decompose the potential $V_{{\rm
leading}}$ into a sum of terms which are written as products of a
function of $\hat{X}$ and a function of $\tilde{X}$, where the terms
can be grouped according to the number of Lorentz indices contracted
between the two objects.  With some algebra, we can write this
potential as
\bea
\label{eq:matrix-potential}  
V_{\rm leading} & = & V_{\rm gravity} + V_{\rm electric} + V_{\rm magnetic} \\
V_{\rm gravity} & = & - {15 R^2 \over 4 r^7} \left( \hatt^{IJ} \st_{IJ}
- {1 \over 9} \hatt^I{}_I \st^J{}_J\right) 
\label{eq:v-gravity} \\
V_{\rm electric} & = & - {45 R^2 \over r^7} \hj^{IJK} \sj_{IJK} 
\label{eq:v-electric} \\
V_{\rm magnetic} & = & - {45 R^2\over r^7} \hm^{+-ijkl} \sm^{-+ijkl} 
\label{eq:v-magnetic}
\eea This is, as we shall discuss shortly, precisely the form of the
interactions we expect to see from 11D supergravity in light-front
coordinates, where ${\cal T},\ijj$ and $\imm$ play the role of the
(integrated) stress tensor, membrane current and 5-brane current of
the two objects.  The quantities appearing in this decomposition are
defined as follows.

$\itt^{IJ}$ is a
symmetric tensor with components
\bea
\label{eq:matrix-t}
\itt^{--} & = & {1 \over R} \; \str \frac{{\cal F}}{96}  \\
\itt^{-i} & = & {1 \over R} \;\str \left(\half \dx^i \dx^j \dx^j +
{1 \over 4} \dx^i F^{jk} F^{jk} 
                  + F^{ij} F^{jk} \dx^k \right) \nonumber \\
\itt^{+-} & = & {1 \over R} \;\str \left(\half \dx^i \dx^i + {1
\over 4} F^{ij} F^{ij} \right) \nonumber \\ 
\itt^{ij} & = & {1 \over R}  \;\str \left( \dx^i \dx^j + F^{ik}
F^{kj} \right) \nonumber \\ 
\itt^{+i} & = & {1 \over R} \;\str \dx^i \nonumber \\
\itt^{++} & = & {N \over R} \nonumber
\eea
$\ijj^{IJK}$ is a totally antisymmetric tensor with components
\bea
\label{eq:matrix-j}
\ijj^{-ij} & = & {1 \over 6 R} \str \left( \dx^i \dx^k F^{kj} -
\dx^j \dx^k F^{ki} - \half \dx^k \dx^k F^{ij} \right.\\ 
& & \qquad \qquad  \left. + {1 \over 4} F^{ij} F^{kl} F^{kl} +F^{ik}
F^{kl} F^{lj} \right) \nonumber \\ 
\ijj^{+-i} & = & {1 \over 6 R} \str \left( F^{ij} \dx^j \right) \nonumber \\
\ijj^{ijk} & = & - {1 \over 6 R} \str \left( \dx^i F^{jk} + \dx^j F^{ki} + \dx^k F^{ij} \right) \nonumber \\
\ijj^{+ij} & = & - {1 \over 6 R} \str F^{ij} \nonumber
\eea
Note that we retain some quantities --- in particular $\ijj^{+-i}$ and
$\ijj^{+ij}$ --- which vanish at finite $N$ (by the Gauss constraint and
antisymmetry of $F^{ij}$, respectively).  These terms
represent BPS charges (for longitudinal and transverse membranes)
which are only present in the large $N$ limit.  We define higher
moments of these terms below which can be
non-vanishing at finite $N$.

$\imm^{IJKLMN}$ is a totally antisymmetric tensor with
\be
\label{eq:matrix-m}
 \imm^{+-ijkl} = {1 \over 12 R} \str \left(F^{ij} F^{kl} + F^{ik}
F^{lj} + F^{il} F^{jk}\right)\,.
\ee
At finite $N$ this vanishes by the Jacobi identity, but we shall
retain it because it represents the charge of a longitudinal 5-brane.
The other components of ${\cal M}^{IJKLMN}$ do not appear in the Matrix
potential.  In principle, we expect another component of the 5-brane
current, ${\cal M}^{-ijklm}$, to be well-defined.  This term arises
from a moving longitudinal 5-brane.  This term does not appear in the
2-body interaction formula because it would couple to the transverse
5-brane charge ${\cal
M}^{+ijklm}$ which, as we have discussed, vanishes in light-front coordinates.
The component ${\cal M}^{-ijklm}$ can, however, be determined from the
conservation of the 5-brane current, and is given by \cite{mvr}
\begin{equation}
{\cal M}^{-ijklm} = \frac{5}{4 R}  {\rm STr}\;
\left( \dot{X}^{[i} F^{jk} F^{lm]} \right).
\end{equation}

Let us compare the interaction potential (\ref{eq:matrix-potential})
with the leading long-range interaction between two objects in 11D
light-front compactified supergravity.  The scalar propagator in 11D is
\begin{equation}
\laplace{}^{-1}(x) = {1 \over 2 \pi R} \sum_n \int{dk^- d^{\,9}k_\perp \over (2 \pi)^{10}}
{e^{-i{n \over R} x^- - i k^- x^+ + i k_\perp \cdot x_\perp} \over
2 {n \over R} k^- - k_\perp^2}
\end{equation}
where $n$ counts the number of units of longitudinal momentum $k^+$.
To compare the leading term in the long-distance potential with matrix
theory we only need to extract the term associated with $n = 0$,
corresponding to interactions mediated by exchange of a supergraviton
with no longitudinal momentum.
\begin{equation}
\label{eq:propagator}
\laplace{}^{-1}(x-y) = {1 \over 2 \pi R} \delta(x^+ - y^+) { - 15
\over 32 \pi^4 \vert x_\perp-y_\perp\vert^7 } 
\end{equation}
Note that the exchange of quanta with zero longitudinal momentum gives
rise to interactions that are instantaneous in light-front time, as
recently emphasized in \cite{Hellerman-Polchinski}.  This is
precisely the type of instantaneous interaction that we find at one
loop in Matrix theory.  Such action-at-a-distance potentials are
allowed by the Galilean invariance manifest in the light-front
formalism.

The graviton propagator can be written in terms of this scalar
propagator as
\begin{equation}
D_{\rm graviton}^{IJ,KL}  =  2 \kappa^2 \left(\eta^{IK} \eta^{JL} +
\eta^{IL} \eta^{JK} - {2 \over 9} \eta^{IJ} \eta^{KL} \right)
\laplace{}^{-1}(x-y)
\end{equation}
where $ 2 \kappa^2 =(2 \pi)^5 R^3$ in string units.  The effective
supergravity interaction between two objects having stress tensors
$\hat{T}_{IJ}$ and $\tilde{T}_{KL}$ can then be expressed as
\begin{equation}
S = - {1 \over 4} \int d^{11} x d^{11} y \,  \hat{T}_{IJ}(x) D_{\rm
graviton}^{IJ,KL}(x-y) \tilde{T}_{KL}(y)
\end{equation}
This interaction has a leading term of precisely the form
(\ref{eq:v-gravity}) if we define $\itt^{IJ}$ to be the integrated
component of the stress tensor
\begin{equation}
\itt^{IJ} \equiv \int dx^- d^{\, 9}
x_\perp T^{IJ}(x).
\end{equation}
It is straightforward to show in a similar fashion that
(\ref{eq:v-electric}) and (\ref{eq:v-magnetic}) are precisely the
forms of the leading supergravity interaction between membrane
currents and 5-brane currents of a pair of objects.

We can calculate the components of the source currents
(\ref{eq:matrix-t}), (\ref{eq:matrix-j}) and (\ref{eq:matrix-m}) for
all the matrix theory objects we have discussed: the supergraviton,
the membrane and the L5-brane.  For all these objects the currents
have the expected values, at least to order $1/N^2$.  For example, the
stress tensor of a graviton can be written in the form
\begin{equation}
\itt^{IJ} = \frac{p^Ip^J}{p^+} 
\end{equation}
where
\begin{equation}
p^+ = N/R,\;\;\;\;\; p^i = p^+ \dot{x}^i,\;\;\;\;\; p^- = p_\perp^2 / 2 p^+
\end{equation}
The stress tensor and membrane current of the membrane can be computed
in the continuum membrane theory from the action
(\ref{eq:bosonic-membrane-Polyakov-general}) for the bosonic membrane
in a general background.  Using the matrix-membrane correspondence
(\ref{eq:mm-correspondence}) it is possible to show that the matrix
definitions above are compatible with the expressions for the stress
tensor and membrane current of the continuum membrane, although the
matrix expressions are not uniquely determined by this correspondence.

We have thus shown that to leading order in the separation distance
the interaction between any pair of objects in supergravity is
precisely reproduced by one-loop quantum effects in matrix theory.  We
have only shown this explicitly in the case of a pair of bosonic
backgrounds, following \cite{Dan-Wati-2}.  The more general case where
fermionic background fields are included is discussed in
\cite{Mark-Wati-3}.  In the following sections we discuss what is
known about the extension of these results to higher order in $1/r$
and to interactions of more than two distinct objects.

%
%

\subsubsection{General 2-body interactions}
\label{sec:2-body-general}

In the previous subsections we have considered only the leading
$1/r^7$ terms in the 2-body interaction potential.  If we consider all
possible Feynman diagrams which might contribute to higher-order
terms, it is straightforward to demonstrate by power counting that the
complete potential can be written as a sum of terms of the form
\begin{equation}
V = \sum_{n, k, l, m, p} V_{n, k, l, m, p, \alpha} R^{n-1}
\frac{X^l D^p F^k \psi^{2m}}{ r^{3n + 2k + l + 3m+ p-4}}.
\label{eq:potential-general}
\end{equation}
where $n$ counts the number of loops in the relevant diagrams and
$\psi$ describes the fermionic background fields.
Each $D$ either indicates a time derivative or a commutator with an
$X$, as in $\psi[X, \psi]$.
The summation over the index $\alpha$ indicates a sum over many
possible index contractions for every combination of $F$'s, $X$'s and
$D$'s and
$\Gamma$ matrices between the $\psi$'s.

For a completely general pair of objects, only terms in the one-loop
effective action have been understood in terms of supergravity.  At
one-loop order, when the fields are taken on-shell by imposing the
matrix theory equations of motion, all terms with $k + m + p < 4$
which have been calculated vanish.  All terms with $k + m + p = 4$
which have been calculated have $m \geq p$ and can be written in the
form
\begin{equation}
 V_{1,  4-m-p, l, m, p, \alpha} 
\frac{X^l  F^{(4-m-p)} \psi^{2(m-p)} (\psi D \psi)^p}{ r^{7 + m-p+ l}}.
\label{eq:potential-one-loop}
\end{equation}
In this expression, the grouping of $\psi$ terms indicates the
contraction of spinor indices---in general, the terms can be ordered
in an arbitrary fashion when considered as $U(N)$ matrices.  The terms
(\ref{eq:potential-one-loop}) have been explicitly determined for $m <
2$ in \cite{Dan-Wati-2,Mark-Wati-3}, where they were shown to
precisely correspond to multipole interaction terms in linearized
supergravity.  We now briefly describe some of those terms which have
been interpreted in this fashion
\vspace{0.15in}

\noindent
$m = p = 0, \; k = 4, \; l = 0:$ These are the leading $1/r^7$ terms in the
interaction potential between a pair of purely bosonic objects
discussed above.  They are precisely equivalent to the leading term in
the supergravity potential between a pair of objects with appropriate
integrated stress tensors, membrane currents and 5-brane currents.
\vspace{0.08in}

\noindent
$m = p =0, \; k = 4, \; l > 0:$ This infinite set of terms was shown in
\cite{Dan-Wati-2} to be equivalent to the higher-order terms in the
linearized supergravity potential arising from higher moments of the
bosonic parts of the
stress tensor, membrane current and 5-brane current.  The simplest
example (discussed in \cite{Mark-Wati}) is the term of the form $F^4 X/r^8$
which appears in the case of a graviton moving in the long-range
gravitational field of a matrix theory object with angular momentum
\begin{equation}
J^{ij} = T^{+ i (j)} -T^{+ j (i)}
\end{equation}
where the first moment of the matrix theory stress tensor component
$T^{+ i}$ is defined through (as discussed in subsection \ref{sec:extended})
\begin{equation}
T^{+ i (j)} = \frac{1}{R}  {\rm Tr}\; \left( \dot{X}^i X^j \right)
\end{equation}
In \cite{Mark-Wati-3} it was shown that terms of the general form $F^4
X^l/r^{7 + l}$ can describe higher-moment membrane-5-brane and
D0-brane-D6-brane interactions as well as membrane-membrane and
5-brane-5-brane interactions, generalizing previous results in
\cite{Berkooz-Douglas,WT-adhere}.
\vspace{0.08in}

\noindent
$m = 1, \; p = 1, \; k = 2, \; l \geq 0:$  The terms of the form
\begin{equation}
\frac{F^2  (\psi D \psi) X^l}{ r^{7 + l}} 
\label{eq:f2l2}
\end{equation}
correspond again to leading and higher-moment interactions in
linearized supergravity, where now the components of the (integrated)
gravity currents have contributions from the fermionic backgrounds as
well as the bosonic backgrounds.
These terms
are also related to linearized supercurrent interactions arising from
single gravitino exchange, as discussed in \cite{bktw,Mark-Wati-3}.
\vspace{0.08in}

\noindent
$m = 2, \; p = 2, \; k = 0, \; l \geq 0:$  The terms of the form
\begin{equation}
\frac{(\psi D \psi) (\psi D \psi) X^l}{ r^{7 + l}} 
\end{equation}
correspond, just like the terms (\ref{eq:f2l2}), to fermionic
contributions to the linearized supergravity interaction arising from
fermion contributions to the integrated supergravity currents.
\vspace{0.08in}

\noindent
$m = 1, \; p = 0, \; k = 3, \; l \geq 0:$  The terms of the form
\begin{equation}
\frac{F^3 \psi \psi X^l}{ r^{8 + l}} 
\end{equation}
have a similar interpretation to the terms (\ref{eq:f2l2}).  In these
terms, however, the dipole moments of the currents have
nontrivial fermionic contributions in which no derivatives act on the
fermions \cite{Mark-Wati-3}.  The simplest example of this is the spin
contribution to the matrix theory angular momentum
\begin{equation}
J^{ij}_{{\rm fermion}} = \frac{1}{4 R}  {\rm Tr}\;
\left( \psi \gamma^{ij} \psi \right)
\label{eq:spin-angular}
\end{equation}
This contribution was first noted in the context of spinning gravitons
in \cite{Kraus-spin}.  This angular momentum term couples to the
component $T^{-i}\sim F^3$ of the matrix theory stress-energy tensor
through terms of the form $\hat{J}^{ij}\tilde{T}^{-i}r^j/r^9$.
\vspace{0.08in}

\noindent
$m > 1, \; k =  4-m, \; l \geq 0:$  The terms of the form
\begin{equation}
\frac{F^2 \psi^4 X^l}{ r^{9 + l}},\;\;\;\;\;
\frac{F \psi^6 X^l}{ r^{10 + l}},\;\;\;\;\;
\frac{\psi^8 X^l}{ r^{11 + l}}, \;\;\;\;\;
\frac{(\psi D \psi) F \psi^2 X^l}{ r^{8 + l}},\;\;\;\;\;
\frac{(\psi D \psi) \psi^4 X^l}{ r^{9 + l}}
\end{equation}
have not been completely calculated or related to supergravity
interactions, although as we will discuss in the following section
these terms are known and agree with supergravity interactions in the
special case $N = 2$.  From the structure which has already been
understood it seems most likely that these terms arise from fermion
contributions to the  higher multipole moments of the supergravity
currents, and that these terms will also agree with the corresponding
supergravity interactions.
\vspace{0.15in}

This is all that is known about the 2-body interaction for a
completely general (and not necessarily supersymmetric) pair of matrix
theory objects.  To summarize, it has been shown that all terms of the
form $F^k(\psi D \psi)^p \psi^{2 (m-p)}$ with $k + p + m = 4$
correspond to supergravity interactions, at least for the terms with
$m < 2$.  It seems likely that this correspondence persists for the
remaining values of $m > 1$, but the higher order fermionic
contributions to the multipole moments of the supergravity currents
have not yet been calculated for a general matrix theory object.  It
is likely that all these terms are protected by a supersymmetric
nonrenormalization theorem of the type discussed in the following
section.  This has not yet been proven, but might follow from
arguments similar to those in \cite{pss}.  The only other known
results are for a pair of gravitons, which we now review.

\subsubsection{General two-graviton interactions}
\label{sec:2-graviton-general}

In the case of a pair of gravitons, the general interaction potential
(\ref{eq:potential-general}) simplifies to
\begin{equation}
V =\sum_{n = 0}^{ \infty}  \sum_{k = 0}^{ \infty} 
\sum_{m = 0}^{4}  V_{n, k, m} R^{n -1}
\frac{v^k \psi^{2m}}{r^{3n + 2k + 3m-4}} 
\label{eq:graviton-potential-general}
\end{equation}

The sum over $m$ is finite since in the $SU(2)$ theory all terms with
fermions can be described in terms of a product of 2, 4, 6 or 8
$\psi$'s.  The leading terms for each value of $m$ have been computed
using the one-loop approach, and agree with supergravity.  The sum of
these terms is (see \cite{psw} and references therein for further
details)
\begin{eqnarray}
V_{(1)} & = & -\frac{15}{16}  \left[
v^4 + 2v^2 v_i D^{ij} \partial_j
+2v_i v_j D^{ik} D^{jl} \partial_k 
\partial_l \right. \\ & &\left.\hspace{0.3in}
+ \frac{4}{9}  v_iD^{ij} D^{km} D^{lm} \partial_j \partial_k\partial_l
+ \frac{2}{63}  D^{in} D^{jn} D^{km} D^{lm}
\partial_i\partial_j \partial_k
\partial_l \right] \frac{1}{r^7}  \nonumber
\end{eqnarray}
where
\begin{equation}
D^{ij} = \psi \gamma^{ij} \psi
\end{equation}
The term with a single $D$ proportional to $1/r^8$ arises from the
spin angular momentum term described in (\ref{eq:spin-angular}).

No further checks have been made on the matrix theory/supergravity
correspondence for terms with nontrivial fermion backgrounds.
Simplifying to the spinless case, the complete effective potential
(\ref{eq:graviton-potential-general}) simplifies still further to
\begin{equation}
V =\sum_{n, k}
 V_{n, k} R^{n -1}
\frac{v^k}{r^{3n + 2k-4}} .
\end{equation}
Following \cite{bbpt}, we write these terms in matrix form
\begin{equation}
 \begin{array}{cccccccccc}
V & = &\frac{1}{R}V_{0,2} \;v^2 & &  & & & & & \\
 & & + & V_{1,4} \frac{v^4}{r^7}  & + &
 V_{1,6} \frac{v^6}{r^{11}}  & + &
 V_{1,8} \frac{v^8}{r^{15}}  & + &\cdots\\
 & & + & R \;V_{2,4}  \;\frac{v^4}{r^{10}}  & + &
R\; V_{2,6} \; \frac{v^6}{r^{14}}  & + &
R\; V_{2,8} \; \frac{v^8}{r^{18}}  & + &\cdots\\
 & & + &R^2 \;V_{3,4} \; \frac{v^4}{r^{13}}  & + &
R^2\; V_{3,6} \; \frac{v^6}{r^{17}}  & + &
R^2\; V_{3,8} \; \frac{v^8}{r^{21}}  & + &\cdots\\
 && + & \vdots & + & \vdots & + & \vdots & + &\ddots
\end{array}
\label{eq:matrix-terms}
\end{equation}
where each row gives the contribution at fixed loop order.
We will now give a brief review of what is known about these
coefficients.  First, let us note that in Planck units this potential
is (restoring factors of $\alpha' = l_{11}^3/R$ by dimensional analysis)
\begin{equation}
V =\sum_{n, k}
 V_{n, k} \frac{l_{11}^{3n+ 3k - 6} }{ R^{k-1}} 
\frac{v^k}{r^{3n + 2k-4}} .
\end{equation}
Since the gravitational coupling constant is $\kappa^2 = 2^7
\pi^8l_{11}^9$ we only expect terms with
\begin{equation}
n + k \equiv 2 \; ({\rm mod} \; 3)
\end{equation}
to correspond with classical supergravity interactions, since all
terms in the classical theory have integral powers of $\kappa$.
Of the terms explicitly shown in (\ref{eq:matrix-terms}) only the
diagonal terms satisfy this criterion.  By  including factors of
$\hat{N}$ and $ \tilde{N}$ for semi-classical graviton states with finite
momentum $P^+$ and comparing to supergravity, one can argue that the
terms on the diagonal are precisely those which should correspond to
classical supergravity.  The terms beneath the diagonal should vanish
for a naive agreement with supergravity at finite $N$, while the terms
above the diagonal correspond to quantum gravity corrections.
It was argued in \cite{bbpt} that the sum of diagonal terms
corresponding to the effective classical supergravity potential
between two gravitons should be given by
\begin{equation}
V_{{\rm classical}} = \frac{2}{15 R^2} \left( 1-\sqrt{1-\frac{15 R}{
2}v^2 }\right).
\label{eq:two-classical}
\end{equation}

Now let us discuss the individual terms in (\ref{eq:matrix-terms}).
As we have discussed, the one-loop analysis gives a term
\begin{equation}
V_{1, 4} = -\frac{15}{16} 
\end{equation}
which agrees with linearized supergravity.  The analysis of DKPS
\cite{DKPS} can be extended to the remaining one-loop terms.  
The next one-loop term vanishes
\begin{equation}
V_{1,  6} = 0.
\end{equation}
Some
efforts have been made to relate the higher order terms $V_{1, 8},
\ldots$ to quantum effects in 11D supergravity, but so far this
interpretation is not clear.  For some discussion of this issue see
\cite{Esko-Per-short,Beckers-graviton} and references therein.  The term
\begin{equation}
V_{2, 4} = 0
\end{equation}
was computed by Becker and Becker \cite{Becker-Becker}.  As expected,
this term vanishes.  The term
\begin{equation}
V_{2, 6} = \frac{225}{32} 
\end{equation}
was computed in \cite{bbpt}.  This term agrees with the expansion of
(\ref{eq:two-classical}).
A general expression for the two-loop effective potential given by the
second line of (\ref{eq:matrix-terms}) was given in
\cite{Becker-complete}.

It was shown in \cite{pss,pss2} by Paban, Sethi and Stern that there
can be no higher-loop corrections to the $v^4$ and $v^6$ terms on the
diagonal.  Their demonstration of these results follows from a
consideration of the terms with the maximal number of fermions which
are related to the $v^4$ and $v^6$ terms by supersymmetry.  For
example, this is the $\psi^8/r^{11}$ term in the case $v^4$.  They
show that the fermionic terms are uniquely determined by
supersymmetry, and that this in turn uniquely fixes the form of the
bosonic terms proportional to $v^4$ and $v^6$ (see also \cite{hks-2}
for more about the case of $v^4$).
Thus, they have shown that that
\begin{equation}
V_{(n > 1), 4} = V_{(n > 2), 6} = 0.
\end{equation}
This nonrenormalization theorem was originally conjectured by BFSS in
analogy to similar known theorems for higher-dimensional theories
\cite{Dine-Seiberg}.

This completes our summary of what is known about 2-body interactions
in matrix theory.  The complete set of known terms is given by
\begin{equation}
 \begin{array}{ccccccccccc}
V & = &\frac{1}{2R} v^2 & &  & & & & & &\\
& & & + & -\frac{15}{16}  \frac{v^4}{r^7}  & + &
0& + &({\rm known})  & \rightarrow &\\
& & & + & 0 & + &
 \frac{225}{32} R  \frac{v^6}{r^{14}}   & + &
({\rm known})  & \rightarrow &\\
& & & + &0 & + &
0 & + &
?  & + &\cdots\\
& &&  & \downarrow &  & \downarrow & + & \vdots & + &\ddots
\end{array}
\label{eq:matrix-terms-2}
\end{equation}

It has been proposed that for arbitrary $N$ the analogues of the
higher-loop diagonal terms should naturally take the form of a
supersymmetric Born-Infeld type action
\cite{Chepelev-Tseytlin2,Esko-Per2,Balasubramanian-gl,Chepelev-Tseytlin3},
which would give rise in the case $N = 2$ to a sum of the form
(\ref{eq:two-classical}).  There is as yet, however, no proof of this
statement beyond two loops.  One particular obstacle to calculating
the higher-loop terms in this series is that it is necessary to
integrate over loops containing propagators of massless fields.  These
propagators can give rise to subtle infrared problems with the
calculation.  Some of these difficulties can be avoided by trying to
reproduce higher-order supergravity interactions from interactions of
more than two objects in matrix theory, the subject to which we will
turn in section \ref{sec:N-body}.

\subsubsection{The Equivalence Principle in matrix theory}
\label{sec:equivalence}

We have seen that the form of the linearized theory of 11D
supergravity is precisely reproduced by a one-loop calculation in
matrix theory.  This equivalence follows provided that the expressions
in (\ref{eq:matrix-t}-\ref{eq:matrix-m}), as well as the higher
moments of these expressions and related expressions for the fermion
components of the supercurrent are interpreted as definitions of the
stress tensor, membrane current and other supercurrent components of a
given matrix theory object.  It is perhaps somewhat surprising given
that this correspondence holds exactly at finite $N$ to observe that
Einstein's Equivalence Principle  breaks down at finite $N$, even in
the linearized theory \cite{Dan-Wati-2}.

The Equivalence Principle essentially states that given a background
gravitational field produced by some source matter configuration, any
two objects which are small compared to the scale of variation in the
metric and which have the same initial space-time velocity vector
$\dot{x}^I$ will follow identical trajectories through space-time.  
This follows from the fact that objects which are moving in the
influence of a gravitational field follow geodesics in space-time.
Of
course, this result is only valid if the objects are not influenced by
any other fields in the theory such as an electromagnetic 1-form or
3-form field.

To see a simple example of a case where the equivalence principle is
violated in matrix theory, consider a source at the origin consisting
of a single graviton with $p^+ =\tilde{N}/R$ and $\tilde{v}^i = 0$.
This source produces a long-range gravitational field and no 3-form or
gravitino field.  Now consider a probe object at a large distance $r$.
We take the probe to be a small membrane sphere, initially stationary,
of radius $r_0$ and with longitudinal momentum $p^+ = {N}/R$.  It is
straightforward to calculate the energy $p^-$ of the membrane; we find
that the 11-momentum of the membrane has the light-front components
\[
p^+ = N/R \hskip 1.0 cm p^i = 0 \hskip 1.0 cm p^- = {8 r_0^4 \over R N^3} c_2 \,.
\]
The initial velocity of the membrane is then
\[
\dot{x}^+ = 1 \hskip 1.0 cm \dot{x}^i = 0 \hskip 1.0 cm
\dot{x}^- = {p^- \over p^+} = 8 r_0^4 {c_2 \over N^4}\,.
\]
According to the equivalence principle, any two membrane spheres with
different values of $r_0$ but the same value of $\dot{x}^-=r_0^4
c_2/N^4$ should experience precisely the same acceleration.  Using the
general formula for the 2-body interaction potential in matrix theory,
however, it is straightforward to calculate
\[
\ddot{x}^i = - {R \over N} \, {\partial V_{\rm matrix} \over \partial x^i} =
- 1680 R \tilde{N} {x^i \over \vert x \vert^9} \,\, 
{r_0^8 \over N^8} \left(c_2^2 - {1 \over 3} c_2 \right) \,.
\]
The leading term in an expansion in $1/N$ of this acceleration is
indeed a function of $\dot{x}^-$.  Thus, in the large $N$ limit the
equivalence principle is indeed satisfied.  The subleading term,
however, has a different dependence on $r_0$ and $N$.  Thus, the
equivalence principle is not satisfied at finite $N$.

This result implies that even if finite $N$ matrix theory is
equivalent to DLCQ M-theory, this theory does not seem to be related
to a smooth theory of Einstein-Hilbert gravity, even on a compact
space and with restrictions on longitudinal momentum.  This is not a
problem if one only takes seriously the large $N$ version of the
conjecture.  If one wishes to make sense of the finite $N$ theory in
terms of some theory with a reasonable classical limit, however, it
may be necessary to consider some new ideas for what this theory may
be.  It is tempting to think that the theory at finite $N$ might be
some sort of theory of classical gravity on a noncommutative space.
Since the equivalence principle in the form we have been using it
depends upon the geodesic equations, which are defined only on a
smooth commutative space, it is natural to imagine that this principle
might have to be corrected at finite $N$ when the space has nontrivial
noncommutative structure.

\subsection{The N-body problem}
\label{sec:N-body}

So far we have seen that in general the linearized theory of
supergravity is correctly reproduced by an infinite series of terms
arising from one-loop calculations in matrix theory.
We have also discussed 2-loop calculations of two-graviton
interactions which seem to agree with supergravity.  If matrix theory
is truly to reproduce all of classical supergravity, however, it must
reproduce all the nonlinear effects of the fully covariant
gravitational theory.  The easiest way to study these nonlinearities
is to consider N-body interaction processes.  For example, following
\cite{Dine-Rajaraman}
let us
consider a probe body at position $r_3$ in the long-range
gravitational field produced by a pair of bodies at positions $r_1 =
0, r_{2}\ll r_3$.  We can consider a perturbative expansion of Einstein's
equations.  At leading order we have the linearized theory which gives
a long-range field satisfying (schematically, dropping indices)
\[
\partial^2 h   \sim T
\]
where $T$ is a matter source.  At the next order we have
\[
\partial^2 h + h \partial^2 h + (\partial h)^2 \sim T + Th,
\]
which we can rewrite in the form
\begin{equation}
\partial^2 h \sim T + Th+ h \partial^2 h + (\partial h)^2
\label{eq:nonlinear-gravity-schematic}
\end{equation}
The action of a probe object in the long-range field produced by
objects 1 and 2 can be written in a double expansion in the inverse
separations $r_3$ and $r_2$ as
\begin{equation}
T_3 h_{12} \sim \frac{T_3 (T_1 + T_2)}{r_3^7}  + \frac{T_3T_{(12)}}{r_3^7 r_2^7}  + \cdots
\label{eq:general-3-expansion}
\end{equation}
where $T_{(12)}$  is an interaction term contributing through the
quadratic terms on the RHS of (\ref{eq:nonlinear-gravity-schematic}).
On the matrix theory side, an apparently analogous calculation can be
performed by first doing the one-loop calculation we have already
described to find the linearized interaction between the 3rd object
and the 1-2 system, and then doing a further loop integration to
evaluate the quantum corrections to the long-range field generated by
the first two sources, giving an expression of the form
\[
\frac{T_3 \langle T_{1+2}\rangle}{r_3^7}.
\]
We expect quantum corrections to the expectation value of the
schematic form
\[
\langle T_{1 + 2} \rangle \sim T_1 + T_2 + \frac{T_{(12)}}{ r_2^7}  + \cdots
\]
which roughly conforms to the structure expected from
(\ref{eq:general-3-expansion}).  Thus, in principle, it seems like it
should be possible to make a correspondence between the double power
series expansions computed in the two theories, given the results of
the one-loop expansion for a completely general pair of objects such
as was calculated in \cite{Mark-Wati-3}.
Indeed, a simple subset of terms were shown to correspond in this way
in \cite{Mark-Wati-2}.
The terms considered in  that paper were the terms in the 3-graviton
interaction potential proportional to $v_3^4/r_3^7$.  Considering the
form discussed above for the components of the matrix stress tensor,
it is clear that such terms only arise in the part of the interaction
potential corresponding to
\begin{equation}
\frac{ v_3^4 \langle T^{+ +}_{1+2}\rangle}{r_3^7}.
\label{eq:v4-term}
\end{equation}
But the stress tensor component
\[
\langle T^{+ +}_{1 + 2}\rangle= \frac{N_1 + N_2}{ R} 
\]
is a constant which suffers no quantum corrections in matrix theory.
This is a conserved charge: the total longitudinal momentum of the 1-2
system, and is responsible for the long-range component $h^{+ +}$ of
the metric.  It is therefore easy to see that this term is correctly
reproduced by matrix theory.  The terms corresponding to other powers
of $v$ are more complicated, however, as the relevant components of
$T_{1 + 2}$ are corrected by quantum effects.

In addition to the practical complications of the calculation, there
are several conceptual subtleties in using the approach we have just
described to making a concrete correspondence between the matrix
theory and supergravity descriptions of a general 3-body interaction process.
The first subtlety arises, as was pointed out by Okawa and Yoneya in
\cite{Okawa-Yoneya}, from the fact that the complete gravity action is not
simply the probe-source term  (\ref{eq:general-3-expansion}), but also
contains terms cubic in the gravitational field $(\int h^3)$.  These
terms have a more complicated structure than the simple probe-source
terms considered above, and  it is more complicated to relate them to
the results of the matrix theory calculation.  The second subtlety
which arises is that the precise choice of gauge made in the matrix
theory calculation has a very strong impact on the form of the
expressions found in the resulting effective action.  Of course, for
any physical quantity such as an S-matrix element, the result of a
complete calculation will be independent of gauge choice.
Nonetheless, to compare terms in the fashion we are suggested here
will require a careful choice of gauge in matrix theory to match the
appropriate gauge chosen in the supergravity theory.  From this point
of view, it is somewhat remarkable that in the calculation of the
leading-order terms the natural gauge choices in the two theories
(background field gauge in matrix theory and linearized gauge in
supergravity) give rise to results which can be easily compared.

In any case, one might hope to navigate through these complications in
the general 3-body problem, although this clearly would involve a
substantial amount of work.  In a very impressive pair of papers by
Okawa and Yoneya \cite{Okawa-Yoneya,Okawa-Yoneya-2} (see also the more
recent work \cite{hpsw}), the full S-matrix
calculation was carried out for the interaction between 3 gravitons in
both matrix theory and in supergravity, and it was shown that there
was a precise agreement between all terms.  Unlike other work on this
problem, Okawa and Yoneya did not use the double expansion to simplify
the problem but simply carried out the complete calculation.

One would naturally like to extend these results beyond the 3-body
problem to the general N-body problem.  The hierarchy of scales
leading to the double expansion discussed above can be generalized, so
that one has a different scale for each distance in the problem.  This
organizes the large number of terms in the N-body interaction into a
more manageable structure.  To date, however there has been very
little work done on the problem of understanding higher order
nonlinearities in the theory beyond those involved in the 3-body problem.

A very intriguing paper by Dine, Echols and Gray \cite{deg2} attempts
to find a matrix-supergravity correspondence for some special terms in
the general N-body interaction potential.  Although they find that
some terms agree, they also find some terms which appear in the matrix
theory potential which have the wrong scaling behavior to correspond
to supergravity terms.  We briefly describe these terms here in the
language we have been using of stress tensor components.

For a 3-graviton system the term (\ref{eq:v4-term}) is associated with
an infinite series of higher-moment terms, as described in subsection
\ref{sec:2-body-general} and in more detail in
\cite{Mark-Wati,Dan-Wati-2}.  The first of these higher moment terms is
\begin{equation}
v_3^4 \langle T^{+ + (ij)} \rangle_{12} \partial_i \partial_j
\frac{1}{r_3^7} 
\label{eq:v-moments}
\end{equation}
This expectation value is given by
\[
\langle X^i X^j \rangle \sim \frac{\delta_{ij}}{r_2}  +
\frac{v_2^iv_2^j + \delta^{ij}v^2_2}{r_2^5}  + \cdots
\]
The contribution to (\ref{eq:v-moments}) from the first delta function
vanishes since $\partial^2 r^{-7} = 0$ away from the origin in the
9-dimensional transverse space.  The second term gives rise to a term
in the 3-body potential of the form
\[
V_a \sim \frac{v_3^4 (v_2 \cdot \partial)^2}{ r_2^5}  \frac{1}{r_3^7} 
\]
Dine, Echols and Gray argue that such a term should also be found in
supergravity, giving an example of an agreement between two-loop
matrix theory and tree level supergravity in the $U(3)$ theory
at order $v^6/r^{14}$.  This argument can be
repeated by taking a higher moment of this term in a 4-body system
\[
V_a \sim v_4^4 (v_3 \cdot \partial)^2  \frac{1}{r_4^7} 
\langle X^i X^j \rangle_{12} \partial_i \partial_j \frac{1}{r_3^5} 
\]
This time, however, the first term in the expectation value does not
give 0, so that matrix theory predicts a term of the form
\[
V_a \sim v_4^4 \left( (v_3 \cdot \partial)^2  \frac{1}{r_4^7}  \right)
\left(\partial^2 \frac{1}{r_3^5}  \right) \frac{1}{r_2} 
\]
As argued by Dine, Echols and Gray, this term has the wrong scaling to
correspond to a classical supergravity interaction.  Indeed, this term
is of the form $v^6/r^{17}$, corresponding to a term ``below the
diagonal'', which is expected to vanish.

The appearance of this term in the matrix theory perturbation series
is troubling.  It seems to indicate that there may be a breakdown of
the correspondence between matrix theory and even classical
supergravity.  This is the first concrete calculation where the two
perturbative expansions have been shown to contain terms which may
disagree.  On the other hand, there are subtleties in this calculation
which may need be resolved.  For one thing, there are the issues of
gauge choices mentioned above.  This calculation implicitly assumes a
gauge which may not be appropriate for comparison to the 4-body
interaction terms being considered in supergravity.  There are also
issues of infrared divergences which may lead to unexpected
cancellations.  In any case, clearly more work is needed to determine
whether this indeed represents a breakdown of the relationship between
matrix theory and classical supergravity which works so well for lower
order terms.

\subsection{Longitudinal momentum transfer}
\label{sec:longitudinal}

In this section we have so far concentrated on interactions in matrix
theory and supergravity where no longitudinal momentum is transferred
from one object to another.  A supergravity process in which
longitudinal momentum is transferred from one object to another is
described in the IIA theory by a process where one or more D0-branes
are exchanged between coherent states consisting of clumps of
D0-branes.  Such processes are exponentially suppressed since the
D0-branes are massive, and thus are not relevant for the expansion of
the effective potential in terms of $1/r$ which we have been
discussing.  In the matrix theory picture, this type of exponentially
suppressed process can only appear from nonperturbative effects.
Clearly, however, for a full understanding of interactions in Matrix
theory it will be necessary to study processes with longitudinal
momentum transfer in detail and to show that they also correspond
correctly with processes in supergravity and M-theory.  Some progress
has been made in this direction.  Polchinski and Pouliot have
calculated the scattering amplitude for two 2-branes for processes in
which a 0-brane is transferred from one 2-brane to the other
\cite{Polchinski-Pouliot}.  In the Yang-Mills picture on the
world-volume of the 2-branes, the incoming and outgoing configurations
in this calculation are described in terms of an $U(2)$ gauge theory
with a scalar field taking a VEV which separates the branes.  The
transfer of a 0-brane corresponds to an instanton-like process where a
unit of flux is transferred from one brane to the other.  The
amplitude for this process was computed by Polchinski and Pouliot and
shown to be in agreement with expectations from supergravity.  This
result suggests that processes involving longitudinal momentum
transfer may be correctly described in Matrix theory.  It should be
noted, however, that the Polchinski-Pouliot calculation is not
precisely a calculation of membrane scattering with longitudinal
momentum transfer in Matrix theory since it is carried out in the
2-brane gauge theory language.  In the T-dual Matrix theory picture
the process in question corresponds to a scattering of 0-branes in a
toroidally compactified space-time with the transfer of membrane
charge.  Processes with 0-brane transfer and the relationship between
these processes and graviton scattering in matrix theory have been
studied further in \cite{dkm,BFSS2,hks,Esko-Per-momentum}.

\section{Matrix theory in a general background}
\setcounter{equation}{0}
\label{sec:matrix-general-background}

So far we have only discussed matrix theory as a description of
M-theory in infinite flat space.
In this section we consider the possibility of extending the theory
to compact and curved spaces.  
As a preliminary to the discussion of compactification, we give an explicit
description of T-duality in gauge theory language in subsection
\ref{sec:T-duality}.  We then discuss the
compactification of the theory on tori in subsection \ref{sec:tori}.
Following the discussion of matrix theory
compactification, we turn in subsection \ref{sec:curved-background} to
the problem of using matrix theory methods to describe M-theory in a
curved background space-time.

\subsection{T-duality}
\label{sec:T-duality}

In this subsection we briefly review how T-duality may be understood
from the point of view of super Yang-Mills theory.  For more details
see \cite{WT-compact,WT-Trieste}.

In string theory, T-duality is a symmetry which relates the type IIA
theory compactified on a circle of radius $R_9$ with type IIB theory
compactified on a circle with dual radius $\hat{R}_9 = \alpha'/R_9$.
In the perturbative type II string theory, T-duality exchanges winding
and momentum modes of the closed string around the compact direction.
For open strings, Dirichlet and Neumann boundary conditions are
exchanged by T-duality, so that Dirichlet $p$-branes are mapped under
T-duality to Dirichlet $(p \pm 1)$-branes \cite{dlp}.

\begin{figure}
\begin{center}
\epsfig{figure=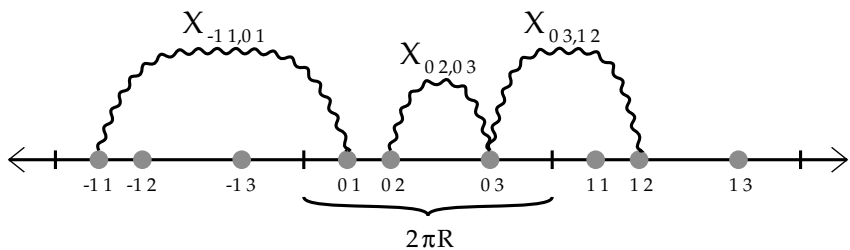, width=0.8\textwidth}
\end{center}
\caption{D0-branes on the cover of  $S^1$ are indexed
by two integers\label{fig:cover}}
\end{figure}

It was argued by Witten \cite{Witten-bound} that the low-energy theory
describing a system of $N$  parallel D$p$-branes in flat space is the
dimensional reduction of ${\cal N} = 1$, $(9 + 1)$-dimensional super
Yang-Mills theory to $p + 1$ dimensions.  In the case of $N$
D0-branes, this gives the Lagrangian (\ref{eq:D0-Lagrangian}).  To
understand T-duality from the point of view of this low-energy field
theory, we consider the simplest case of $N$ D0-branes moving in a
space which has been compactified in a single direction by identifying
\begin{equation}
x^9 \approx x^9 + 2 \pi R^9.
\label{eq:compact-quotient}
\end{equation}
To interpret this equivalence in terms of the matrix degrees of
freedom of the D0-branes it is natural to pass to the covering space
$\br^{9, 1}$, where the $N$ D0-branes are each represented by an
infinite number of copies labeled by integers $n \in \bz$.  We can
thus describe the dynamics of $N$ D0-branes on $\br^{8, 1}\times S^1$
by a set of infinite matrices $M^i_{ma, nb}$ where $a, b \in\{1,
\ldots, N\}$ are $U(N)$ indices and $m, n \in\bz$ index copies of each
D0-brane which differ by translation in the covering space (See
Figure~\ref{fig:cover}).  In terms of this set of infinite matrices, the
quotient condition (\ref{eq:compact-quotient}) becomes a set of
constraints on the allowed matrices which can be written (dropping the
$U(N)$ indices $a, b$) as
\begin{eqnarray}
X^i_{mn} & = & X^i_{(m-1)(n-1)},\;\;\;\;\;  i < 9 \nonumber\\
X^9_{mn} & = & X^9_{(m-1)(n-1)},\;\;\;\;\;  m\neq n\label{eq:constraints}\\
X^9_{nn} & = & 2 \pi R_9 \identity +X^9_{(n-1)(n-1)}\ . \nonumber
\end{eqnarray}

From the structure of the constraints (\ref{eq:constraints})  it is
natural to interpret the matrices $X^i_{mn}$ in terms of the $(n-m)$th
Fourier modes of a theory on the dual circle.  The infinite matrix
$X^9$ becomes a covariant derivative operator
\begin{equation}
X^9 \rightarrow (2 \pi \alpha') (i \partial_9 + A_9)
\label{eq:T-duality}
\end{equation}
in a $U(N)$ Yang-Mills theory on the dual torus, while $X^i$ for $i < 9$
becomes  an adjoint scalar field.  The fermionic fields in the theory
can be interpreted similarly.

This gives a precise equivalence between the low-energy world-volume
theory of a system of $N$ D0-branes on $S^1$ and a system of $N$
D1-branes on the dual circle.  The relationship between winding
modes $X^i_{mn}$ in the D0-brane theory and modes with $n-m$ units of
momentum in the dual theory  corresponds precisely to the mapping from
winding to momentum modes in the closed string theory under T-duality.

This argument can easily be generalized to a system of multiple
D$p$-branes transverse to a torus $T^d$, which are equivalent to a
system of wrapped D$(p + d)$-branes on the dual torus.  When we
compactify in multiple dimensions,  the possibility arises of having a
topologically nontrivial gauge field configuration on the dual torus.
To discuss this possibility it is useful to use a slightly more
abstract language to describe the T-duality.

The constraints (\ref{eq:constraints}) can be formulated by saying
that there exists a translation operator $U$ under which the infinite
matrices $X^a$  transform as
\begin{equation}
U X^a U^{-1} = X^a + \delta^{a9} 2 \pi R_9\identity.
\label{eq:constraint2}
\end{equation}
This relation is satisfied formally by the operators
\begin{equation}
X^9 = i \partial_9 + A_9, \;\;\;\;\; \;\;\;\;\;
U = e^{2 \pi i\hat{x}^9 R_9}
\end{equation}
which correspond to the solutions discussed above.  In this
formulation of the quotient theory, the operator $U$   generates the
group $\Gamma =\bz$ of covering space transformations.  Generally,
when we take a quotient theory of this type, however, there is a more
general constraint which can be satisfied.  Namely, the translation
operator only needs to preserve the state up to a gauge
transformation.  Thus, we can consider the more general constraint
\begin{equation}
U X^a U^{-1} = \Omega (X^a + \delta^{a9} 2 \pi R_9\identity) \Omega^{-1}.
\label{eq:generalconstraint}
\end{equation}
where $\Omega \in U(N)$ is an arbitrary element of the gauge group.
This relation is satisfied formally by
\begin{equation}
X^9 = i \partial^9 + A_9, \;\;\;\;\; \;\;\;\;\;
U = \Omega \cdot e^{2 \pi i\hat{x}^9 R_9} 
\end{equation}
This is precisely the same type of solution as we have above; however,
there is the additional feature that the translation operator now
includes a nontrivial gauge transformation.  On the dual circle
$\hat{S}^1$ this corresponds to a gauge theory on a bundle with a
nontrivial boundary condition in the compact direction 9.  

A similar story occurs when several directions are compact.  In this
case, however, there is a constraint on the translation operators in
the different compact directions.  For example, if we have
compactified on a 2-torus in dimensions 8 and 9, the generators $U_8$
and $U_9$ of a general twisted sector must generate a group isomorphic
to $\bz^2$ and therefore must commute.  The condition that these
generators commute can be related to the condition that the boundary
conditions in the dual gauge theory correspond to a well-defined
$U(N)$ bundle over the dual torus.  For compactifications in more than
one dimension such boundary conditions can define a topologically
nontrivial bundle.  It is interesting to note that this construction
can even be generalized to situations where the generators $U_i$ do
not commute.  Physically, such a configuration is produced when there
is a background NS-NS $B$ field.  This construction leads to a dual
theory which is described by gauge theory on a noncommutative torus
\cite{cds,Douglas-Hull,hww}.  A description of this scenario along the
lines of the preceding discussion is given in \cite{Cheung-Krogh}.
The connection between nontrivial background field configurations and
noncommutative geometry has been a subject of much recent interest
\cite{Seiberg-Witten-noncommutative}.

\subsection{Matrix theory on tori}
\label{sec:tori}

{}From the discussion in the previous section, it follows that the matrix
theory description of M-theory compactified on a torus $T^d$ becomes
$(d + 1)$-dimensional super Yang-Mills theory.  The argument of
Seiberg and Sen in \cite{Seiberg-DLCQ,Sen-DLCQ} is valid in this
situation, so that $U(N)$ super Yang-Mills theory on $(T^d)^*$ should
describe M-theory compactified on $T^d$.  When $d \leq 3$ the quantum
super Yang-Mills theory is renormalizable so this is a sensible way to
approach the theory.  As the dimension of the torus increases,
however, the matrix description of the theory develops more and more
complications.  In general, the super Yang-Mills theory on the
$d$-torus encodes the full U-duality symmetry group of M-theory on
$T^d$ in a rather nontrivial fashion.

Compactification of the theory on a two-torus was discussed
by Sethi and Susskind \cite{Sethi-Susskind}.  They pointed out that
as the $T^2$ shrinks, a new dimension appears whose quantized momentum
modes correspond to magnetic flux on the $T^2$.  In the limit where
the area of the torus goes to 0, an $O (8)$ symmetry appears.  This
corresponds with the fact that IIB string theory appears as a limit of
M-theory on a small 2-torus
\cite{Aspinwall-duality,Schwarz-multiplet}.  

Compactification of the theory on a three-torus was discussed in
\cite{Susskind-duality,grt}.  In this case, M-theory on $T^3$ is
equivalent to $(3 + 1)$-dimensional super Yang-Mills theory on a
torus.  This theory is conformal and finite.  M-theory on $T^3$ has a
special type of T-duality symmetry under which all three dimensions of
the torus are inverted.  In the matrix description this is encoded in
the Montanen-Olive S-duality of the 4D super Yang-Mills theory.

When more than three dimensions are toroidally compactified, the
theory undergoes even more remarkable transformations \cite{fhrs}.
When compactified on $T^4$, the manifest symmetry group of the theory
is $SL(4,Z)$.  The expected U-duality group of M-theory compactified
on $T^4$ is $SL(5,Z)$, however.  It was pointed out by Rozali
\cite{Rozali} that the U-duality group can be completed by
interpreting instantons on $T^4$ as momentum states in a fifth compact
dimension.  This means that Matrix theory on $T^4$ is most naturally
described in terms of a (5 + 1)-dimensional theory with a chiral $(2,
0)$ supersymmetry.  This unusual $(2, 0)$ theory with 16
supersymmetries \cite{Seiberg-16} appears to play a crucial role in
numerous aspects of the physics of M-theory and 5-branes, and has been
studied extensively in recent years.

Compactification on tori of higher dimensions than four continues to
lead to more complicated situations, particularly when one gets to
$T^6$, when the matrix theory description seems to be as complicated
as the original M-theory.  A significant amount of literature has been
produced on this subject, to which the reader is referred to further
details (see \cite{Obers-Pioline,Banks-TASI} for reviews and further
references).  Despite the complexity of $T^6$ compactification,
however, it was suggested by Kachru, Lawrence and Silverstein
\cite{kls} that compactification of Matrix theory on a more general
Calabi-Yau 3-fold might actually lead to a simpler theory than that
resulting from compactification on $T^6$.  If this speculation is
correct and a more explicit description of the theory on a Calabi-Yau
compactification could be found, it might make matrix theory a
possible approach for studying realistic 4D phenomenology.

\subsection{Matrix theory in curved backgrounds}
\label{sec:curved-background}

We now consider matrix theory in a space which is infinite but may be
curved or have other nontrivial background fields.  We would like to
generalize the matrix theory action to one which includes a general
supergravity background given by a metric tensor, 3-form field, and
gravitino field which together satisfy the equations of motion of 11D
supergravity.  This issue has been discussed in
\cite{dos,Douglas-curved,Douglas-curved-2,dko,Seiberg-DLCQ,Douglas-Ooguri,Lif-back,Mark-Wati-3}.
In \cite{Seiberg-DLCQ} it was argued that light-front M-theory on an
arbitrary compact or non-compact manifold should be reproduced by the
low-energy D0-brane action on the same compact manifold; no explicit
description of this low-energy theory was given, however.  In
\cite{dko} an explicit prescription was given for the first few terms
of a matrix theory action on a general K\"ahler 3-fold which agreed
with a general set of axioms proposed in \cite{Douglas-curved-2}.  In
\cite{dos} and \cite{Douglas-Ooguri}, however, it was argued that no
finite $N$ matrix theory action could correctly reproduce physics on a
large K3 surface.  We review here an explicit proposal for a
formulation of matrix theory in an arbitrary background geometry
originally presented in \cite{Mark-Wati-3}.

If we assume that matrix theory is a correct description of M-theory
around a flat background,  then there is a large class of curved backgrounds
for which we know it is possible to construct a matrix theory action
for $N \times N$ matrices.  This is the class of backgrounds which can
be produced as long-range fields produced by some other supergravity matter
configuration with a known description in matrix theory.  Imagine that
a background metric $g_{IJ}= \eta_{IJ} + h_{IJ}$, a 3-form field $A_{IJK}$ and
a gravitino field $\psi_I$ of light-front compactified 11-dimensional
supergravity can be produced by a matter configuration described in
matrix theory by matrices $\tilde{X}^i$.  Then the matrix theory action
describing $N \times N$ matrices $X^i$
in this background should be
precisely the effective action found by considering the block-diagonal
matrix configuration
\[
X^i = \left[\begin{array}{cc}
{X}^i & 0\\
0 & \tilde{X}^i
\end{array}\right]
\]
(and a similar fermion configuration) and integrating out the
off-diagonal fields as well as fluctuations around the background
$\tilde{X}$.

{}From the results found in  \cite{Dan-Wati-2,Mark-Wati-3},
we know that for
weak background fields, the first few terms in an expansion of this
effective action in the background metric are given by
\begin{eqnarray}
S_{\rm eff} & = &  S_{{\rm matrix}}  +\int dx \; T^{IJ} (x) h_{IJ} (x) 
+ \cdots\label{eq:first}\\
&=  &  S_{{\rm matrix}}  +\int
dx^+\{T^{IJ} h_{IJ}(0) + T^{IJ(i)}
\partial_i h_{IJ}(0)  + \cdots \} + \cdots
\nonumber
\end{eqnarray}
where $T^{IJ (\cdots)}$ are the moments of the matrix theory
stress-energy tensor, and there are analogous terms for the coupling
of the membrane, 5-brane and fermionic components of the supercurrent
to $A_{IJK}$ and $S_I$.  If the standard formulation of matrix theory
in a flat background is correct, the absence of corrections to the
long-range $1/r^7$ potential around an arbitrary matrix theory object
up to at least order $1/r^{11}$ implies that this formulation must be
correct at least up to terms of order $\partial^4 h$ and $h^2$.  

As we have derived it, this formulation of the effective action is
only valid for certain background geometries which can be produced by
well-defined matrix theory configurations.  It is natural, however, to
suppose that this result can be generalized to an arbitrary
background.  Thus, it is proposed in \cite{Mark-Wati-3} that up to
nonlinear terms in the background, the general form of the matrix
theory action in an arbitrary but weak background is given by
\begin{eqnarray}
S_{\rm weak}  & = & \int d \tau \;
\sum_{n = 0}^{\infty}  \sum_{i_1, \ldots, i_n}\frac{1}{n!} 
\left(
T^{IJ (i_1 \cdots i_n)}  \partial_{i_1} \cdots \partial_{i_n}  h_{IJ}
\right.\nonumber\\
& &\hspace{1.2in} \left.
+J^{IJK (i_1 \cdots
i_n)}
 \partial_{i_1} \cdots \partial_{i_n} A_{IJK} 
\right. \label{eq:general-background}\\
& &\hspace{1.2in} \left.
+M^{IJKLMN (i_1
\cdots i_n)}  \partial_{i_1} \cdots \partial_{i_n}  A^D_{IJKLMN} 
\right.\nonumber\\
& &\hspace{1.2in} \left.
+ {\rm fermion \; terms} \right) \nonumber
\end{eqnarray}
Let us make several comments about this action.  First, this
formulation is only appropriate for backgrounds with no explicit $x^-$
dependence, as we do not understand how to encode higher modes in the
compact direction in the components of the supergravity currents.
Second, note that the coupling to $A^D$ is free of ambiguity since the
net 5-brane charge must vanish for any finite matrices, so that only
first and higher derivatives of $A^D$ appear in the action.  Third,
note that though we only have explicit expressions for the fermion
terms in the zeroeth and
some of the first moments of $T$, we may in principle generalize the
calculations of \cite{Dan-Wati-2,Mark-Wati-3} to determine all the
fermionic contributions from higher order terms in the one loop matrix
theory potential.

The linearized couplings in the action (\ref{eq:general-background})
are motivated by the results of one-loop calculations in matrix
theory.  In principle, it may be possible to extend the formulation of
matrix theory in weak background fields to higher order by performing
general higher-loop calculations in matrix theory.  For example, a
complete description of the 2-loop interaction in matrix theory
between an arbitrary 3 background configurations would suggest the
form of the coupling between one object considered as a probe and the
quadratic terms in the background produced by the other pair of
objects.  Generally, knowing the full $n$-loop interaction between $n
+ 1$ matrix theory objects would suggest the $n$th order coupling of
the matrix degrees of freedom to the background fields.
Unfortunately, as we have discussed such calculations are
rather complicated.  In addition to the technical difficulties of
doing the general 2-loop calculation, there are subtleties related to
the gauge choice and possible infrared divergences.  Furthermore,
finite $N$ calculations will only help us to learn the higher-order
couplings to the background if the results of these calculations are
protected by supersymmetric nonrenormalization theorems, and as we
have discussed there is no strong reason to believe that such
nonrenormalization theorems hold for the general $n$-loop $SU(N)$ calculation.
Thus, to write a completely general coupling of matrix theory to a
nontrivial supergravity background, it is probably necessary to find a
new general principle, such as a matrix version of the principle of
coordinate invariance.

Another approach which one might take to define matrix theory in a
general background geometry is to follow the original derivation of
matrix theory as a regularized membrane theory, but to include a
general background geometry instead of a flat background as was used
in \cite{Goldstone-Hoppe,dhn}.  The superspace formulation of a
supermembrane theory in a general 11D supergravity background was
given in \cite{bst}.  In principle, it should be possible to simply
apply the matrix regularization procedure to this theory to derive
matrix theory in a general background geometry.  Unfortunately,
however, the connection between superspace fields and component fields
is not well-understood in this theory.  Until recently, in fact, the
explicit expressions for the superspace fields were only known up to
first order in the component fermion fields $\theta$
\cite{Cremmer-Ferrara}.  In \cite{dpp}, this analysis was extended to
quadratic order in $\theta$ with the goal of finding an explicit
formulation of the supermembrane in general backgrounds in terms of
component fields, to which the matrix regulation procedure could be
applied to generate a general background formulation of matrix theory.
These results can be compared with the proposal just described for the
linear couplings to the background.  The two formulations seem to be
completely compatible \cite{Millar-Taylor}, although extra terms appear
in the matrix theory action which cannot be predicted from the form of
the continuous membrane theory.

In \cite{Douglas-curved-2}, Douglas proposed that any formulation of
matrix theory in a curved background should satisfy a number of
axioms.  All these axioms are satisfied in a straightforward fashion
by the proposal in \cite{Mark-Wati-3}, except one: this exception is
the axiom that states that a pair of D0-branes at points $x^i$ and
$y^i$ should correspond to diagonal $2 \times 2$ matrices where the
masses of the off-diagonal fields should be equal to the geodesic
distance between the points $x^i$ and $y^i$ in the given background
metric.  In \cite{Mark-Wati-4} it was shown that the linearized terms
in the action (\ref{eq:general-background}) are consistent with this
condition and that the linear variation in geodesic distance between a
pair of D0-branes is correctly reproduced by coupling the matrix
theory stress tensor to the background metric through a 
combinatorial identity which follows from the particular ordering
implied by the symmetrized trace form of the multipole moments of the
stress tensor.  The fact that this condition can be satisfied at
linear order provides hope that it might be possible to extend the
action to all orders in a consistent way.  In \cite{dko}, it was indeed shown
by Douglas, Kato and Ooguri that a set of some higher order terms for
the action on a Ricci-flat
K\"ahler manifold can be found which are consistent with the geodesic
length condition, but these authors also found that this condition did
not uniquely determine most of the terms in the action so that a more
general principle is still needed to construct the action to all orders.

We synopsize the discussion in this section as follows:
(\ref{eq:general-background}) seems to be a consistent proposal for
the linearized couplings between matrix theory and weak supergravity
background fields.  The expressions for the higher moments of the
supergravity currents which couple to the derivatives of the
background fields are known up to terms quadratic in the fermions, and
the remaining terms can be found from a one-loop matrix theory
computation.  This proposal can be generalized to $m$th order in the
background fields, where matrix expressions are needed for quantities
which can be determined from an $m$-loop matrix theory calculation.
Whether these terms can be calculated and sensibly organized into
higher-order couplings of matrix theory to background fields depends
on whether higher-loop matrix theory results are protected by
supersymmetric nonrenormalization theorems.  It is worth
emphasizing that the definitions of the matrix theory currents we have
described here depend upon gauge choices for the propagating
supergravity fields.  For a given gauge choice, the theory is only
defined for backgrounds compatible with the gauge condition.  Making
the appropriate gauge choices represents another obstacle to carrying
out this analysis to higher order.

\section{Outlook}
\setcounter{equation}{0}
\label{sec:matrix-summary}

We conclude with a brief review of the connection between matrix
theory and M-theory, and a short discussion of the current state of
affairs and the outlook for further developments in matrix theory.

We have discussed two complementary ways of thinking about matrix
theory: first as a quantized regularized theory of a supermembrane,
which naturally describes a second-quantized theory of objects moving
in an 11-dimensional target space, and second as the DLCQ of M-theory
which is equivalent to a simple limit of type IIA string theory
through the Seiberg-Sen limiting argument.

Using matrix degrees of freedom, it is possible to describe pointlike
objects which have many of the physical properties of supergravitons.
It is also possible to use the matrix degrees of freedom to describe
extended objects which behave like the supermembrane and 5-brane of
M-theory.  For supergravitons and membranes this story seems fairly
complete; for 5-branes, only a few very special geometries have been
described in matrix language, and a complete description of dynamical
(longitudinal) 5-branes, even at the classical level, is still
lacking.

As we have discussed, to date all perturbative calculations except the
3-loop calculation of Dine, Echols and Gray indicate that matrix
theory correctly reproduces classical 11D supergravity.  It has been
suggested that the agreement between the theories at 1-loop and 2-loop
orders is essentially an accident of supersymmetry, however there is
little understanding of how to interpret or organize higher-loop
terms.  There is also very little understanding at this point of how
quantum corrections to the supergravity theory can be understood in
terms of matrix theory, although there is evidence
\cite{Esko-Per-short,Beckers-graviton,hpsw} that quantum gravity
effects are not reproduced by perturbative calculations in matrix
theory but will require a better understanding of the large $N$ limit
of the theory.

At
this point there are essentially 4 possible scenarios for the validity
of the matrix theory conjecture:

\noindent {\bf i)} Matrix theory is correct, and DLCQ supergravity is
reproduced at finite $N$ by perturbative matrix theory calculations.

\noindent {\bf ii)} Matrix theory is correct in the large $N$ limit,
and noncompact supergravity is reproduced by a naive large $N$ limit
of the standard perturbative matrix theory calculations.

\noindent {\bf iii)} Matrix theory is correct in the large $N$ limit,
but to connect it with supergravity, even at the classical level, it
is necessary to deal with subtleties in the large $N$ limit.  (i.e.,
there are problems with the standard perturbative analysis at higher
order)

\noindent {\bf iv)} Matrix theory is simply wrong, and further terms
need to be added to the dimensionally reduced super Yang-Mills action
to find agreement with M-theory even in the large $N$ limit.

\vspace{0.1in}

Now let us examine the evidence:

\noindent $\bullet$
The breakdown of the Equivalence Principle seems incompatible with
(i), but compatible with all other possibilities.

\noindent $\bullet$
If the result of Dine, Echols and Gray in \cite{deg2} is correct, and
has been correctly interpreted, clearly (i) and (ii) are not possible.
The fact that the methods of Paban, Sethi and Stern for proving
nonrenormalization theorems  in the $SU(2)$ theory break down for
$SU(3)$ at two loops and at higher loop order \cite{Sethi-Stern-2}
also hints that (ii) may not be correct.

\noindent $\bullet$
The analysis of Seiberg and Sen seems to indicate that
one of the possibilities (i)-(iii) should hold.

It seems that (iii) is the most likely possibility, given this limited
evidence.  There are several issues which are extremely important in
understanding how this problem will be resolved.  The first is the
issue of Lorentz invariance.  If a theory contains linearized gravity
and is Lorentz invariant, then it is well known that it must be either
the complete generally covariant gravity theory or just the pure
linearized theory.  Since we know that matrix theory has some
nontrivial nonlinear structure which reproduces part of the
nonlinearity of supergravity, it would seem that the conjecture must
be valid if and only if the theory is Lorentz invariant.
Unfortunately, so far there is no complete understanding of whether
the quantum theory is Lorentz invariant (classical Lorentz invariance
was demonstrated in \cite{dmn}).  It was suggested by Lowe in
\cite{Lowe-constraints} that the problems found in \cite{deg2} might
be related to a breakdown of Lorentz invariance and that in fact extra
terms must be added to the theory to restore this invariance; this
would lead to possibility (iv) above.

Another critical issue in understanding how the perturbative matrix
theory calculations should be interpreted is the issue of the order of
limits.  In the perturbative calculations discussed here we have
assumed that the longitudinal momentum parameter $N$ is fixed for each
of the objects we are taking as a background, and we have then taken the
limit of large separations between each of the objects.  Since the
size of the wavefunction describing a given matrix theory object will
depend on $N$ but not on the separation from a distant object, this
gives a systematic approximation scheme in which the bound state and
wavefunction effects for each of the bodies can be ignored in the
perturbative analysis.  If we really are interested in the large $N$
theory, however, the correct order of limits to take is the opposite.
We should fix a separation distance $r$ and then take the large $N$
limit.  Unfortunately, in this limit we have no systematic
approximation scheme.  The wavefunctions for each of the objects
overlap significantly as the size of the objects grows.  Indeed, it
was argued recently by Polchinski \cite{Polchinski-M-theory} that the
size of the bound state wavefunction of $N$ D0-branes will grow at
least as fast as $N^{1/3}$.  As emphasized by Susskind in
\cite{Susskind-flat}, this overlap of wavefunctions makes the theory
very difficult to analyze.  Indeed, if possibility (iii) above is
correct, it may be very difficult to use matrix theory to reproduce
all the nonlinear structure of classical supergravity, let alone to
derive new results about quantum supergravity.  On the other hand, it
may be that whatever mechanism allows the one-loop and two-loop matrix
theory results to correctly reproduce the first few terms in
supergravity and to evade the problem of wavefunction overlap may
persist at higher orders.  Indeed, one of the must important
outstanding questions regarding matrix theory is to understand
precisely which terms in the naive perturbative expansion of the
quantum mechanics will agree with classical supergravity, and more
importantly, {\it why} these terms agree.  As mentioned in the last
section, one of the other main outstanding problems in matrix theory
is understanding how the matrix quantum mechanics theory behaves when
M-theory is compactified on a curved manifold.  In order to use matrix
theory to make new statements about corrections to classical
supergravity in phenomenologically interesting models such as M-theory
on compact 7-manifolds or orbifolds, it will be necessary to solve
both of these problems.  In each case, a certain amount of luck will
be needed for it to be possible to probe physically interesting
questions using existing computational techniques.

In these lectures we have focused on understanding some basic aspects
of matrix theory: the definitions of the theory in terms of the
membrane and DLCQ of M-theory, and the construction of the objects and
supergravity interactions of M-theory using matrix degrees of
freedom.  We conclude with a few brief words about some of the topics
we have not discussed.  

In addition to the matrix model of M-theory,
there have been numerous related models suggested in the literature in
the last few years.  Some of these which have received particular
attention are the $(0 + 0)$-dimensional matrix model of  IIB string
theory suggested by Ishibashi, Kawai, Kitazawa and Tsuchiya
\cite{IKKT}, the $(1 + 1)$-dimensional matrix string theory of
Dijkgraaf, Verlinde and Verlinde \cite{DVV} and the family of AdS/CFT
conjectures proposed by Maldacena \cite{Maldacena-AdS}.
All these proposals relate a particular limit of string theory or
M-theory in a fixed background to a field theory.  Many connections
between these models have been made, and in fact most of these
proposals are related by a duality symmetry to the matrix theory we
have discussed here.  A fundamental question at this point, however,
is how we may move away from a fixed background and discuss questions
of cosmological significance.

Even within the framework of the matrix model of M-theory we have
discussed in these lectures, there are many very interesting
directions and particular applications which have been pursued which
we did not have time to review here in any detail.  These include
questions about black holes in matrix theory (see, {\it e.g.},
\cite{bfks,Kabat-Lifschytz} and references therein), higher
dimensional compactifications and the matrix model of the (2, 0)
theory which arises upon compactification on $T^4$ (\cite{Rozali}, see
\cite{Banks-TASI} for a review and further references), the detailed
structure of the $N = 2$ bound state (see {\it e.g.},
\cite{Halpern-Schwartz,fghh} and references therein), and many other
directions of recent research.

In closing, it seems that matrix theory has achieved something which
just a few years ago would have been deemed virtually impossible to
accomplish in such a simple fashion: it gives a well-defined framework
for M-theory and quantum gravity which reduces any problem, at least
in light-front coordinates, to a computation which can in principle be
defined and fed into a computer.  Thus, in some sense this may be the
first concrete answer to the problem of finding a consistent theory of
quantum gravity.  Unfortunately, even though this theory is a simple
quantum mechanics theory, and not even a field theory, it is
computationally intractable at this point to ask many of the really
interesting questions about M-theory using this model.  It is clearly
a very interesting problem to try to find better ways of doing
interesting M-theory calculations using the matrix model.  But even if
matrix theory is never able to give us a computational handle on some
of the subtle aspects of M-theory, it certainly has given us a new
perspective on how to think about a microscopic theory of quantum
gravity.  One of the most interesting aspects of the matrix picture is
the appearance of dynamical higher-dimensional extended objects from a
system of ostensibly pointlike degrees of freedom, as discussed in
Section \ref{sec:matrix-objects}.  It seems likely that this feature
of matrix theory may play a key role in future attempts to describe a
more covariant or background-independent microscopic model for
M-theory, string theory or quantum gravity.

\section*{Acknowledgments}

I would like to thank L.\ Thorlacius and the other organizers of the
``Quantum Gravity'' NATO Advanced Study Institute for putting together
an excellent summer school and for inviting me to participate.  Thanks
to the students and other lecturers for providing a stimulating
atmosphere and for many interesting discussions and questions.  I
would also like to thank N.\ Prezas and M.\ Van Raamsdonk for pointing
out errors in a preliminary version of these lecture notes and
suggesting constructive changes.

This work is supported in part by the A.\ P.\ Sloan Foundation and in
part by the DOE through contract \#DE-FC02-94ER40818.  

Some of the material in these lectures was previously presented as
part of a course at MIT: physics 8.871, Fall 1998.  Some of this
material was also previously presented in lectures at the Korean
Institute for Advanced Study in October 1998 and at the Komaba '99
workshop in Tokyo, 1999.

\bibliographystyle{plain}


\end{document}